\documentclass[11pt]{article}
\PassOptionsToPackage{}{geometry}
\usepackage[a4paper,margin=3cm]{geometry}
\pdfoutput=1
\usepackage[utf8]{inputenc}
\usepackage{bm}
\usepackage{mathrsfs}
\usepackage{empheq}
\usepackage{verbatim}
\usepackage{bibentry}
\usepackage{hyperref}
\usepackage{physics}

\usepackage{slashed}
\usepackage{comment}
\usepackage{subcaption}
\usepackage{multirow}
\usepackage{tikz}
\usepackage{tikz-cd}
\usepackage{color}
\usepackage{tikz}
\usepackage{array}
\usepackage{tablefootnote}
\usepackage{latexsym}
\usepackage{amssymb}
\usepackage{amsmath}
\usepackage{mathtools}
\usepackage{stackrel}
\usepackage{youngtab}
\usepackage{cancel}
\usepackage[nosort]{cite}
\usepackage{graphicx}
\usepackage{epsfig}
\usepackage{tensor}
\usepackage{braket}
\usepackage{ulem}

\newcommand{\nin}{\noindent}
\usepackage{tocloft} 
\usepackage{ulem}
\usepackage{soul,xcolor}
\setstcolor{red}

\newcommand{\nn}{\nonumber}
\newcommand{\be}{\begin{equation}}
\newcommand{\ee}{\end{equation}}

\usepackage{bm}
\usepackage{mathrsfs}
\usepackage{empheq}
\usepackage{verbatim}
\usepackage{bibentry}
\usepackage{hyperref}
\usepackage{slashed}
\usepackage{subcaption}
\usepackage{multirow}
\usepackage{color}
\usepackage{tikz}
\usepackage{array}
\usepackage{tablefootnote}
\usepackage{latexsym}
\usepackage{float}
\usepackage{amssymb}
\usepackage{amsmath}
\usepackage{mathtools}
\usepackage{stackrel}
\usepackage{youngtab}
\usepackage{cancel}
\usepackage[nosort]{cite}
\usepackage{graphicx}
\usepackage{epsfig}
\usepackage{tensor}
\usepackage{physics}
\usepackage{hyperref}
\hypersetup{
    colorlinks=true,
    urlcolor=blue,
    linkcolor=blue,
    citecolor=blue
}

\usepackage[customcolors]{hf-tikz}
\usepackage{tikz,tikz-3dplot,tkz-fct}
\usetikzlibrary{shapes.multipart,backgrounds,fit,positioning,calc,arrows,automata,backgrounds,calendar,chains,matrix,mindmap,patterns,petri,shadows,shapes.geometric,shapes.misc,spy,trees,shadings}
\usepackage{hyperref}
\usepackage{mathrsfs}

\newcommand{\ud}{\underline}

\usepackage{ulem}

\usepackage{tocloft} 
\usepackage{ulem}
\usepackage{soul,xcolor}
\setstcolor{red}
\usepackage{dsfont}

\numberwithin{equation}{section}

\usepackage{authblk}

\title{\sf Ladder Operators and Fermionic Tensor Fields\\ on Maximally Symmetric Spaces }
\author[1,2] 
{Facundo L. Cruz$^{1,2}$,  Mat\'\i as N. Semp\'e$^{3}$
and Guillermo A. Silva}

\affil[1]
{\small\it Instituto de F\'isica de La Plata - CONICET/UNLP

Diagonal 113 e/ 63 y 64, 1900 - La Plata, Argentina}
\affil[2]
{\it Departamento de F\'isica, Universidad Nacional de La Plata  

C.C. 67, 1900 - La Plata, Argentina}

\affil[3] 
{\small \it Instituto de Astronomía y Física del Espacio - CONICET/UBA

Ciudad Universitaria, 1428 - CABA, Argentina}

\hypersetup{
    colorlinks=true,
    linkcolor=blue,
    citecolor=blue,
    urlcolor=blue,
    linktoc=page
}

\begin{document}
\maketitle

\begin{abstract}

We construct first-order ladder operators for spin-$\frac{1}{2}$ Dirac fields and transverse, $\gamma$-traceless spin-$\frac{3}{2}$ Rarita--Schwinger fields on maximally symmetric spaces using non-isometric closed conformal Killing vectors. For both spins, we find three distinct operators: two of them, $\mathcal{D}$ and $\mathcal{D}^{s}$, shift the conformal label as $\Delta \to \Delta \pm 1$, while a third operator, $\widetilde{\mathcal{D}}$, reverses the sign of the Dirac eigenvalue at fixed $\Delta$. The latter exists in arbitrary dimensions and reduces to the standard infinitesimal conformal transformation of a primary spinor when acting on massless spin-$\frac{1}{2}$ fields. On $S^N$, the ladder operators relate neighboring fermionic harmonics and generate the spinor tower from Killing-spinor seeds. In Lorentzian signature, we study their action on de Sitter mode spaces. In $dS_4$, the spin-$\frac{3}{2}$ ladders connect the zero-Dirac-mass sector with the fermionic gauge points $M=\pm i/\ell$, while $\widetilde{\mathcal{D}}$ extends to arbitrary mass the conformal-like transformation previously identified for the gauge field. We explicitly present the spin-$\frac{1}{2}$ and spin-$\frac{3}{2}$ fermionic harmonics on spheres as well as the de Sitter mode solutions. Finally, we derive the  Casimir operators on $S^3$, $dS_3$, and $dS_4$ corresponding to SO(4), SO(3,1) and SO(4,1), and relate their eigenvalues for UIRs to the allowed masses in the fermionic field equations. These results provide a unified geometric framework relating conformal Killing geometry, fermionic Dirac-type spectra, and the representation theory of maximally symmetric spaces.

\end{abstract}

\vfill

{\footnotesize
emails: ~
facundo.cruz@fisica.unlp.edu.ar,
msempe@iafe.uba.ar,
silva@fisica.unlp.edu.ar
}

\newpage

\thispagestyle{empty}
\tableofcontents
\newpage

\section{Introduction}

The importance of geometric structures in analyzing field equations can hardly be overstated. Historically, these structures have been exploited, particularly in maximally symmetric spaces, to define quantum numbers that facilitate the separation of variables. In this paper, we follow this approach by studying spin-$\frac{1}{2}$ and spin-$\frac{3}{2}$ fermionic field equations on the sphere and in de Sitter space. Our motivation stems from the central role of de Sitter space as the natural background for modeling both the present observed accelerated expansion of the Universe and early-stage inflation \cite{bou,StrVol,AnMus,bau,slei,bros,Ak,Anous2}.

As realized by Wigner, quantum physics in highly symmetric instances is naturally organized  through the representation theory of the spacetime isometry group. One-particle states are identified with unitary irreducible representations of the spacetime symmetry group and  representation theory has played a central role in the classification of relativistic quantum fields. In flat spacetime, this viewpoint leads to the familiar distinction between massive and massless particles and determines the possible spin and helicity content of the corresponding fields. More generally, the equations of motion, constraints, and gauge redundancies of a field can be understood as different realizations of the underlying representation-theoretic data. Reviews and treatments of this perspective for relativistic quantum field theory can be found in \cite{Wigner, Wein,HinterGrav, Rahman}. In de Sitter space, however, the representation theory is considerably richer. In addition to the massive and massless representations, there exist partially massless representations, which arise at special values of the mass where the corresponding de Sitter representations become reducible and acquire gauge invariances that remove only a subset of helicities \cite{Di,FR,dob,HiguPhD,Higu2,DW1,Boers,zimo,hint}. These representations   interpolate, in a representation-theoretic sense, between massive and massless sectors and provide a particularly striking connection between gauge symmetry, unitarity, and spacetime geometry (see \cite{seng,Anous1,schau,Pethy,AGS,pene,Anninos,dio+silva,Higu+lets,Boul,dS4meta,MaldadS,GiombiSun} for a flavor of recent work). Despite the extensive study of de Sitter representations, the discovery of new gauge systems and partially massless theories continues to expose structures beyond the standard massive and massless classification. In particular, higher-spin and conformally invariant gauge theories suggest that the space of consistent de Sitter representations and their field-theoretic realizations remains far from fully understood \cite{vasiliev,Basile, Barnich,AnniDenef,gaz}.

The representation-theoretic description does not merely classify fields, it also suggests explicit differential operations relating different realizations of the same underlying symmetry structure. A familiar manifestation of this idea is the use of ladder operators, which in quantum mechanics map eigenstates of an operator to those with shifted eigenvalues. Similar structures permeate the theory of differential equations and special functions, where first-order differential operators intertwine equations whose parameters differ by fixed amounts. From a field-theoretic perspective, such operators are particularly compelling because they map solutions of one field equation directly to those of another, thereby instantiating algebraic relations between representations at the level of the fields themselves \cite{perelo}.

Such structures naturally arise for fields propagating on curved backgrounds. In particular, conformal Killing vectors (CKVs) are well known to enhance the symmetries in the case of massless fields \cite{BD,Iorio,freedman}. Furthermore, part of the present authors recently showed that CKVs can be used to construct first-order differential operators relating solutions of field equations with different parameters. For integer-spin fields, ladder-operator constructions based on closed conformal Killing vectors (CCKVs) were established in maximally symmetric spacetimes \cite{LSS}. These constructions reveal a direct connection between background geometry and algebraic relations among field equations: the existence of a CCKV provides precisely the geometric structure required for a first-order operator to intertwine the relevant equations of motion. Relevant previous works are \cite{Cardo,Muck,let24,letUnconven}

The aim of the present paper is to   extend the program discussed in the previous paragraph to half-integer spin fields. Fermionic fields are governed by Dirac-type operators, and the action of a spacetime vector on a spinor cannot be obtained simply by applying the ordinary Lie derivative, an spinor lie derivative is defined instead \cite{kos,bT,FoF,ortin2,ortin} (see also \cite{bourg} for related work). Indeed, the geometry of the spin bundle enters explicitly, and the compatibility between the Dirac operator and the conformal Killing structure introduces additional terms that have no analogue for scalar or tensor fields. Consequently, although one might expect the integer-spin construction to extend straightforwardly to fermions, the resulting operators have a more subtle structure that their bosonic counterpart.

This issue is particularly relevant for higher-spin fermionic fields. The Rarita-Schwinger field provides the simplest example of a field carrying both spinorial and tensorial degrees of freedom, and  its connection to the  de Sitter spacetime representation theory has been investigated recently \cite{letsios3/2,vasilirarita}. These studies have revealed that a unitary gauge-invariant realization of the Rarita--Schwinger field exists only in four spacetime dimensions.
The works also uncovered, in four dimensions, an additional conformal-like symmetry whose generators close into an $\mathrm{SO}(4,2)$ algebra. Interestingly, however, the corresponding operator does not coincide with the standard conformal transformation of a primary spin-$\frac{3}{2}$ field. This raises the question of whether the conformal-like transformation found in four dimensions is part of a more general structure. 
In this work, we show that this is indeed the case by constructing a generalization of the conformal-like operator,  demonstrating that it is naturally related to the ladder-operator construction. In particular, we find a first-order differential map that, in arbitrary spacetime dimension, relates the solutions corresponding to the two possible signs of the Dirac eigenvalue. Moreover, in even dimensions, the operator translates into an additional symmetry that commutes with the Dirac operator and therefore preserves its eigenvalue. These results reveal a dimension-dependent structure underlying the conformal-like and ladder symmetries of half-integer spin fields in de Sitter spacetime.

The purpose of the present work is to investigate this structure systematically. We revisit and extend the construction of first-order differential operators acting on half-integer spin fields in the presence of a closed conformal Killing vector, showing that additional terms should be taken into account to construct the operator. Our aim is not only to construct the operators explicitly, but also to clarify their geometric interpretation and their relation to the corresponding field equations. In particular, we show how  CCKVs determine  differential transformations that intertwine Dirac-type equations  with different values of the mass. Concomitantly, our operators map between different representations of the exceptional series as in the bosonic case. In this sense, the operators provide a direct differential realization of the  relations between different representations of the de Sitter symmetry group.

In summary, we develop a geometric framework for first-order differential transformations of half-integer spin fields in de Sitter spacetime. Starting from the existence of closed conformal Killing vectors, we construct the corresponding ladder and conformal-like operators and determine their action on the relevant Dirac-type equations. We use this framework to clarify previous results for fermionic fields and to expose the differences between the bosonic and fermionic cases. More broadly, our results illustrate how the enhanced geometric structure of maximally symmetric spacetimes can be used to uncover non-trivial relations among quantum fields that are not manifest from their equations of motion alone.

\subsection*{Outline of the paper}

The paper is organized as follows. In Section 2, we introduce the notation and definitions for conformal primaries in conformally flat spaces, focusing   on fermionic fields. For a more detailed review of their properties, we refer the reader to \cite{LSS}.

In Section 3, we construct ladder operators for spin-$\frac{1}2$ and spin-$\frac{3}2$ fields, showing how they relate eigenfunctions of the Dirac operator with different eigenvalues. We further demonstrate that, for any mass and dimension, there exists an additional operator that relates eigenmodes with opposite-sign eigenvalues.

In Section 4, we study the action of  ladder operators on Dirac (spin-$\frac{1}{2}$) and Rarita-Schwinger (spin-$\frac{3}{2}$) UIR solutions. We present explicit examples on spheres of several dimensions. Since these operators are useful when the eigenvalues of the Dirac operator are discrete, we also discuss their action  in  Lorentzian signature  on discrete and exceptional dS$_4$ UIR modes.  We conclude by showing that the ladder operators constructed from conformal Killing vectors on $S^{3}$ provide a useful framework for understanding the transformation properties of massless spin-$\frac{1}{2}$ modes under boost symmetries in dS$_4$.

In appendix \ref{ckvPot} we review the embedding space description of curved maximally symmetric spaces. In appendices \ref{B} and \ref{spin dS}, we summarize the fermionic modes on the sphere and in the de Sitter space for both Dirac and Rarita–Schwinger fields. In Appendix \ref{casimirsS3} we calculate the two  Casimir operators of SO(4) and SO(3,1)  for spin-$\frac{1}{2}$ and $\frac{3}{2}$. This provides an explicit demonstration of the inequivalence of the UIRs associated with positive- and negative-eigenvalue spinor harmonics on odd-dimensional spheres. In Appendix \ref{apendixE}, we compute both Casimir operators of SO(4,1) for spin-$\frac{1}{2}$ fields in $dS_4$ and show that Dirac spinors with positive and negative eigenvalues furnish equivalent irreducible representations. 
For the spin-$\frac{3}{2}$ field, we compute the quadratic Casimir and comment on the structure of the quartic Casimir.

\section{Primary Fields and Conformal Flatness}

In this section, we outline our conventions for conformal primary fields on conformally flat geometries.

 \subsection{Conformally Flat Geometries}

\nin An $N$-dimensional conformally flat (CF) metric $g_{\mu\nu}$ can  be expressed in Cartesian coordinates 
$x^\mu$  as
\be
ds^2=a^2(x)dx^2=\underbrace{a^2(x)\eta_{\mu\nu}}_{g_{\mu \nu}}dx^\mu dx^\nu\,\,,
\label{met}
\ee
where $a(x)$ is a smooth positive function.

Conformal transformations (CTs) are diffeomorphisms 
\be
x \mapsto x' (x)\,\,,
\label{diff}
\ee
under which the metric transforms as
$$g_{\mu\nu}(x)\mapsto\Upsilon^2(x) g_{\mu\nu}(x)\,\,.$$
Using the form \eqref{met} for the metric, the complete set of local CTs can be written explicitly. Indeed, they coincide with the well-known conformal transformations of flat space \cite{osb,Alv}. Concretely, one finds \cite{LSS}
\be
a^2(x')dx'^2=a^2(x')\Omega(x)^2  dx^2=\Upsilon^2(x)  a^2(x) dx^2~~~\leadsto~~~\Upsilon (x)=\frac{a(x')\Omega (x)}{a(x)}\,\,.
\label{ups}
\ee
where $\Omega(x)$ denotes the flat-space conformal factor associated with the coordinate change \eqref{diff}. Hence, an Euclidean conformally flat metric admits locally the  conformal algebra so$(N + 1, 1)$, just as flat space does. On the other hand, Lorentzian conformally flat spaces possess SO$(N,2)$ conformal group.

Flat-space CTs connected to the identity take the form
$$x'^\mu(x)\approx x^\mu+\zeta^\mu(x)+...,~~~  |\zeta|\ll1$$
using that \be
g_{\mu\nu}=a^2(x)\eta_{\mu\nu}\,\,,
\label{CF}
\ee
one finds  that the vector field $\zeta_\mu:=g_{\mu\nu}\,\zeta^\nu$ satisfies \cite{Alv,LSS}
\be
\text{\sf Conformal Killing Vector:} ~~~\nabla_\mu\zeta_\nu+\nabla_\nu\zeta_\mu=\frac2N(\nabla_\rho\zeta^\rho)g_{\mu\nu}\,.
\label{CKV}
\ee
Here, as appropriate to tensor fields, $\nabla_\mu=\partial_\mu+\Gamma_\mu$ is the covariant derivative with $\Gamma_\mu=\Gamma_\mu(g)$ the Levi-Civita connection associated to the CF metric \eqref{CF}.

CKV fields define infinitesimal expressions of the conformal factor $\Upsilon(x)$ and Lorentz transformation $\Lambda^\mu{}_\nu(x)$  as \cite{LSS} (see also \cite{osb})
$$\Lambda^\mu{}_\nu(x)\approx\delta^\mu_\nu+\omega^\mu{}_\nu(x)+... ~~  ~~\text{and}~~~~
\Upsilon(x)\approx1+ \sigma(x) +...,$$
where the generators $\omega _{\mu\nu}:=g_{\mu\alpha}\,\omega^\alpha{}_\nu$ and $\sigma$ are  given  by
\begin{align}
\omega _{\mu \nu}(x) &:=\frac12(\nabla_\nu\zeta_\mu-\nabla_\mu\zeta_\nu)\,\, ,\nn\\
\sigma(x)&:=\frac1N(\nabla_\rho\zeta^\rho) \,\,.
\label{infi}
\end{align}
The finite form expressions for  $\Lambda^\mu{}_\nu(x)$ and $\Upsilon(x)$ are obtained by solving the differential equations  (2.16)-(2.18) in \cite{osb}.

\subsection{Conformal Primary Fields}

The conformal primary fields $  \Phi$ in CF geometries are characterized by the set $(\Delta,s)$ known as the scaling dimension  and  spin, respectively. These weights encapsulate the transformation properties of $  \Phi$ under CTs.  

\vspace{2mm}

\nin{\sf Dirac}: A spin-$\frac12$ conformal primary $\psi_\alpha$
transforms as
\begin{equation}
    \psi_\alpha \to \psi'_\alpha(x')=\frac{1}{\Upsilon(x)^\Delta} (e^{\frac{1}{2} \omega^{\mu \nu}  \,\frac{[\gamma_{\mu}, \gamma_\nu]}{4}})^{\ \beta}_\alpha \,\psi_\beta(x) \,\,,
    \label{Prim}
\end{equation}
with $ \omega^{\mu \nu}(x)$ given by  \eqref{infi}, and $\gamma_\mu :=e^{\underline a}_\mu(x)\gamma_{\underline a}$ the curved Dirac gamma matrices that satisfy 
$\{\gamma_\mu,\gamma_\nu\}=2 g_{\mu\nu}$. Here, $\alpha,\beta,..$ denote spinorial indices.

At the infinitesimal level, one obtains\footnote{The variation is defined as $\delta  \Phi(x):=  \Phi'(x)- \Phi(x)$. The spinor covariant derivative $\nabla_\mu = \partial_\mu + \omega_\mu$ arises because when comparing spinor fields at different spacetime points, parallel transport along the spin bundle is required. Explicitly, $ \psi'(x') \simeq  \psi'(x) + \zeta^\mu\nabla_\mu \psi'(x) + \cdots$, where $\zeta^\mu := x'^\mu - x^\mu$.}
\begin{align}
    \delta \psi_\alpha&=- \left(\zeta^\rho \nabla_\rho-\tfrac{1}{4}  \omega^{\mu\nu}\gamma_{\mu\nu}  +\Delta\, \sigma \right)_\alpha{}^\beta\, \psi_\beta\,\,,\nn
\end{align}
which we write, suppressing the indices, as
\begin{align}
    \delta \psi&=-\big( {\cal L}_\zeta + \Delta \,\sigma  \big)  \psi \,\,.
    \label{Dirac}
\end{align}
The covariant derivative acting on spinors reads $\nabla_\mu:=\partial_\mu+\frac14\omega_\mu^{\underline{ab}}\,\gamma_{\underline{ab}}$ with $\omega_\mu$ the spin connection, $\gamma_{\mu\nu}:=\tfrac12[\gamma_\mu,\gamma_\nu]$. The spinor Lie derivative $\mathcal{L}_\zeta$ along a vector field $\zeta$ is then given by 
\be
{\cal L}_\zeta\psi_{\alpha}:= \zeta^\rho \nabla_\rho \psi_\alpha  +\frac{1}{4} \nabla_\mu \zeta _\nu \,(\gamma^{\mu\nu})_\alpha{}^\beta \psi_\beta \,\,.
\label{SLD}
\ee 
The spinorial Lie derivative in \eqref{SLD} forms a closed Lie algebra representation only for conformal Killing vectors \cite{bT,ortin,ortin2}. For arbitrary vector fields, the homomorphism fails because the commutator acquires an anomaly proportional to the symmetric shear tensor (see \cite{bT,Miz} for explicit computations).   See Appendix \ref{B} for further notation and conventions.

\vspace{2mm}

\nin{\sf Rarita-Schwinger}: A spin-$\tfrac32$ conformal primary $\psi_{\mu\alpha}$ 
transforms  as
\begin{equation}
    \psi_{\mu\alpha}  \to  \psi'_{\mu\alpha}(x')= \frac{1}{\Upsilon(x)^\Delta} (e^{\frac{1}{4}\omega  ^{\rho \sigma}\, \gamma_{\rho \sigma}})_\alpha{}^\beta \Lambda_\mu{}^{\nu}(x) \psi_{\nu\beta}(x)\,\,.
\end{equation}

At the infinitesimal level one finds
\begin{equation}
    \delta \psi_{\mu\alpha}=-\big( {\cal L}_\zeta + (\Delta-1) \,\sigma \big) \psi_{\mu\alpha} \,\,.
\end{equation}    
The gravitino Lie derivative ${\cal L}_\zeta$ reads
\begin{align} 
{\cal L}_\zeta\psi_{\mu\alpha}&:= \zeta^\rho \nabla_\rho \psi_{\mu\alpha}+ \nabla_\mu \zeta^\nu \,\psi_{\nu\alpha}+\frac{1}{4} \nabla_\rho \zeta _\sigma \,(\gamma^{\rho\sigma})_\alpha{}^\beta \psi_{ \mu\beta} \,\,,
\label{RSLD}
\end{align}  
where $\nabla_\mu=\partial_\mu+\omega_\mu+\Gamma_\mu$ is  the full (affine plus Lorentz) covariant derivative when acting on RS fields (see \cite{kos,FoF,ortin, ortin2} for further details).

---

\nin In the following sections, we will study curved Maximally Symmetric Spaces (MSS), i.e. spheres and hyperbolic spaces in Euclidean and de Sitter and anti-de Sitter spaces in Lorentzian signature. In all cases,  the $N+1$ non-isometric conformal Killing vectors (CKVs), which satisfy $\nabla_\rho c^{(I) \,\rho}\ne0 ~(I=1,...,N+1)$, can be written in terms of scalar potentials  as $c^{(I)}_{\mu}=\partial_\mu X^I$ (see app.\ref{ckvPot}).  Non-isometric CKVs satisfy the following identities, which we will use repeatedly \cite{LSS}
\be
\nabla_\mu c_\rho=\sigma_c \,g_{\mu\rho}\,\,,~~~~c_\mu=-\ell^2\nabla_\mu\sigma_c \,\,,
\label{pCKV}
\ee
here $\ell^2>0$ for positive curvature spaces. Consequently, using $\nabla_\mu c^\nu = \sigma_c\, \delta_\mu^\nu$, the spinor and gravitino Lie derivatives for   non-isometric CKV $c^\mu$ in MSS simplify to
\begin{align*}
{\cal L}_c\psi_{\alpha}&= c^\rho \nabla_\rho \psi_\alpha \,\,, \\
 {\cal L}_c\psi_{\mu\alpha} &= c^\rho \nabla_\rho \psi_{\mu\alpha}+ \sigma_c\,\psi_{\mu\alpha} \,\,. 
\end{align*}

\section{Fermionic ladder operators} 
\label{laddersferm}

\subsection{Dirac Spinor}

We start by assuming an eigenfunction $\psi$ of the Dirac operator in a curved  $N$-dimensional maximally symmetric spacetime with  radius $\ell$ 
\begin{equation}
  \slashed{\nabla} \psi=  \frac{i\lambda}{\ell} \,    \psi \,\,,~~~~\lambda=\frac{N-1}2- \Delta \,\,,
    \label{Dirac1}
\end{equation}
where we introduced the \emph{scaling dimension} $\Delta$ to parametrize the  eigenvalue. We  define the shadow dimension as
\begin{equation}
\text{\sf Shadow dimension}:~~~~\Delta^s:=N-1-\Delta\,\,,~~\leadsto~~\lambda= \frac12(\Delta^s-\Delta)\,\,.
\label{shadow}
\end{equation}
Our aim is to extend previous work on ladder operators \cite{Cardo, Muck, LSS} to   fermions. 

In terms of a non-isometric CKV $c^\mu$ of the MSS, it is possible to define three ladder operators acting on a Dirac eigen-spinor satisfying \eqref{Dirac1}: they are,   
\begin{align}
    {\cal{D}} \psi&:= c^\mu \big(\nabla_\mu-\frac{i}{2\ell}\gamma_\mu\big) \psi+ (\tfrac12+\Delta)\, \sigma_c \psi \,\,,\nn \\
    {\cal{D}}^s \psi&:= c^\mu \big(\nabla_\mu+\frac{i}{2\ell}\gamma_\mu\big) \psi+ (\tfrac12+\Delta^s)\, \sigma_c \psi \,\,,\nn \\
    {\cal{\tilde{D}}} \psi&:= c^\mu \nabla_\mu \psi+ \tfrac{N-1}{2} \sigma_c \psi-\tfrac i\ell\big( \tfrac{N-1}2- \Delta\big) c^\mu \gamma_\mu \psi \,\,.
    \label{ladd}
\end{align}
Indeed, their action on Dirac eigenmodes \eqref{Dirac1} generate new eigenmodes $\psi',\psi''$ and $\tilde\psi$,
\begin{align}
    \psi':={\cal{D}}\psi\,\,,~~~~
    \psi'' :={\cal{D}}^s \psi\,\,,~~~~
    \tilde\psi :=\tilde{\cal D}\psi\,\,,
\end{align}
which satisfy \eqref{Dirac1} but with shifted scaling dimension
\begin{align}
    \slashed{\nabla} \psi'&=\frac i\ell\big(\tfrac{N-1}2-\Delta'\big)\, \psi' \ \ \text{~with}\ \ \Delta'=\Delta+1 \,\,,
\label{DeltaEucl}\\
    \slashed{\nabla} \psi''&=\frac i\ell\big(\tfrac{N-1}2-\Delta''\big)\, \psi'' \ \ \text{with}\ \ \Delta''=\Delta-1\,\,,\nn\\
    \slashed{\nabla} \tilde \psi&=-\frac i\ell\big(\tfrac{N-1}2-\tilde\Delta\big)\, \tilde\psi \ \ \text{ with}\ \ \tilde\Delta=\Delta\,\,. \nn
\end{align}
We can rephrase the action of the ladder operators  in terms of anti-commutators with the Dirac operator as
\begin{align}
    \{\ell\slashed{\nabla},{\cal{D}}\} \psi&= {2i} \big(\tfrac{N-2}2- \Delta\big)\,{\cal{D}}\psi \,\,,\nn\\
    \{\ell\slashed{\nabla},{\cal{D}}^s\}\psi&=-{2i} \big(\tfrac{N-2}2- \Delta^s\big) \,{\cal{D}}^s\psi \,\,,\nn\\
    \{\slashed{\nabla},\tilde{{\cal{D}}}\}\psi&=0\,\,.
\end{align}
Notice the similarity of these expressions to those found  in terms of commutators  for the bosonic case \cite{LSS}.

---

\nin {\small{\bf Proof}:  assume that \eqref{Dirac1} holds  and consider the ansatz, built from a non-isometric CKV $c^\mu$,
\begin{equation}
    {\cal{D}} \psi=c^\rho \nabla_\rho \psi +  \alpha\, \sigma_c  \psi+ \frac{i\beta}\ell c^\mu \gamma_\mu \psi\,\,.
\end{equation}
The  coefficients  $\alpha,\beta$ are fixed by demanding ${\cal{D}} \psi$ to be an eigenfunction of the Dirac operator
\begin{equation}
    \slashed{\nabla}({\cal D }\psi)=\frac1\ell(-i \lambda+i\delta \lambda)\, {\cal D }\psi \,\,. 
\label{eM}
\end{equation}
This last equation can be rewritten as
\begin{equation}
    \{\slashed{\nabla},{\cal{D}} \}\psi=\frac i\ell\delta\lambda\, {\cal{D}}\psi \,\,.
    \label{demand}
\end{equation}
We now proceed to the explicit computation of the anti-commutator. The first term in ${\cal D}\psi$ gives, using \eqref{pCKV}\footnote{For MSS the commutator of covariant derivatives action on a spinor give   $[\nabla_\mu,\nabla_\nu]\psi=\frac{1}{2\ell^2} \gamma_{\mu\nu}\psi$.}
\begin{align}
    \{\slashed{\nabla}, c^\rho \nabla_\rho\} \psi 
    &=\frac{i\lambda}\ell  \sigma  \psi+ \frac{2i\lambda}\ell  c^\rho \nabla_\rho \psi+\frac{1}{\ell^2}\frac{N-1}{2}  c^\rho \gamma_\rho \psi \,\,.
\end{align}
For  the second term we have
\begin{align}
    \{\slashed{\nabla},  \sigma_c \} \psi
    &=-\frac{1}{\ell^2} c^\mu\gamma_\mu  \psi+\frac{2i\lambda}\ell \sigma_c  \psi \,\,,
\end{align}
and for the last term we obtain
\begin{align}
    \{\slashed{\nabla},  c^\rho \gamma_\rho \} \psi 
    &= N  \sigma_c \psi+2  c^\rho \nabla_\rho \psi\,\,.
\end{align}
Collecting all terms and comparing to \eqref{demand} we get
\begin{align}
    \{\ell\slashed{\nabla},{\cal{D}} \} \psi
    &= \underbrace{( {2i\lambda} +2i\beta)}_{i\delta \lambda}  c^\rho \nabla_\rho \psi + \underbrace{(i\lambda+2i\alpha\lambda +iN \beta)}_{i\alpha\, \delta\lambda}  \sigma_c \psi  + \underbrace{\frac{1}{\ell}(\tfrac{N-1}{2}-\alpha)}_{-\beta\, \delta\lambda/\ell}  c^\rho \gamma_\rho \psi\,\,.
\end{align}
The under-braces highlight the expected result for each coefficient, as required by \eqref{demand}. This implies three equations
\begin{align}
2\lambda+2\beta&=\delta\lambda \,\,,\nn\\
    \lambda+2\lambda\alpha+N \beta&=\alpha\, \delta\lambda \,\,,\nn\\
      \frac{N-1}{2}-\alpha &=-\beta\, \delta\lambda\,\,.
\end{align}
The system admits  three non-trivial solutions to these equations. They are
\begin{align}
\left(\alpha,\beta,\delta\lambda \right)=\left\{\begin{array}{ll}
       \text{I}.\, &\left(\frac{N}{2}-\lambda,-\frac{1}{2},2\lambda-1\right) \\
       \text{II}.~  &\left(\frac{N}{2}+\lambda,\frac{1}{2},2\lambda+1 \right)\\
     \text{III}.~  &\left(\frac{N-1}{2},-\lambda,0 \right)
     \end{array}\right.\,\,.
     \label{options}
\end{align}
These lead to the ladder operators displayed in \eqref{ladd}.
}

---

We conclude this section by recalling previous work on spinorial Lie derivatives, CKVs, and massless fermions. It is straighforward to see that on  a general spacetime admitting a CKV $\zeta$, with conformal factor $\sigma_\zeta = \frac{1}{N}\nabla_\mu \zeta^\mu$, the modified spin-$\frac{1}{2}$ Lie derivative
\begin{equation}
    \widehat{\mathcal{L}}_\zeta \psi := \mathcal{L}_\zeta \psi + \Delta\, \sigma_\zeta \psi \,\,,
    \label{Lhat}
\end{equation}
forms a closed representation of the CKV Lie algebra for any constant conformal weight $\Delta$. This means that for $\zeta,\xi$ CKV
$$[\widehat{\mathcal{L}}_\zeta,\widehat{\mathcal{L}}_\xi]=\widehat{\mathcal{L}}_{[\zeta,\xi]}.$$
Moreover, one also finds \cite{Miz}
\begin{equation}
    [\widehat{\mathcal{L}}_\zeta, \slashed{\nabla}] \psi = -  \sigma\, \slashed{\nabla} \psi+\frac{N-1-2\Delta }{2} (\nabla_\mu\sigma_\zeta)\gamma^\mu\psi\,\,.
    \label{bourgi}
\end{equation}
The significance of $\widehat{\mathcal{L}}_\zeta$ lies in the fact that for $\Delta = \frac{N-1}{2}$, which corresponds to the classical scale dimension of a fermion in $N$-dimensions, the commutator simplifies to \cite{bT, Miz}
$$[\widehat{\mathcal{L}}_\zeta, \slashed{\nabla}] \psi = -  \sigma_\zeta\, \slashed{\nabla} \psi \,\,. $$ 
Thus, if $\psi$ satisfies the massless Dirac equation ($\slashed{\nabla}\psi = 0\,\leftrightarrow\,\Delta=\frac{N-1}2$), the modified Lie derivative action $\widehat{\mathcal{L}}_\zeta \psi$ is also a solution, showing that $\widehat{\mathcal{L}}_\zeta$ maps solutions to solutions and generates symmetries of the massless Dirac operator. It is amusing to see that when applied to massless Dirac fields, 
\be
\tilde{\mathcal{D}}\rfloor_{M=0}=\widehat{\mathcal{L}}_\zeta\rfloor_{\Delta=(N-1)/2}\,.
\label{new}
\ee
The specific choice $\Delta = \frac{N-1}{2}$ is expected from the well-known conformal invariance of the massless Dirac action under combined CKV transformations and associated Weyl rescalings \cite{BD,Iorio} (see \cite{spindel} for related work).

\subsection{Rarita-Schwinger field}

We now proceed to construct ladder operators for transverse and $\gamma$-traceless vector-spinor fields (also called Rarita-Schwinger fields) on MSS spaces  satisfying
\begin{equation}
    \slashed{\nabla} \psi_\mu=\frac{i\lambda}\ell \psi_\mu\,\, ,~~~~\gamma^\mu \psi_\mu=\nabla_\mu \psi^\mu=0\,\,,  ~~~~\lambda=\frac{N-1}2- \Delta\,\,.
    \label{RSef}
\end{equation}
Again we parametrize the eigenvalue by the scaling dimension $\Delta$. Three ladder operators can be constructed from non-isometric CKV $c^\mu$, they are given by
\begin{align}
    {\cal{D}} \psi_\mu&:= \mathcal{L}_{  c} \psi_\mu+(\Delta-\tfrac12) \sigma_c \psi_\mu-\frac{i}{2\ell} c^\rho \gamma_\rho \psi_\mu-\frac{2}{1+2 \Delta} \big(\nabla_\mu-\tfrac{i}{2\ell} \gamma_\mu\big) c^\rho \psi_\rho \,\,,
    \label{raritaladders}\\
    {\cal{D}}^s \psi_\mu&:= \mathcal{L}_{  c} \psi_\mu+(\Delta^s-\tfrac12) \sigma_c \psi_\mu+\frac{i}{2\ell} c^\rho \gamma_\rho \psi_\mu-\frac{2}{1+2 \Delta^s} (\nabla_\mu+\tfrac{i}{2\ell} \gamma_\mu) c^\rho \psi_\rho \,\,, \nonumber
    \\
    {\cal{\tilde{D}}} \psi_\mu&:= \mathcal{L}_{ c} \psi_\mu+\frac{N-3}{2} \sigma_c \psi_\mu-\frac{i\lambda}\ell c^\rho \gamma_\rho \psi_\mu+\frac i\ell\frac{ 2\lambda-1}{ N-1 } \gamma_\mu c^\rho \psi_\rho-\frac{2}{N-1} (\nabla_\mu-\tfrac{i}{2\ell} \gamma_\mu) c^\rho \psi_\rho \,\, . \nn 
\end{align}
The actions of these operators on eigenfunctions \eqref{RSef} define
$$\psi'_\mu:={\cal D} \psi_\mu\,\,,~~~~\psi''_\mu:= {\cal D}^s  \psi_\mu\,\,,~~~~\tilde \psi _\mu:= {\cal \tilde{D}}  \psi_\mu \,\,.$$
which are again eigenfunctions of the Dirac operator, albeit with shifted  scaling dimension
\begin{align}
     \slashed{\nabla} \psi'_\mu&=\frac i\ell \big(\tfrac{N-1}2-\Delta'\big)\, \psi'_\mu \nn\ \ \text{~with}\ \ \Delta'=\Delta+1 \,\,,\\
\slashed{\nabla} \psi''_\mu&=\frac i\ell \big(\tfrac{N-1}2-\Delta''\big)\, \psi''_\mu\nn \ \ \text{with}\ \ \Delta''=\Delta-1\,\,,\\
    \slashed{\nabla} \tilde \psi_\mu&=-\frac i\ell \big(\tfrac{N-1}2-\tilde\Delta\big)\, \tilde\psi_\mu \ \ \text{ with}\ \ \tilde\Delta=\Delta \,\,.
\end{align}
Equivalently we have the following anti-commutators
\begin{align}
     \{\ell\slashed{\nabla},{\cal{D}}\} \psi_\mu&=2i\big(\tfrac{N-2} 2-\Delta\big)\,{\cal{D}}\psi_\mu \,\,,\nn\\
    \{\ell\slashed{\nabla},{\cal{D}}^s\}\psi_\mu&=-2i\big(\tfrac{N-2} 2-\Delta^s\big)\, {\cal{D}}^s\psi_\mu\,\,,\nn \\
     \{\ell\slashed{\nabla},{\tilde{\cal{D}}\}}\psi_\mu&=0\,\,.
\end{align}

---

\nin {\small{\sf Proofs}: we assume the vector-spinor $\psi_\mu$ is an eigenfunction of the Dirac operator \eqref{RSef} and consider the ansatz
\begin{equation}
    {\cal{D}} \psi_\rho= c^\mu \nabla_\mu \psi_\rho+\alpha\,  \sigma_c \psi_\rho+\frac{i\beta}\ell  c^\mu \gamma_\mu \psi_\rho+ \epsilon c^\mu \nabla_\rho \psi_\mu+\frac{i\eta}\ell  \gamma_\rho c^\mu \psi_\mu\,\,.
    \label{anRS}
\end{equation}
The coefficients  $\alpha,\beta,\epsilon,\eta$ are fixed by imposing 
\begin{equation}
    \{\ell\slashed{\nabla},{\cal{D}} \}\psi_\mu=i\delta\lambda\, {\cal{D}}\psi_\mu \,\,.
    \label{27}
\end{equation}
We consider the anti-commutator of the Dirac operator on each of the terms  in \eqref{anRS}. The first term gives\footnote{In going to the second line we use that for MSS we have
$  
    [\nabla_\mu,\nabla_\nu] \psi_\rho=\frac{1}{\ell^2}(\frac12   \gamma_{\mu\nu}  \psi_\rho+ g_{\mu \rho} \psi_\nu- g_{\nu \rho} \psi_{\mu} )\,\,.
$ 
}
\begin{align}
    \{\gamma^\mu \nabla_\mu,  c^\nu \nabla_\nu\} \psi_\rho &= \gamma^\mu  \sigma_c \nabla_\mu \psi_\rho + \gamma^\mu  c^\nu \nabla_\mu \nabla_\nu \psi_\rho +\frac{i\lambda}\ell  c^\nu \nabla_\nu  \psi_\rho \nn\\ 
    &= \frac{i\lambda}\ell  \sigma_c \psi_\rho+\frac{2i \lambda}\ell  c^\nu \nabla_\nu  \psi_\rho + \frac1{\ell^2}\big( \tfrac{N-1}2c^\nu\gamma_\nu\psi_\rho+\gamma_\rho c^\mu\psi_\mu -c_\rho\cancel{\gamma^\mu   \psi_\mu} \big)\,\,.
\end{align}
The second term gives, using \eqref{pCKV},
\begin{align}
    \{\gamma^\mu \nabla_\mu, \sigma_c\} \psi_\rho&= \gamma^\mu (\nabla_\mu \sigma_c) \psi_\rho+\frac{2i\lambda}\ell \sigma_c \psi_\rho  =-\frac{1}{\ell^2}   c^\mu\gamma_\mu   \psi_\rho+\frac{2i\lambda}\ell  \sigma_c \psi_\rho \,\,.
\end{align}
The third simplifies to
\begin{align}
    \{\gamma^\mu \nabla_\mu,  c^\nu \gamma_\nu\} \psi_\rho &=  N  \sigma_c  \psi_\rho +  c^\nu\gamma^\mu  \gamma_\nu \nabla_\mu\psi_\rho+ \frac{i\lambda}\ell  c^\nu \gamma_\nu \psi_\rho \nn\\
    &=N  \sigma_c \psi_\rho+2c^\mu\nabla_\mu\psi_\rho-\cancel{\frac{i\lambda}\ell c^\nu \gamma_\nu  \psi_\rho}+\cancel{  \frac{i\lambda}\ell  c^\nu \gamma_\nu \psi_\rho}  \,\,,
\end{align}
and the fourth term results in
\begin{align}
    \{\gamma^\mu \nabla_\mu,  c^\nu \nabla_\rho\} \psi_\nu &= \sigma_c \nabla_\rho( \cancel{\gamma^\mu \psi_\mu})+ c^\nu \gamma^\mu \nabla_\mu \nabla_\rho \psi_\nu+\frac{i\lambda}\ell  c^\nu \nabla_\rho \psi_\nu \nn \\
    &=\frac{2i\lambda}\ell c^\nu \nabla_\rho \psi_\nu+  c^\nu \gamma^\mu  \frac{1}{\ell^2}\Big(\tfrac12 \gamma_{\mu\rho}  \psi_\nu + g_{\mu \nu} \psi_\rho- g_{ \rho\nu} \psi_{\mu} \Big) \nn \\
    &=\frac{2i\lambda}\ell c^\nu \nabla_\rho \psi_\nu+ \frac{N-1}{2\ell^2}  c^\nu     \gamma_\rho\psi_\nu +\frac1{\ell^2}( c^\mu\gamma_\mu \psi_\rho- c_\rho\,\cancel{\gamma^\mu \psi_{\mu} } )\,\,, 
\end{align}
and the last one
\begin{align}
    \{\gamma^\mu \nabla_\mu,  c^\nu \gamma_\rho\} \psi_\nu &=\gamma^\mu  \sigma_c \gamma_\rho \psi_\mu+ \gamma^\mu c^\nu \gamma_\rho \nabla_\mu \psi_\nu + \frac{i\lambda}\ell   c^\nu \gamma_\rho \psi_\nu \nn\\
    &=2  \sigma_c \psi_\rho+2  c^\nu \nabla_\rho \psi_\nu-\cancel{\frac{i\lambda}\ell   \gamma_\rho c^\nu\psi_\nu}+ 
    \cancel{\frac{i\lambda}\ell   c^\nu \gamma_\rho \psi_\nu} \,\,. 
\end{align}
Collecting terms we get
\begin{align}
    \{\ell\slashed{\nabla},{\cal{D}}\}\psi_\rho&=\overbrace{2i(  \lambda +   \beta )}^{i\delta\lambda}c^\nu \nabla_\nu \psi_\rho +\overbrace{i( \lambda+2 \lambda\alpha+ \beta N+2 \eta)}^{i\alpha\,\delta\lambda}  \sigma_c \psi_\rho+\overbrace{\frac1\ell\big(\tfrac{N-1}{2}-\alpha+\epsilon )}^{-\beta\,\delta\lambda/\ell} c^\mu \gamma_\mu \psi_\rho \nn\\
    &~+ \underbrace{2i( \lambda\epsilon+ \eta)}_{i\epsilon\,\delta\lambda}c^\mu \nabla_\rho \psi_\mu  +\underbrace{\frac1\ell\big(1+\epsilon\tfrac{N-1}2\big)}_{-\eta\,\delta\lambda/\ell}\gamma_\rho c^\mu \psi_\mu\,\,.
\end{align}
Imposing \eqref{27} we obtain three solutions 
\begin{align}
    (\alpha,\beta,\epsilon, \eta,\delta\lambda)&=\left\{\begin{array}{ll}
    \text{I}.&\big( \tfrac{(  N-2\lambda)^2-4}{2( N -2\lambda)},-\tfrac12,-\tfrac2{ N -2\lambda},\tfrac1{ N -2\lambda},2\lambda-1\big)\\
    \text{II}.&\big( \tfrac{(  N+2\lambda)^2-4}{2( N +2\lambda)},\tfrac12,-\tfrac2{ N +2\lambda},-\tfrac1{ N +2\lambda},2\lambda+1\big)\\
    \text{III}.~&\big(\tfrac{(N+1)(N-3)}{2(N-1)},-\lambda,-\tfrac2{N-1},\tfrac{2\lambda}{N-1},0\big)
    \end{array}\right. \,\,. 
\end{align}
These choices of parameters give the ladder operators in \eqref{raritaladders}. 


}
\section{Action of ladder operators}
 \label{laddersapp}
 
In the present section we study the action of the ladder operators \eqref{ladd} and \eqref{raritaladders} on  Dirac (spin-$\frac12$) and Rarita–Schwinger (spin-$\frac32$) UIR solutions.

\subsection{Spin-$\frac12$: Dirac spinors}

\subsubsection{Euclidean signature: $S^{N}$}

The relation between $\Delta$ in \eqref{Dirac1} and the eigenvalues $L_N$ in  \eqref{DiracN} is
\begin{align}
\Psi_{(+)L_N...}^{(\vec a)}&=\psi_{\Delta_+}~~~\text{with}~~ \Delta_+=-(L_N+\tfrac12)\,\,,\nn\\
\Psi_{(-)L_N...}^{(\vec a)}&=\psi_{\Delta_-}~~~\text{with}~~ \Delta_-= L_N+N-\tfrac12 \,\,.
\label{rel1}
\end{align}
Notice that the scaling dimensions for positive and negative eigenvalues of the Dirac operator are related by the shadow transformation \eqref{shadow}
$$(\Delta_\pm)^s=\Delta_\mp \,\,.$$
The $S^N$ metric is written as ($\ell=1$)
\be
ds^2 = d\theta_N^2 + \sin^2 \theta_N \, \tilde{ds}^{2} \,\,,
\end{equation}
with $\tilde{ds}^{2}$ the metric on the $S^{N-1}$ sphere (see app. \ref{s1/2Con} for conventions). Although the construction described above works for any CCKV $\bm c$ in positive curvature MSS, simple  expressions are found by choosing \cite{LSS} 
\begin{equation}
{\bm c}= \sin \theta_N \, \partial_{\theta_N} ~~ \Rightarrow~~ \sigma_c = \frac{1}{N} \nabla_\mu c^{\mu} =  \cos \theta_N \,\,.
    \label{ckvtheta}
\end{equation}
The explicit expressions for the ladder operators are
\begin{align}
    {\cal{D}} \psi_\Delta&= \sin \theta_N \left(\partial_{\theta_N}-\frac{i}{2}\gamma_{\underline N}\right) \psi_\Delta+ \cos \theta_N (\tfrac12+\Delta)   \psi_\Delta\,\,,\nn \\
    {\cal{D}}^s \psi_\Delta &= \sin \theta_N \left(\partial_{\theta_N}+\frac{i}{2}\gamma_{\underline N}\right) \psi_\Delta+ \cos \theta_N(\tfrac12+\Delta^s)\,   \psi_\Delta \,\,,\nn \\
    {\cal{\tilde{D}}} \psi_\Delta&= \sin \theta_N \left( \partial_{\theta_N}-  i \big( \tfrac {N-1}2 -  \Delta\big) \gamma_{\underline N}\right)\psi_\Delta+ \cos \theta_N\tfrac{N-1}{2}   \psi_\Delta \,\,. 
    \label{ladd22}
\end{align}
The action of  $\mathcal{D}^{s},\mathcal{D} $ with this particular CCKV  solely changes the principal quantum number $L_N$. Explicitly one obtains, setting $C_N=1$ in \eqref{psi}-\eqref{phi},
\begin{align}
\mathcal{D}^{s}\Psi^{(\vec a)}_{(+)L_N L_{N-1}... }&= (N+L_N +L_{N-1})\Psi^{(\vec a)}_{(+)L_{N}+1\,L_{N-1}... }\nn\\ 
\mathcal{D}\Psi^{(\vec a)}_{(+)L_N L_{N-1}... }&= -(L_N -L_{N-1})\Psi^{(\vec a)}_{(+)L_{N}-1\,L_{N-1}... }\nn\\
\mathcal{D}^{s}\Psi^{(\vec a)}_{(-)L_N L_{N-1}... }& = -(L_N-L_{N-1}) \Psi^{(\vec a)}_{(-)\, L_N-1\, L_{N-1}... } \nn\\
\mathcal{D}\Psi^{(\vec a)}_{(-) L_{N} L_{N-1}... } &= (N+L_N+L_{N-1}) \, \Psi^{(\vec a)}_{(-)\,L_N+1\,  L_{N-1}... }\,\,.\label{ladd1}
\end{align}
On the other hand, the action of $\mathcal{\tilde D}$ flips the sign of the eigenvalue.  
One obtains\footnote{The $\tilde{\mathcal{D}}$ action can be  written as  (see comment below \ref{casis} and eqn. \ref{oddgamma5})
\begin{align}
\text{\sf N even}:&~~    \mathcal{\tilde D} = \sigma_3\otimes i\cancel{ \nabla}_{S^{N-1}} =\left(
\begin{array}{cc}
 i \cancel \nabla_{S^{N-1}}& 0\\
0 & -i \cancel\nabla_{S^{N-1}}\\ 
\end{array}
\right) \,\,,\nn \\
 \text{\sf N odd}:&~~    \mathcal{\tilde D}= - \gamma^{\ud N}\cancel{  \nabla}_{S^{N-1}}\,\,.
 \label{tildeLow}
\end{align}}
\begin{align}
\text{\sf N even}:~~~\tilde{\mathcal{D}}\Psi^{(\uparrow,a_{n-1}...a_1)}_{(\pm)L_N L_{N-1}... } &= -\left(L_{N-1} + \frac{N-1}{2} \right)  \Psi^{(\uparrow,a_{n-1}...a_1)}_{(\mp)L_N L_{N-1}... }  \nn \\
\tilde{\mathcal{D}}\Psi^{(\downarrow,a_{n-1}...a_1)}_{(\pm)L_N L_{N-1}... } &= \left(L_{N-1} + \frac{N-1}{2} \right)  \Psi^{(\downarrow,a_{n-1}...a_1)}_{(\mp)L_N L_{N-1}... } \nn\\
\text{\sf N odd}:~~~ \tilde{\mathcal{D}}\Psi^{(\vec a )}_{(\pm)L_N L_{N-1}... } &= \left(L_{N-1} + \frac{N-1}{2} \right)  \Psi^{(\vec a)}_{(\mp)L_N L_{N-1}... } \,.
\label{Dtil1/2}
\end{align}

\vspace{2mm}

\nin{\sf Killing spinors and   spin-$\frac12$ harmonics:} a Killing spinor on the $N$-sphere is defined by\footnote{We set the de Sitter radius $\ell=1$.} 
\begin{equation}
  \text{\sf Killing Spinor Equation}:~~  \nabla_\mu \epsilon_{(\pm)} =\pm \frac{i}{2 } \gamma_\mu \epsilon_{(\pm)}\,\,.
  \label{KSEp}
\end{equation}
The presence of the imaginary unit on the rhs follows from the integrability condition of the equation (see chapter 22 in \cite{freedman}), and the full
solution  for all $S^{N}$ is explicitly known  \cite{pope}.\footnote{Their form in the standard sphere coordinates $\Omega=(\theta_N,...,\theta_2,\theta_1)$ is \cite{pope} (see app.\ref{B} for conventions)
\begin{equation}
    \epsilon_{(+)}(\Omega)=e^{\frac{i}{2}\theta_N \gamma_{\ud{N}}}\prod_{j=1}^{N-1}e^{\frac{1}{2} \theta_j \gamma_{\ud{j}}\gamma_{ \ud{j+1}}} \,\epsilon_0 \, \, ,
    \label{pope}
\end{equation}
where $\epsilon_0$ is an arbitrary constant complex spinor.}
Contracting \eqref{KSEp} with $\gamma^{\mu}$ yields
\begin{equation}
\slashed{\nabla}\epsilon_{(\pm)}= \pm i\frac{N}{2 }  \epsilon_{(\pm)} \,\,,
\label{KSEeucl}
\end{equation}
which is precisely the equation for the lowest spin-$\frac12$ harmonic spinor $L_{N}=0$ 
(cf. \ref{DiracN}).  Since $L_N=0$ implies that all lower quantum numbers vanish $L_{N-1}=...=L_2=\ell=0$ (cf.   \ref{regu}), the degeneracy of the solution is parametrized by  $\vec a$. This degeneracy precisely coincides with the dimension of $\epsilon_0$ in \eqref{pope}. We can therefore conclude that the space of solutions of the Killing spinor equation (KSE) is isomorphic to the $L_N=0$ space,  
\be 
\text{span}\{\epsilon_{(\pm)}\}\sim\text{span}\big\{ \Psi_{(\pm)L_N=0\,...\, \ell=0}^{(\vec a)}\big\} \,\, .
\label{KSEsph}
\ee
See also section 5 of \cite{Higu+lets} for an alternative derivation using the sphere inner product.

Equation \eqref{ladd1} implies that in Euclidean signature Killing spinors serve as seeds to generate the full tower of spinor harmonics. Indeed, these can be obtained by repeated action of $\mathcal{D}^{s}$ on $\epsilon_{(+)}$  (similarly for the negative harmonics acting with $\cal D$ on $\epsilon_{(-)}$). 

We can also turn the argument around. If we assume that a $\Delta=-\frac12$ eigen-spinor $\Psi$ is annihilated by $\cal D$ for all  non-isometric CKV,  then,  since the set $\{c^\mu_{(I)}\}$ of non-isometric CKV on $S^N$ span the full tangent space at any point, we have
$$c_{(I)}^\mu V_\mu = 0~\forall~c_{ ( I)}~~\Rightarrow ~V_\mu=0\,\,.$$   
Thus,  
$${\cal D}\Psi_\Delta=0~~\forall\text{ non-isometric CKV }c_\mu~\&~{\Delta=-\frac12}~~\Rightarrow~~\nabla_\mu \Psi_{\Delta } =  \frac{i}{2} \gamma_\mu \Psi_\Delta \, \,.$$
We conclude by reminding the reader that the importance of Killing spinors stems from the fact that Killing and Conformal Killing vectors on the sphere and de Sitter space can be constructed from Killing spinor bilinears  as $\bar\epsilon'\gamma_\mu\epsilon$ (see \cite{Anous2,pope,Higu+lets,Fujii}).


\subsubsection{Lorentzian signature: de Sitter$_{N}$}

The relation between the scaling dimension $\Delta$ in \eqref{Dirac1} and the Dirac mass $M$ in\footnote{We set the de Sitter radius $\ell=1$ and without loss of generality, we write the Dirac equation with $+M$ in the rhs. Although two sets of solutions are possible we restrain from adding $(\pm)$ subindex related to \emph{positive/negative energy} solutions associated to the short wavelength as discussed in \eqref{CompCh}. } 
$$ \slashed{\nabla}  \Psi_{M...}^{(\vec a)}  = M \Psi_{M...}^{(\vec a)} \,\,, $$ is
\begin{align}
 \Delta=\frac{N-1}2+iM ~~\leadsto~~\Psi_{ML_{N-1}...}^{(\vec a)} 
&=\psi_{\Delta} \,\, .
\label{rel111}
\end{align}
The shadow transform \eqref{shadow} is equivalent to $M\to-M$ and the action of the ladder operators expressed in \eqref{DeltaEucl} is equivalent to 
\be
{\cal D},{\cal D}^s\text{ action}:~~~M\to M\mp i \,\,. 
\label{DirM}
\ee
Since spin-$\frac12$ UIRs for $\text{Spin}(N,1)$  exist only for \cite{dio+silva,zimo,hint}
\be
\Delta=\frac{N-1}2+iM ~~ \text{with}~~M\in\mathbb R \,\,,
\label{UIRdirac}
\ee
ladder operators generally destroy the unitary property of the representation. A notable exception is $\tilde{\cal D}$, which maps $\Psi_M\to\Psi_{-M}=\tilde{\cal D}\Psi_M$. The existence of $\tilde D$ is not surprising for $M=0$, since spin-$\frac12$ massless fermions are well known to be conformal invariant \cite{BD,DeserNepo,Iorio}. Indeed, for $M=0$ we have $\Delta=\frac{N-1}2$ and $\tilde{\cal D}$ reduces to the infinitesimal transformation for a primary field (cf.\eqref{Prim})
$$\tilde{\cal D}\psi_{\Delta_{\rfloor_{M=0}}}=\big( {\cal L}_\zeta + \Delta \,\sigma  \big)  \psi_\Delta=-\delta\psi_\Delta\,\,.$$
In the massive case $\tilde{\cal D}$  generalizes to
arbitrary mass  the operators $\mathbb T_V$ defined for Killing spinors in Section 5 of \cite{Higu+lets}. 

The space of Killing spinor solutions furnishes a representation of the de Sitter symmetry algebra. This is a consequence of  
\be
[{\cal L}_k,\nabla_\mu]=[{\cal L}_k,\gamma_\mu]=0 \,\,,
\label{Kill}
\ee
for $\bm k$ a de Sitter Killing vector.  Since the space of solutions is finite dimensional, span$\{\epsilon\}$ is necessary a nonunitary representation of Spin$(N,1)$. 

We can alternatively understand the non-unitary of the Killing spinor solution space by studying the Dirac equation satisfied by the Killing spinors. Contracting the KSE with $\gamma^\mu$ one finds\footnote{For notational convenience we consider a positive sign in the KSE. The results for the opposite sign are straightforward.}
$$\gamma^\mu\cdot\left(\nabla_\mu \epsilon = \frac{i}{2} \gamma_\mu \epsilon\right)~~\Rightarrow~~\slashed{\nabla}  \epsilon  =i\tfrac N2\epsilon \,\,. $$
This means that the Killing spinor satisfies the Dirac equation with   pure imaginary mass $M=i\frac N2$. Since the space of solutions is finite dimensional while Spin$(N,1)$ is non-compact, we conclude that the Killing spinor space of solutions   furnish a non-UIR.\footnote{The fact that Killing spinors carry a non-unitary representation is also clear from the finite dimensionality of the solution space, on which $Spin(N,1)$ acts irreducibly. A well-known theorem states that all non-trivial UIRs of  non-compact groups must be infinite-dimensional. Hence, any non-trivial finite-dimensional representation of $Spin(N,1)$ is necessarily non-unitary.}   

For completeness we now summarize  the general solution to the KSE in de Sitter. Writing the de Sitter metric as (we set the de Sitter radius   $\ell=1$)
\begin{equation}
    ds^2= -dt^2 + \cosh^2 t \, d\Omega_{N-1}^2 \,\,.
\end{equation}
The solution to the Killing spinor equation 
$$\nabla_\mu \epsilon = \frac{i}{2} \gamma_\mu \epsilon \,\,,$$
can be found by performing a Wick rotation on the $L_N=0$ harmonics of $S^N$  \cite{Higu+lets}. 
Therefore in Lorentzian signature
we have 
$$\text{span}\{\epsilon  \}\sim\text{span}\big\{ \Psi_{  M\, L_{N-1}=0\,...\, \ell=0}^{(\vec a)}\big\}_{\rfloor M=i\frac N2} \,\,.$$
This relation constitutes the Lorentzian counterpart to the Euclidean result in \eqref{KSEsph}.

\vspace{2mm}
\nin{\sf Ladders, Discrete series and Killing spinors}: Although the action of $\mathcal{D}$ and $\mathcal{D}^s$ does not preserve unitarity, as seen when comparing \eqref{DirM} and \eqref{rel111}, an interesting result is that ${\cal D},{\cal D}^s$ relate  (Lorentzian) massless fermion UIRs to  Killing spinors.\footnote{Recall that massless Lorentzian fermions possess half as many independent states relative to massive UIRs.}

We discuss 4d de Sitter, extensions to other even dimensional de Sitter spaces are straightforward. Consider the Wick rotated version of the non-isometric CKV given in \eqref{ckvtheta}
\begin{equation}
    {\bm c} = \cosh t \,  \partial_t \leadsto \sigma_c= \sinh t  \,\,.
    \label{lorCKV}
\end{equation}
Setting $C_N=1$ in \eqref{fg1/2dS}, the following results then follow
\begin{align}
\mathcal{D}\Psi^{(\vec{a})}_{ML_3...\ell\rfloor _{M=i}}&=0 \,\,,\nn\\
\mathcal{D}^{s}\Psi^{(\vec{a})}_{ML_3...\ell\rfloor_{M=0}}&= -(L_3+2)\Psi^{{(\vec a)}}_{ML_3...\ell\rfloor_{M=i}} \,\,,\nn\\
\mathcal{D}^{s}\mathcal{D}^{s}\Psi^{(\vec{a})}_{ML_3...\ell\rfloor_{M=0}}&=(L_3+3)(L_3+2) \Psi^{(\vec{a})}_{ML_3...\ell\rfloor_{M=2i}} \,\,,\nn\\
...&=... \,\,.
\end{align}
Hence, acting twice with $\mathcal{D}^{s}$ on a massless fermion solution yields a spinor that satisfies the Dirac equation with $M=2i$\footnote{Alternatively, an $M=-2i$ fermion can be obtained by acting with $\cal D$ on a $M=0$ fermion solution.}
\begin{equation}
\tilde\Psi=({\cal D}^s)^2\Psi^{(\vec{a})}_{ML_3...\ell\rfloor_{M=0}}~~\leadsto~~\slashed{\nabla}\tilde\Psi=2i\tilde\Psi \,\,.
\label{M0KS}
\end{equation}
Since the arguments following  \eqref{KSEeucl} remain valid in Lorentzian signature, by choosing the massless seed fermion solution $\Psi^{(\vec{a})}_{M=0\;L_3...\ell }$ to have vanishing quantum numbers ($L_3=L_2=\ell=0$) yields a 4d de Sitter Killing spinor. 

We summarize the action of the ${\cal D},\tilde{\cal D}$ operators in the following diagram (suppressing the $L_i=0$ quantum numbers)  
\begin{equation}
...\stackrel[\mathcal{D}]{\mathcal{D}^{s}}{\xrightleftharpoons{\hspace{4mm}}}\Psi^{(\vec{a})}_{M=-2i} \stackrel[\mathcal{D}] {}{\longleftarrow}\Psi^{(\vec{a})}_{M=-i}
\underset{~\mathcal{D} }{\longleftarrow}  \Psi^{(\vec{a})}_{M=0}
\overset{\mathcal{D}^{s}}{\longrightarrow}  \Psi^{(\vec{a})}_{M=i} \stackrel[ ]{\mathcal{D}^{s}}{\longrightarrow}\Psi^{(\vec{a})}_{M=2i}\stackrel[\mathcal{D}]{\mathcal{D}^{s}}{\xrightleftharpoons{\hspace{4mm}}}...
\label{dig1/2}
\end{equation}
As discussed above the vector index $\vec a$ accounts for the degeneracy of the Killing spinor.

\subsubsection{dS$_4$ massless Dirac fermion and $S^3$ ladders}
\label{ds4massless}

In this section, we focus on $\text{dS}_4$, though extensions to other dimensions are straightforward. We show that the SO(4,1)  transformation laws for massless spinors in $\text{dS}_4$ naturally involve the ladder operators $\mathcal{D}$ and $\mathcal{D}^s$ constructed from the non-isometric CCKVs of $S^3$. 

---

\nin We work in global conformal de Sitter coordinates, where the metric takes the form
\begin{equation}
    ds^2 = \frac{-dt^2 + d\Omega_3^2}{\sin^2 T} =  {\rho}(T)^2 \left( -dT^2 + d\Omega_3^2 \right),
    \label{ESU}
\end{equation}
where $ {\rho}(t) = 1/\sin T$ is the conformal factor, and
\begin{equation}
    d\Omega_3^2 = d\theta_3^2 + \sin^2 \theta_3 \left( d\theta_2^2 + \sin^2 \theta_2 \, d\phi^2 \right) \,\,,
\end{equation}
denotes the standard metric on the $3$-sphere. From the covariant spinor derivatives 
\begin{equation}
    \nabla_T \Psi= \partial_T \Psi \,\,, \quad \nabla_{\theta_3}\Psi= \partial_{\theta_3} \Psi + \frac{1}{2} \cot T \, \gamma^{\ud{03}} \Psi \,\,,\nn
\end{equation}
\begin{equation}
    \nabla_{\theta_2} \Psi = \partial_{\theta_2} \Psi +\frac{1}{2}\cot T \sin \theta_3 \, \gamma^{\ud {02}}\,\Psi- \frac{1}{2}\cos \theta_3 \, \gamma^{\ud{32}} \,\Psi \,\,,\nn
\end{equation}
\begin{equation}
    \nabla_{\phi} \Psi = \partial_\phi \Psi + \frac{1}{2}\cot T \sin \theta_2 \sin \theta_3 \, \gamma^{\ud{01}}\Psi - \frac{1}{2}\cos \theta_3 \sin \theta_2  \, \gamma^{\ud{31}}\Psi - \frac{1}{2}\cos \theta_2 \,\gamma^{\ud{21}}\Psi \,\,.
\end{equation}
we obtain the Dirac operator 
\begin{align} \gamma^{\mu}\nabla_\mu\Psi&= \left[ \sin T \, \gamma^{\ud 0} \nabla_T + \sin T \, \gamma^{\ud 3} \nabla_{\theta_3} + \frac{\sin T}{\sin \theta_3} \gamma^{\ud 2} \nabla_{\theta_2} + \frac{\sin T}{\sin \theta_3 \sin \theta_2} \gamma^{\ud{1}} \nabla_\phi \right]\Psi \nn\\
&= \sin T \Bigg[ \gamma^{\ud 0} \left( \partial_T - \frac{3}{2}\cot T\right) + \gamma^{\ud 3} ( \partial_{\theta_3} + \cot \theta_3 )  \nn \\
&    \hspace{2cm} 
+ \frac{\gamma^{\ud 2}}{\sin \theta_3}\left( \partial_{\theta_2} + \frac{1}{2} \cot \theta_2\right) 
+ \frac{\gamma^{\ud 1}}{\sin \theta_3 \sin \theta_2} \partial_\phi \Bigg] \Psi \,\,.
\end{align}
The Dirac equation for a massless spinor in the chiral representation \eqref{chirL} reads
\begin{equation}
 \slashed{\nabla}\Psi = \sin T\left[    \gamma^{\ud{0}}\left( \partial_{T} - \frac{3}{2} \cot T \right) \Psi +  \left(
\begin{array}{cc}
 0 & i \slashed{\nabla}_{S^{3}} \\
-i\slashed{\nabla}_{S^{3}} & 0 \\ 
\end{array}
\right)\Psi\right] =0 \,\,.
\end{equation}
A Weyl transformation of the spinor under the conformal factor in \eqref{ESU} 
\be
\text{\sf Weyl transformation}:~~~~\Psi(T,\Omega) =\frac1{\rho(T)^\Delta}\tilde\Psi(T,\Omega) = (\sin T)^{3/2} \tilde \Psi(T,\Omega) \,\,,
\label{WT}
\ee
reduces the massless Dirac equation to that in Einstein Static Universe (cf. \eqref{ESU}). Notice that the power of the conformal factor in  \eqref{WT} coincides with the scaling dimension defined in \eqref{rel111}. The decoupled upper and lower spinor components now satisfy
\begin{equation}
 \left(
\begin{array}{cc}
 0 & i (\partial_T+\slashed{\nabla}_{S^{3}}) \\
i(\partial_T-\slashed{\nabla}_{S^{3}}) & 0 \\ 
\end{array}
\right)\tilde \Psi = 0  \,\,,
\label{DiracESU}
\end{equation}
which is easily solved by separating variables.  The solutions are labeled by $S^3$ quantum numbers $(a,L_3,L_2,\ell)$, their 4d chirality parametrized by $\uparrow,\downarrow$, and   a $(\pm)$ sign associated to the energy eigenvalue in ESU, 
\begin{equation}
\tilde \Psi^{(\downarrow,a)}_{(\mp)ML_{3}L_2\ell}(T,\Omega)\rfloor_{ M=0} = e^{\mp i \omega T}  \left(
\begin{array}{c}
\bm\chi^{(a)}_{(\mp)L_{3}L_2\ell} (\Omega) \\
  0
\end{array}
\right) = e^{\mp i \omega T } \Phi^{(\downarrow,a)}_{(\mp)L_3L_2\ell} (\Omega)\,\,,\nn
\end{equation}
and
\begin{equation}
\tilde \Psi^{(\uparrow,a)}_{(\mp)L_{3}L_2\ell}(T,\Omega)\rfloor_{ M=0} =e^{\mp i \omega T}   \left(
\begin{array}{c}
  0 \\
   \bm\chi^{(a)}_{(\pm)L_3 L_2\ell} (\Omega) \\
\end{array}
\right)=e^{\mp i \omega T} \,\Phi^{(\uparrow,a)}_{(\pm)L_3L_2\ell}(\Omega)\,\,,
\label{M=0dS}
\end{equation}
with $S^3$ spinor harmonics given by 
\begin{equation}
     \bm \chi_{(\pm)L_3L_2\ell}^{(a)}(\theta_3, \theta_2, \phi) = \big(g(\theta_3)\mp f(\theta_3)\big)\tilde{\bm \chi}^{(a)}_{(-)L_2 \ell}(\theta_2,\phi) + i\big(g(\theta_3) \pm f(\theta_3)\big)\tilde{\bm\chi}^{(a)}_{(+)L_2\ell}(\theta_2, \phi)\,\,.
\end{equation}
The functions $f,g$ are displayed in \eqref{sols3}. Inserting \eqref{M=0dS} in \eqref{DiracESU} we find that the energy of the modes depends solely on the principal angular momenta on $S^3$ 
\be
\omega(L_3)=L_3+\frac32 \,\,.
\label{dSEner}
\ee
The massless \emph{positive/negative conformal frequency} modes in de Sitter read
\be
\Psi^{(a_2,a_1)}_{(\mp)ML_{3}L_2\ell}(T,\Omega)\rfloor_{ M=0}=(\sin T)^{3/2}{}e^{\mp i\omega T}\Phi^{(a_2,a_1)}_{(s)L_{3}L_2\ell}(\Omega)\,\,.
\label{final}
\ee
The 4d spinors $\Phi^{(\uparrow,a)}_{(\pm)L_3L_2\ell},\Phi^{(\downarrow,a)}_{(\pm)L_3L_2\ell} $ in \eqref{M=0dS} and \eqref{final} are built from 3d spinors $\bm \chi_{(\pm)L_3L_2\ell}^{(a)}$ with the $\uparrow,\downarrow$-index indicating the 4d chirality. In the chiral representation we are working on,  see \eqref{chirL}, it relates to whether they are inserted in  the upper or lower components. The value of the subindex $s=\pm$ in $\Phi_{(s)...}^{(a_2,a_1)}$, related to the spinor harmonic on $S^2$, is determined by $a_2$ and the sign of the energy $(\pm)$, see  \eqref{M=0dS}.

We now study the action of the SO(4,1) de Sitter isometry group on the modes \eqref{M=0dS}. In even dimensions the chiral matrix commutes with the spinor Lie derivative, thus, chirality is preserved. We verify this below explicitly in the chiral representation \eqref{chirL} by showing that the upper and lower spinor components do not mix under the Lie derivative  \eqref{SLD}. We find the appearance of our ladder operators in the computation.

We now proceed to show that massless helicity spinors do not mix under the isometry group. To this end, the relevant case to consider is that corresponding to a non-compact de Sitter boost generators (i.e. involving the $t$-coordinate) spanning a so(1,1) $\subset$ so(4,1). Without loss of generality we can take it to be
\begin{equation}
     \bm D=\sin T  \cos \theta_3 \,  \partial_T + \cos T \sin \theta_3\,   \partial_{\theta_3} \,\,.
\end{equation}
Its action on spinors is given by \eqref{SLD} which explicitly reads
\begin{equation}
\mathcal{L}_{ D} \Psi = \sin T \cos \theta_3 \, \partial_T \Psi + \cos T \sin \theta_3 \, \partial_{\theta_3} \Psi - \frac{1}{2} \gamma^{\ud{03}} \sin T \sin \theta_3 \Psi \,\,.
    \label{boost}
\end{equation}
Notice that in the chiral representation \eqref{chirL} the last term is diagonal in spinor indices $\gamma^{\ud{ 03}} =\sigma_3\otimes\sigma_3= \text{diag}(1,-1,-1,1) $.   

The action of \eqref{boost} on  the $\Psi^{(\downarrow,a)}_{(+)M=0\,L_3...} $ solution yields\footnote{The opposite energy modes $\Psi^{(\downarrow,a)}_{(-)M=0\,L_3...} $ yield an analogous result.}
\begin{align}
\mathcal{L}_{  D}\Psi^{(\downarrow,a)}_{(+)M=0\,L_3...} = (\sin T )^{3/2}  \Bigg\{ &\frac{e^{i (\omega-1)T}}{2} \Big[ \left( \frac{3}{2}-\omega \right) \cos \theta_3 + \sin \theta_3 \partial_{\theta_3} - \frac{i}{2}\sin \theta_3 \sigma_3 \Big]\Phi^{(\downarrow,a)}_{(+)L_3...} \nn\\
&+ \frac{e^{i(\omega +1)T}}{2} \Big[  \left(\frac{3}{2}+\omega \right) \cos \theta_3 + \sin \theta_3 \partial_{\theta_3} +\frac{i}{2} \sin \theta_3 \sigma_3\Big]  \Phi^{(\downarrow,a)}_{(+)L_3...} \Bigg\}\,\,.
    \label{op}
\end{align}
Since the  4d fermion $\Phi_{(+)L_3...}^{(...)}$ involves the 3d harmonic  $\bm \chi_{(+)L_3...}^{(...)}$, its associated scaling dimension is $\Delta_+=-(L_3+\frac12)$. Using the relation \eqref{dSEner} we can write 
$$\frac{3}{2}-\omega=\frac{1}{2}+\Delta_+~\text{  and }~~ \frac{3}{2}+\omega=\frac{5}{2}-\Delta_+ \,\,.$$
It is now amusing to see that the expressions inside the brackets in \eqref{op} acting on $  \Phi$ are precisely the ladder operators built out from the $S^3$ CKV  \eqref{ckvtheta}, i.e. $  {\bm c }=\sin \theta_3\,  \partial_{\theta_3}.$
Their explicit form is
\begin{equation}
    \mathcal{D}^{s}= \sin \theta_3 \partial_{\theta_3} + \left(\frac{5}{2}-\Delta_+ \right) \cos \theta_3 + \frac{i}{2} \sin \theta_3 \sigma_3\,\,,
\end{equation}
\begin{equation}
    \mathcal{D}=  \sin \theta_3 \partial_{\theta_3} + \left(\frac{1}{2}+\Delta_+ \right) \cos \theta_3 - \frac{i}{2} \sin \theta_3 \sigma_3\,\,.
\end{equation}
Using the relations  \eqref{ladd1}, the action of the boost on the negative frequency solutions is 
\begin{align}
\mathcal{L}_{  D} \Psi^{(\downarrow,a)}_{(+)ML_3L_2\ell}&(T,\Omega)\rfloor_{M=0} \nn\\
&= (\sin T)^{3/2} \Bigg[       \frac{e^{i (\omega-1)T}}{2} \left(
\begin{array}{c}
\mathcal{D}\bm\chi^{(a)}_{(+)L_{3}L_2\ell} (\Omega) \\
  0
\end{array}
\right) + \frac{e^{i (\omega +1)T}}{2} \left(
\begin{array}{c}
\mathcal{D}^{s}\bm\chi^{(a)}_{(+)L_{3}L_2\ell} (\Omega) \\
  0
\end{array}
\right) \Bigg]\nn \\
&= (\sin T) ^{3/2} \Bigg[ -\frac{e^{i(\omega-1)T}}{2} (L_3-L_2)  \left(
\begin{array}{c}
\bm\chi^{(a)}_{(+)L_{3}-1\,L_2\ell} (\Omega) \\
  0
\end{array}
\right) \nn\\
&\hspace{3cm}\quad + \frac{e^{i(\omega+1)T}}{2} (3+L_3+L_2) \left(
\begin{array}{c}
\bm\chi^{(a)}_{(+)L_3+1\,L_2\ell} (\Omega) \\
  0
\end{array}
\right) \Bigg] \,\,.
\end{align}
We can rewrite this result as
\begin{equation}
\mathcal{L}_{ D} \Psi^{(\downarrow,a)}_{(+)M=0\,L_3L_2\ell}  = \frac{3+L_3+L_2}2  \Psi^{(\downarrow,a)}_{(+)M=0\,L_3+1\,L_2\ell} -   \frac{L_3-L_2}2\Psi^{(\downarrow,a)}_{(+)M=0\,L_3-1\,L_2\ell}  \,\,,
\label{transf}
\end{equation}
reproducing the results of \cite{lets1/2}. 

We conclude this section by noting that \eqref{transf} is particularly significant. While for massive spin-$\frac12$ fermions the up/down mode labels mix under the action of the connected isometry group \cite{lets1/2}, for massless spin-$\frac12$ fermions the $(\downarrow,a)$ and $(\uparrow,a)$ states transform under independent irreducible representations. Consequently, the total solution space fully decomposes into two decoupled sectors\footnote{In \eqref{M0f} we refer to \emph{positive} energy modes, a similar splitting happens for the charge conjugate modes.}
\be
\text{\sf de Sitter}_4:~~~\text{span}\big\{ \Psi_{  M\, L_{3}\,...}^{(a_2,a_1)}\big\}_{\rfloor M\ne0}~\rightarrow~~\text{span}\big\{ \Psi_{  M\, L_{N-1}\,...}^{(\uparrow,a_1)}\big\}_{\rfloor M=0}\oplus~\text{span}\big\{ \Psi_{  M\, L_3\,...}^{(\downarrow,a_1)}\big\}_{\rfloor M=0} \,\,.
\label{M0f}
\ee
This halving of the representation space, when compared to massive representations, is the defining hallmark of Discrete or Exceptional representations. It provides the de Sitter analog to the chirality splitting and multiplet shortening familiar from flat even dimensional Minkowski space.\footnote{Strictly speaking, the decoupling occurs only for the connected component of the de Sitter group,  SO$_0$(4,1). If discrete symmetries such as parity are included to form the full group O(4,1) (or the Spin(1,4) group for fermions), the two sectors are mapped into one another and must be combined to form a single irreducible representation of O(4,1).}$^{,}$\footnote{In addition, we also have the complex conjugate solutions which upon quantization describe anti-particles.}

\subsection{Spin$-\frac32$: Rarita-Schwinger spinors}
\subsubsection{Euclidean signature: $S^{N}$}

The relation between the scaling dimension $\Delta$ in \eqref{RSef} and the  principal quantum number $L_N$ in \eqref{diracrarita}  is identical to that of the spin-$\frac{1}{2}$ eigenmodes. For TT-modes, we have
\begin{align}
\Psi_{\mu\,(+)L_N...}^{...}&=\psi_{\mu\,\Delta_+}~~~\text{with}~~ \Delta_+=-(L_N+\tfrac12) \,\,,\nn\\
\Psi_{\mu\,(-)L_N...}^{...}&=\psi_{\mu\,\Delta_-}~~~\text{with}~~ \Delta_-= L_N+N-\tfrac12 \,\,.
\label{rel2}
\end{align}
The $\Delta$ subscripts on the right-hand side of \eqref{rel2} emphasize the specific values of the scaling dimension that yield unitary irreducible representations (UIRs) (see Appendix \ref{B}).

We begin by noting that the Lie derivative of a vector-spinor \eqref{RSLD} along the CKV \eqref{ckvtheta} reduces  to 
\begin{align}
\mathcal{L}_{  c} \Psi_{\theta_N} &= \left(\sin \theta_N \partial_{\theta_N} + \cos\theta_N\right)\Psi_{\theta_N}\,\,, \\
\mathcal{L}_{  c} \Psi_{\theta_i} &= \sin \theta_N \partial_{\theta_N} \Psi_{\theta_i} \,\,.
\end{align}
Inserting these expressions into the ladder operators
\eqref{raritaladders} and acting on type I modes we find, setting $C_N=1$,
\begin{align}   \mathcal{D}\Psi^{{\sf I}\,(\vec{a})}_{\mu\,(+)  L_N  L_{N-1}...} &=-\frac{(L_N-L_{N-1})(1+L_N)}{L_N} \Psi^{{\sf I}\,(\vec{a})}_{\mu\,(+) L_N-1 \, L_{N-1}... }\nn\\
\mathcal{D}^{s}\Psi^{{\sf I}\,(\vec{a})}_{\mu\,(+)  L_N  L_{N-1}... }&=\frac{(L_N+N-1)(N+L_N+L_{N-1})}{L_N+N}  \Psi^{{\sf I}\,(\vec{a})}_{\mu\,(+)  L_N+1\,  L_{N-1}... }\nn\\
\mathcal{D}^{s}\Psi^{{\sf I}\,(\vec{a})}_{\mu\,(-)  L_N  L_{N-1}... } &=-\frac{(L_N-L_{N-1})(1+L_N)}{L_N} \Psi^{{\sf I}\,(\vec{a})}_{\mu\,(-)  L_N-1\,  L_{N-1}... }\nn\\
\mathcal{D}\Psi^{{\sf I}\,(\vec{a})}_{\mu\,(-)  L_N  L_{N-1}... } &=\frac{(L_N+N-1)(N+L_N+L_{N-1})}{L_N+N}  \Psi^{{\sf I}\,(\vec{a})}_{\mu\,(-)  L_N+1\,  L_{N-1}... }
\end{align}
For type-II modes, we find ($C_N=1$)
\begin{align}
\mathcal{D}\Psi^{{\sf II}\,(\vec{a})}_{\mu\,(+)  L_N  L_{N-1}... } &= -(L_N-L_{N-1})\Psi^{{\sf II}\,(\vec{a})}_{\mu\,(+)  L_N-1\,  L_{N-1}... }\nn\\
\mathcal{D}^{s} \Psi^{{\sf II}\,(\vec{a})}_{\mu\,(+)  L_N  L_{N-1}... } &= (N+L_N+L_{N-1})\Psi^{{\sf II}\,(\vec{a})}_{\mu\,(+)  L_N+1\,  L_{N-1}... }\nn\\
\mathcal{D}^{s}\Psi^{{\sf II}\,(\vec{a})}_{\mu\,(-)  L_N  L_{N-1}... }  &= -(L_N-L_{N-1}) \Psi^{{\sf II}\,(\vec{a})}_{\mu\,(-)  L_N-1\,  L_{N-1}... }  \nn\\
 \mathcal{D}\Psi^{{\sf II}\,(\vec{a})}_{\mu\,(-)  L_N  L_{N-1}... }  &= (N+L_N+L_{N-1}) \, \Psi^{{\sf II}\,(\vec{a})}_{\mu\,(-)  L_N+1\,  L_{N-1}... } 
\end{align}
Notice that the ladder operators act on type-II modes in exactly the same way as they act on Dirac spinors (cf. eqn. \eqref{ladd1}). 

It remains to find the action  of $\mathcal{\tilde D}$. As in the case of spin-$\frac12$ we  distinguish between even and odd dimensions. In even dimensions,  the  $\mathcal{\tilde D}$-action gives a sign depending on whether it acts on  $a_n =\,\uparrow $ or $a_n=\,\downarrow $ modes, whereas in odd dimensions no such distinction appears. We elaborate the  $N=4$ and $N=5$ cases for concreteness; the results extend naturally  to other dimensions. 

In $N=4$ dimensions the action  on type I modes yields 
\begin{align}
   \text{\sf N = 4 (even dim)}:~~~     \mathcal{\tilde D} \Psi^{{\sf I}\,(\uparrow,a)}_{\mu\,(\pm)  L_4  L_3 L_2 \ell} &= -\frac{1}{3}\left(\frac{3}{2}+L_3 \right)\Psi^{{\sf I}\,(\uparrow,a)}_{\mu\,(\mp)  L_4  L_3 L_2 \ell} \,\,,\nn\\
\mathcal{\tilde D} \Psi^{{\sf I}\,(\downarrow,a)}_{\mu\,(\pm)  L_4  L_3 L_2 \ell} &= \frac{1}{3}\left(\frac{3}{2}+L_3 \right)\Psi^{{\sf I}\,(\downarrow,a)}_{\mu\,(\mp)  L_4  L_3 L_2 \ell}\,\,.
\end{align}
This implies that, for the vector-spinor case, we can summarize the action as (cf. \eqref{tildeLow})
$$\mathcal{\tilde D} = \frac{i}{3} \sigma_3\otimes \cancel {\tilde \nabla} .$$ 
In odd dimensional spacetimes,  as in the spinor case,  the result is independent of the values of $(a_2,a)$. The operator acts as 
\begin{equation}
  \text{\sf N = 5 (odd dim)}:~~~       \mathcal{\tilde D} \Psi^{{\sf I}\,(  \vec{a})}_{\mu\,(+)L_5 L_4... } =\frac{1}{2}(L_4+2) \Psi^{{\sf I}\,(  \vec{a})}_{\mu\,(+)L_5 L_4... } \,\,.
\end{equation}
Using \eqref{oddgamma5} we can write
$$\mathcal{\tilde D}=- \frac{1}{2}\gamma^{\ud 5}\cancel{\tilde \nabla}\,\,.$$  
We conclude by  noticing  that the action of 
$\mathcal{\tilde D}$ on type-II modes is identical to its action in the spin-$\frac12$ case,  see \eqref{ladd1}.\eqref{Dtil1/2}.

\subsubsection{Lorentzian signature: de Sitter$_{N}$}

The relation between the scaling dimension $\Delta$ in \eqref{RSef} and the mass $M$ in the TT-modes equation\footnote{Equation \eqref{RSDir} is not the  Rarita-Schwinger equation as originally formulated in \cite{RSpr} (see also \cite{freedman}), 
\be
\gamma^{\mu\nu\rho}\left(\nabla_\nu-\frac M2\gamma_\nu\right)\Psi_\rho=0 \,\,,
\label{RSeq}
\ee
with $M\in\mathbb{R}$. However, since the transverse and gamma-traceless conditions $\nabla^\mu\Psi_\mu=\gamma^\mu\Psi_\mu=0$ follow directly from the equation of motion for $M\ne0$, a straightforward manipulation shows that \eqref{RSeq} can be recast into the equivalent form \eqref{RSDir} (see \cite{dio+silva} for a recent discussion).} 
\be
\text{\sf TT-modes}:~~~~\slashed{\nabla}  \Psi_{\mu\, M...}^{(\vec a)}  = M \Psi_{\mu\,M...}^{(\vec a)}~~\oplus~~\nabla^\mu\Psi_{\mu\, M...}^{(\vec a)}=\gamma^\mu\Psi_{\mu\, M...}^{(\vec a)}=0 \,\,,
\label{RSDir}
\ee
is again (cf. \eqref{rel111})
\begin{align}
 \Delta=\frac{N-1}2+iM ~~\leadsto~~\Psi_{\mu\,ML_{N-1}...}^{(\vec a)} 
&=\psi_{\mu\,\Delta} \,\,.
\end{align}

In the following we discuss 4-dimensional de Sitter space because the spin-$\frac32$ gauge field is exceptional: it furnishes a UIR only in $N=4$. For generic values of the mass $M \in \mathbb{R}$, type-I and type-II modes mix under the action of the de Sitter isometry group. However, at the exceptional values $M = 0$ (zero Dirac mass) and $M= \pm i$ (fermionic gauge field), the representations become reducible under the isometry group. In particular, for  zero Dirac mass the solution space  decomposes into two distinct helicity sectors (cf. \eqref{M0f}) \cite{letsios3/2,vasilirarita},
\begin{align}
\text{\sf Massless}:~~~&\text{span}\big\{ \Psi_{\mu\,  M\,...}^{{\sf I}\,(a_2,a_1)},\Psi_{\mu\,  M\,...}^{{\sf II}\,(a_2,a_1)}\big\}_{\rfloor M\ne0,\pm i}\nn\\
&\rightarrow~~\text{span}\big\{ \Psi_{\mu\,  M\,...}^{{\sf I}\,(\uparrow,a_1)},\Psi_{\mu\,  M\,...}^{{\sf II}\,(\uparrow,a_1)}\big\}_{\rfloor M=0}~ \oplus~\text{span}\big\{ \Psi_{\mu\,  M\,...}^{{\sf I}\,(\downarrow,a_1)},\Psi_{\mu\,  M\,...}^{{\sf II}\,(\downarrow,a_1)}\big\}_{\rfloor M=0} \,\,. 
\end{align}
While the spin-$\frac32$ field space at $M=0$ is fully decomposable into two irreducible representations, the gravitino at the gauge points $M=\pm i$ forms an indecomposable representation due to the isometry-induced mixing of Type II modes with Type I pure gauge modes
\begin{align}
\text{\sf Gauge field}:&~~~\text{span}\big\{ \Psi_{\mu\,  M\,...}^{{\sf I}\,(a_2,a_1)},\Psi_{\mu\,  M\,...}^{{\sf II}\,(a_2,a_1)}\big\}_{\rfloor M\ne0,\pm i} \rightarrow~~
\text{span}\big\{ \underbrace{\Psi_{\mu\,  M\,...}^{{\sf I}\,(a_2,a_1)}}_{\rm pure ~gauge},\Psi_{\mu\,  M\,...}^{{\sf II}\,(a_2,a_1)}\big\}_{\rfloor M=\pm i} \,\,.
\end{align}
This is the well known characteristic behavior of gauge fields: under symmetry transformations $g$ the indecomposability of the fermionic gauge fields solution space $V=V^{\sf I}\oplus V^{\sf II}$ means that $g\cdot v^{\sf I}=v'^{\sf I}$ and $g\cdot v^{\sf II}=v'^{\sf II}+v^{\sf I}$.

Unlike in the  spin-$\frac12$ case, the spin-$\frac32$   operators \eqref{DirM} connect distinct discrete UIRs of the de Sitter isometry group. For spin-$\frac{3}{2}$, the ladder operators map the zero Dirac mass solution space into the   fermionic gauge field space. In the notation of Table 2 in \cite{dS4meta} our ladder operators map  
$${\cal D,D}^s:~~D_{\frac32,\frac12}^\pm\to D_{\frac32,-\frac12}^\pm.$$ 
Using the expressions  \eqref{raritaladders}, their action  on the Type II modes can be explicitly computed and gives \footnote{The Lie derivative expression \eqref{RSLD} evaluated for the  CCKV \eqref{lorCKV} gives
\begin{align}
    \mathcal{L}_c \Psi_t =  \cosh t \, \partial_t \Psi_{t} + \sinh t \, \Psi_t ,~~~~~~
\mathcal{L}_c \Psi_{\theta_i}    = \cosh t \, \partial_t \Psi_{\theta_i  }
\end{align}}
\begin{equation}
...\stackrel[\mathcal{D}]{\mathcal{D}^{s}}{\xrightleftharpoons{\hspace{4mm}}}\Psi ^{{\sf II} }_{\mu\,M=-2i} \stackrel[\mathcal{D}]{\mathcal{D}^{s}}{\xrightleftharpoons{\hspace{4mm}}}\boxed{\Psi^{{\sf II} }_{\mu\,M=-i}
\underset{~\mathcal{D} }{\longleftarrow}  \Psi^{{\sf II}\, }_{\mu\,M=0}
\overset{\mathcal{D}^{s}}{\longrightarrow}  \Psi^{{\sf II}  }_{\mu\,M=i} }
\stackrel[\mathcal{D}]{\mathcal{D}^{s}}{\xrightleftharpoons{\hspace{4mm}}}\Psi^{{\sf II}}_{\mu\,M=2i}\stackrel[\mathcal{D}]{\mathcal{D}^{s}}{\xrightleftharpoons{\hspace{4mm}}}...\nn
\end{equation}
We have boxed here the set of solutions giving UIRs. See  Appendix \ref{discretes} for further details. The missing arrows connecting $M=\pm i \to M=0$ are absent due to the fermionic gauge eigenmodes being annihilated by the ladder operators,
\be
\mathcal{D}\Psi_{M=i}^{{\sf II}}=0 \,\,, ~~~~\mathcal{D}^s\Psi_{M=-i}^{{\sf II}}=0\,\,,~~~~~\forall\, L_3\,\,. 
\label{zm3/2}
\ee

Finally,  it is interesting to recognize that the operator $\mathcal{\tilde D}$, when built from any non-isometric CKV  agrees with the operator $T_V$ introduced in eqn. (6.6) in \cite{let24} for $M=\pm i$ spin-$\frac32$ fields in dS$_4$. For fermionic gauge fields, \cite{letsios3/2,vasilirarita} demonstrated that the $\tilde {\cal D}$ operator  
is a symmetry of the field equations. Furthermore, for $M=\pm i/\ell$ the commutator algebra of $\tilde {\cal D}$ operators for \emph{all} de Sitter CKVs gives  a representation of the conformal algebra  so$(2,4)$. Our $\mathcal{\tilde D}$ extends the action to arbitrary mass solutions, however we have not been able to identify a closed algebraic structure in the generic case.

For completeness, we close this section quoting the action of the ladder operators on zero Dirac mass type I modes in dS$_4$. The action is  summarized in the following diagram 
\begin{align}
...\stackrel[\mathcal{D}]{\mathcal{D}^{s}}{\xrightleftharpoons{\hspace{4mm}}}\Psi ^{{\sf I} }_{\mu\,M=-2i} \underset{~\mathcal{D} }{\longleftarrow} & \Psi^{{\sf I} }_{\mu\,M=-i}
\underset{~\mathcal{D} }{\longleftarrow}   \Psi^{{\sf I}\, }_{\mu\,M=0} 
\overset{\mathcal{D}^{s}}{\longrightarrow}  \Psi^{{\sf I}  }_{\mu\,M=i} 
\overset{\mathcal{D}^{s}}{\longrightarrow}  \Psi^{{\sf I}}_{\mu\,M=2i}\stackrel[\mathcal{D}]{\mathcal{D}^{s}}{\xrightleftharpoons{\hspace{4mm}}}...\nn
\end{align}
In this diagram, as in \eqref{dig1/2}, it is important to recognize that the missing arrows from $M=\pm 2i\to M=\pm i$ are absent because the ladders ${\cal D, D}^s$   are not defined for $M=  2i $ and $M=-2i$, respectively, since the last term in \eqref{raritaladders} blows up for $\Delta=-1/2$ and $\Delta^s=-1/2$ respectively. Relations \eqref{zm3/2} still hold for type I modes. Additional relations can be easily found, examples are
 \begin{equation}
\mathcal{D}\Psi^{\sf I}_{\mu \,M=0\, L_3...}=-i\left(\frac{L_3}{2}+1\right)\Psi^{\sf I}_{\mu \,M=-i\, L_3...}~~~\forall\,L_3\,\,,
\end{equation}
\begin{equation}
    \mathcal{D}\Psi^{\sf I}_{\mu\, M=in\, L_3=n-2}=0 \,\,, \quad n=3,4,... 
\end{equation} 

\section{Conclusions}

In this work, we have constructed ladder operators for fermionic fields of spin-$\frac{1}{2}$ and spin-$\frac{3}{2}$ that relate distinct, discretely labeled unitary irreducible representations (UIRs) of   {Spin}$(N+1)$ on $S^N$ and of $\mathrm{Spin}(N,1)$ on $\mathrm{dS}_N$. In contrast to the bosonic case studied in \cite{LSS}, these transformations cannot be understood solely as the infinitesimal action of an isometry on a primary field. The construction requires additional terms beyond the standard primary transformation law, leading to new differential operators that, to the best of our knowledge, have not previously been identified in the literature. In particular, we have found three distinct  operators,  denoted by $\mathcal{D}$, $\mathcal{D}^{s}$, and $\widetilde{\mathcal{D}}$, that shift the eigenvalue of the Dirac operator and organize the fermionic mode spaces in a manner closely related to, but richer than, their bosonic counterparts.

Two of these operators, $\mathcal{D}$ and $\mathcal{D}^{s}$, play a role analogous to the ladder operators introduced in \cite{LSS}: they shift the conformal label according to $\Delta\rightarrow\Delta\pm1$ and act as eigenvalue-shifting raising and lowering operators for both spin-$\frac{1}{2}$ and spin-$\frac{3}{2}$ modes. Despite this analogy, the fermionic construction exhibits important differences. In particular, while for spin-$\frac12$  we find no analogue with the critical value $\Delta_c$ where the bosonic ladder fails to shift the eigenvalue, we find a critical scaling dimension for spin-$\frac32$  at $M=\pm2i/\ell$. On the other hand, both spin-$\frac12$ and spin-$\frac32$ fermions possess an additional structure encoded by $\widetilde{\mathcal{D}}$ which, in all dimensions, intertwines solutions with opposite Dirac eigenvalues. This should be  contrasted with the familiar action of the chiral matrix $\gamma_{*}$, which provides a similar exchange of Dirac eigenvalues but exists only in even dimensions, or charge conjugation whose existence for sign flipping is also dimension dependent \cite{Stone,tanii}. For the (Lorentzian) massless spin-$\frac12$ case, which corresponds to $\Delta=\Delta^s=(N-1)/2$,  the operator $\tilde {\cal D}$ remarkably reduces to the spin-$\frac{1}{2}$ primary transformation (see \eqref{new}). In  $N=4$ dimensions, our spin-$\frac32$ $\widetilde{\mathcal{D}}$-operators for $M=\pm i/\ell$\footnote{They corresponds to $\Delta=\frac{1}{2}$.} close a SO$(4,2)$ algebra and reduce to the operators $T_V$ found in \cite{letsios3/2}. In the arbitrary mass case, the $\tilde{\cal D}$ operators for both spin-$\frac12$ and spin-$\frac32$ generalize the $\hat{\cal L}_\zeta$ and $T_V$ found in \cite{bT} and \cite{letsios3/2}. However, the commutator algebra  does not close on SO(4,2) and appears to be more involved. We leave this analysis for future work.

For Dirac spinors on $S^N$, the ladder construction provides a particularly direct algebraic method for generating the complete tower of spinor harmonics. Starting from the lowest harmonic set  ($L_N=0$), which spans the Killing spinor  solution space, the full set of higher harmonics can be generated through repeated application of the ladder operators. 

We have also studied the Lorentzian realization of this structure. In $\mathrm{dS}_4$, we explicitly demonstrated how the ladder operators constructed from the conformal Killing vectors of $S^3$ relate to the action of   de Sitter boosts on zero Dirac mass mode solutions. This provides an explicit realization of the ladder structure directly in terms of the de Sitter isometry group. Along the way we showed how the zero Dirac mass solutions furnish a shortened representation. Furthermore, for both spin-$\frac{1}{2}$ and spin-$\frac{3}{2}$ fields we  made explicit the distinction between principal- and discrete-series representations in terms of their corresponding free-mode solutions.

Finally, we showed that the ladder construction provides a direct connection between the $M=\pm i/\ell$ (strictly massless) and the  $M=0$ (zero Dirac mass) spin-$\frac{3}{2}$ modes. For spin-$\frac{1}{2}$, an analogous construction relates the massless Dirac solutions to the space of Killing spinor solutions in de Sitter space. These results indicate that the ladder structure generated by non-isometric conformal Killing vectors 
points to a deeper relation between conformal geometry, fermionic differential operators, and the representation theory underlying de Sitter mode spaces.

Taken together, our results establish a unified ladder-operator framework for fermionic fields on spheres and de Sitter space. The resulting structure provides an explicit bridge between the spectrum of fermionic differential operators, conformal transformations, and the organization of their solutions into UIRs. In particular, the appearance of three distinct operators, rather than a simple fermionic analogue of the two bosonic ladders, suggests that the representation-theoretic structure of fermionic fields contains additional information that is not visible from the primary transformation law alone. This raises the natural question of whether these operators form a larger algebraic structure which we wish to explore in future work.

In appendices \ref{B} and \ref{spin dS} we have collected  the fermionic mode solutions for spin-$\frac{1}{2}$ and spin-$\frac{3}{2}$ fields in Euclidean (sphere) and Lorentzian (de Sitter) signatures. In appendix \ref{casimirsS3}, we derived explicit expressions for the Casimir operators of SO(4)  and SO(3,1)  when acting on fermionic representations.   This explicitly demonstrated the inequivalence of UIRs associated with positive- and negative-eigenvalue spinor harmonics on odd-dimensional spheres in the simplest instance. Finally, we discussed in appendix \ref{apendixE} the SO(4,1) Casimirs in $dS_4$, computing both Casimir operators for spin-$\frac{1}{2}$, showing   that positive- and negative-eigenvalue Dirac modes furnish equivalent irreducible representations (as expected since they are related by $\gamma_*$). For spin-$\frac{3}{2}$, we obtained the quadratic Casimir and discussed the structure of the quartic one. We included these appendices for completeness, since to the best of our knowledge, explicit expressions for both Casimir operators in  fermionic realizations have not been systematically worked out in the literature. We therefore hope that the results collected in the appendices may provide a useful reference for future studies of fermion  fields on maximally symmetric spaces.

In future work, we intend to investigate how the algebra generated by $\mathcal{D}$, $\mathcal{D}^s$, and $\widetilde{\mathcal{D}}$ extends beyond the closed $\mathfrak{so}(4,2)$ algebra established for spin-$\frac{3}{2}$ gauge fields in four dimensions and massless Dirac fermions in arbitrary dimensions. Determining the properties of this algebra in the general massive case could reveal whether these ladder operators generate a larger symmetry structure on the full space of fermionic solutions. This approach would offer a natural framework for generalizing our findings to higher-spin fields,  following the insights found in  \cite{letsios3/2},   illuminating possible hidden conformal symmetries in de Sitter field theory.

\section*{Acknowledgments}

We thank our colleagues and friends D. Anninos, V. Letsios, M. Mizrahi and  A. Waldron for comments, discussions and correspondence. This work was supported by PIP2025-0101210 CONICET.

\appendix

\section{CKV potentials in MSS}
\label{ckvPot}

An $N$-dimensional (curved) MSS can be modeled  as the quadric hypersurface\footnote{The scaling of the space is set to $\ell=1$. }
\be
K(X^0)^2+g_{\mu\nu} X^\mu X^\nu=K \,\,,~~~~K=\pm1 \,\,,
\label{Kt}
\ee
embedded in $(N+1)$-dimensional ambient space
$$ds^2=K(dX^0)^2+g_{\mu\nu} dX^\mu dX^\nu,~~~~\mu=1,...,N.$$
Here $K=+1$ gives positive curvature metrics  (spheres or de Sitter),  whereas $K=-1$ gives negatively curved ones (hyperbolic space or AdS). The Euclidean/Lorentzian character of the MSS is encoded in the metric $g_{\mu\nu}$ used for  contracting the $X^\mu$ indices.  Choosing $g_{\mu\nu}=\delta_{\mu\nu}$ gives spheres $S^N$ or hyperbolic spaces $H^N$, whereas picking $g_{\mu\nu}=\eta_{\mu\nu}=\text{diag}(-+++...)$ gives de Sitter dS$_N$ or anti-de Sitter AdS$_N$ spacetimes.

Solving  \eqref{Kt} as
$$X^0=\frac{r^2-K}{1+K r^2}\,\,,~~~X^\mu=\frac{2x^\mu}{1+Kr^2}\,\,,~~~~r^2=(x^\mu)^2\,\,,$$
leads to 
$$ds^2=\frac4{(1+Kr^2)^2}(dx^\mu)^2 \,\,.$$
The $N+1$ non-isometric CKV of MSS can be written in terms of the embedding   coordinates $X^I$ with $ I=(0,\mu) $ as $c^{(I)}_\mu=\partial_\mu X^I$ with  (see \cite{let24,Allen86} and app.A in \cite{LSS}).

\section{Fermionic Spherical Harmonics}
\label{B}

In this Appendix we present our conventions  and review the recursive method developed in \cite{higu+camp} for cons\-truc\-ting higher-dimensional spinors from lower-dimensional ones. The procedure  naturally accommodates the change in gamma matrix size when transitioning from odd to even dimensions. We explicitly spell out the spinor harmonics in 
$N=2,3$ and $4$ dimensions.

\vspace{2mm}

\nin {\sf Conventions}:
\[ 
\begin{array}{ll}
  \text{.  {\sf Vielbeins}:} &
g_{\mu\nu}=e^{\ud a}_\mu e^{\ud b}_\nu\,\eta_{\ud{ab}}\\ 
 \text{\sf . Full covariant derivative (diffeo+Lorentz)}: &\nabla =\partial+\Gamma+\omega \\
\text{\sf . $n$-bein~postulate}: &0=\nabla_\mu e^{\ud a}_\nu=\partial_\mu e^{\ud a}_\nu -\Gamma_{\mu\nu}^\rho e^{\ud a}_\rho+\omega_\mu{}^{\ud{ab}}e_{\ud b\nu}  \\
\text{\sf .   Flat gammas}: & \{\gamma_{\ud a},\gamma_{\ud b}\}=2\eta_{\ud{ab}} \\
\text{\sf . Curved gammas}: &  \gamma_\mu=e^{\ud a}_\mu\gamma_{\ud a}~\leadsto~ \{ \gamma_{\mu} , \gamma_{\nu}\} =2g_{\mu \nu} \\
\text{\sf .   Spin$-\frac12$ covariant derivative}: &\nabla_\mu\psi=\partial_\mu\psi+\frac14\omega_\mu{}^{\ud{ab}}\gamma_{\ud{ab}}\psi~\\
& ~~~~~~~~~~~\text{  where } ~~~~~~\gamma_{\ud{ab}}=\frac12[\gamma_{\ud a},\gamma_{\ud b}]\\
\text{\sf .   Spin$-\frac32$ covariant derivative}: &\nabla_\mu\psi_\rho=\partial_\mu\psi_\rho+\frac14\omega_\mu{}^{\ud{ab}}\gamma_{\ud{ab}}\psi_\rho-\Gamma_{\mu\rho}^\alpha\psi_\alpha 
\end{array}
\]

\nin {\bf Sphere parametrization}\\ 
The $N$-sphere  coordinates are denoted $x^\mu=(\theta_N, ..., \theta_2, \theta_1)$ with  $\theta_N,\theta_i\in[0,\pi] ~ (i=2,...,N-1)$ and $\theta_1=\phi\in[0,2\pi]$. The metric on the sphere $S^{N}$ reads
\begin{equation}
   \text{\sf N-sphere}:~~~ ds^2 = d\theta_N^2 + \sin^2 \theta_N \, \tilde{ds}^{2}\,\,,
    \label{SN}
\end{equation}
where $\tilde{ds}^{2}$ is the metric on the $S^{N-1}$ sphere. We will  denote objects on $S^{N-1}$ with a tilde. 

\vspace{2mm}

\nin {\bf Christoffel and spin connections}\\
The non-zero Christoffel symbols of the $N$-sphere are
\begin{equation}
\Gamma^{\theta_N}_{\theta_i \theta_j} = - \sin \theta_N \cos \theta_N\, \tilde g_{\theta_i \theta_j} \,\,, \quad \Gamma^{\theta_j}_{\theta_i \theta_N} = \cot \theta_N   \, \delta_i^j  \,\,, \quad \Gamma^{\theta_k}_{\theta_i \theta_j} = \tilde\Gamma^{\theta_k}_{\theta_i \theta_j} \,\,, ~~~~~  i,j,k=1,...,N-1 .
\end{equation}
$N-$beins are
\begin{equation}
    \bm{e}^{\ud N} = \bm d\theta_N \,\,, \quad \bm{e}^{\ud i} = {\sin \theta_N}\,\,\, \bm{\tilde e}^{\ud i} \,\,,~~~ \quad i=1,...,N-1.
\end{equation}
with $\bm{\tilde e}^{\ud i}$  the $n-$beins of  $S^{N-1}$. The solution to
\be
\bm d\bm e^{\ud a}+\bm\omega^{\ud{ab}}\wedge\bm e_{\ud b}=0 \,\,,
\label{torsion}
\ee
for the spin connection components reads \cite{higu+camp,pope,cho}
\begin{equation}
    \bm\omega ^{ \ud{ij}}=   \tilde {\bm \omega} ^{\ud{ij}} \,\,, \quad \bm\omega ^ {\ud{iN}}=  \cos \theta_N\,\tilde {\bm e}^ {\ud i } \,\,, ~~~~\quad i,j =1,...,N-1.
    \label{sPCon}
\end{equation}

\vspace{1mm}

\nin {\bf Spinor conventions}\\
Dirac spinors are $2^{[\frac N 2]}$ dimensional column vectors, where $[...]$ denotes integer part. Flat gamma matrices in $N$-dimensions are $2^{[\frac N 2]}\times 2^{[\frac N 2]}$ satisfying
\begin{equation}
    \{ \gamma^{\ud a},\gamma^{\ud b} \}=2 \delta^{\ud{ab}} \,\,, \quad a,b=1,...N.
\end{equation}

\vspace{1mm} 

\nin {\sf $N$ even}: when the manifold dimension increases from odd to even, the size of the gamma matrices doubles. We decompose them as $\gamma^{\ud a} = (\gamma^{\ud N} , \gamma^{\ud i}) $, and construct them from gamma matrices $\tilde\gamma^{\ud i}$ in one-lower dimension  as
\begin{equation}
\text{\sf Chiral representation}:~~~\gamma^{\ud{N}}=   \left(
\begin{array}{cc}
 0 & \mathds{1} \\
\mathds{1} & 0 \\ 
\end{array}
\right)  \,\,, \quad \gamma^{\ud{i}}=   \left(
\begin{array}{cc}
 0 & i \tilde{\gamma}^{\ud i} \\
-i\tilde{\gamma}^{\ud i} & 0 \\ 
\end{array}
\right) \,\,, \quad i=1,...,N-1.
\label{chir}
\end{equation}
The chiral matrix is defined to be hermitian as
\be
\text{\sf Chiral matrix}:~~~\gamma_*=\left(
\begin{array}{cc}
  -\mathds{1}&0 \\
0&\mathds{1}   \\ 
\end{array}
\right)=c \prod_{\ud a}\gamma^{\ud a}~~\leadsto~~(\gamma_*)^2=\mathds{1}\,\,,
\label{5mat}
\ee
with $c$ a phase factor. In \eqref{chir}-\eqref{5mat}, the identity matrix $\mathds 1$ is $\frac122^{[\frac N 2]}$ dimensional.

\vspace{1mm}

\nin{\sf  $N$ odd}: as we increase the dimension from even to odd, the gamma matrices keep their size. We obtain the odd-dimensional set of gamma matrices appending the  `chiral matrix' to the even-dimensional set\footnote{We need to properly adjust a phase for the new matrix to square to one.}. Writing $\gamma^{\ud a} = (\gamma^{\ud N},\gamma^{\ud{N-1}},\gamma^{\ud i})$ we have
\begin{equation}
\gamma^{\ud N}=\left(
\begin{array}{cc}
 \mathds{1} & 0 \\
0 & -\mathds{1} \\ 
\end{array}
\right)   \,\,, \quad 
\gamma^{\ud{N-1}}=   \left(
\begin{array}{cc}
 0 & \mathds{1} \\
\mathds{1} & 0 \\ 
\end{array}
\right)  \,\,, \quad 
\gamma^{\ud i}=   \left(
\begin{array}{cc}
 0 & i \tilde{\tilde{\gamma}}^{\ud i} \\
-i\tilde{\tilde{\gamma}}^{\ud i} & 0 \\ 
\end{array}
\right) \,\,, \quad i=1,...,N-2.
\label{N odd gamma}
\end{equation}
here $\tilde{\tilde{\gamma}}$ denote gamma matrices in $N-2$ dimensions.

\vspace{2mm}

\nin{\bf Covariant derivatives}\\
In the following $\tilde\nabla_\mu=\partial_\mu+\frac14\tilde \omega_\mu{}^{\ud{ab}}\tilde\gamma_{\ud{ab}},~\mu\in(\theta_1,...\theta_{N-1})$ denote the covariant derivative on $S^{N-1}$ and   $\tilde \gamma_{\theta_i} = \tilde e_{\theta_i}^{\ud i}\tilde\gamma_{\ud i}$ is the curved one-lower dimension gamma matrix with $\tilde e^{\ud i}_{ \theta_i}$   the $n$-beins of    $S^{N-1}$.

\nin {\sf $N$ even}: from \eqref{sPCon}, we have 
\begin{equation}
\nabla_{\theta_N}= \partial_{\theta_N} \,\,, \quad \nabla_{\theta_i} =  \left(
\begin{array}{cc}
 \tilde\nabla_{\theta_i} + \frac{i}{2}\cos \theta_N\, \tilde \gamma_{ \theta_i} & 0 \\
0& \tilde\nabla_{\theta_i} - \frac{i}{2}\cos \theta_N\, \tilde \gamma_{ \theta_i} \\ 
\end{array}
\right)   \,\,, \quad \theta_i \neq \theta_N\,\,.
\label{N even spin}
\end{equation}

\vspace{1mm}

\nin {\sf $N$ odd}:  the spinor covariant derivative now reads
\begin{equation}
    \nabla_{\theta_N}= \partial_{\theta_N}\,\,,~~~~~~~\nabla_{\theta_i}=\tilde\nabla_{\theta_i}- \frac{1}{2}\cos \theta_N\, \gamma^{\ud{N}}\,  \tilde{\gamma_{ \theta_i}}\,\,,~~~~\theta_i\ne\theta_N\,\,. 
\end{equation}

\subsection{Spin$-\frac12$ harmonics on $S^N$}
\label{s1/2Con}

In this section, we summarize how the spin-$\frac1 2$  harmonics on spheres $S^{N}$ are built from lower dimensional ones. When Wick rotated, the solutions turn into the solutions of the Dirac equation in  (global) de Sitter spacetime (see app. \ref{spin dS} and refs. \cite{higu+camp,lets1/2} for further details). 

Spin-$\frac12$ harmonics for the Dirac operator $\slashed{\nabla}:=\gamma^\mu\nabla_\mu$  on $S^N$ satisfy\footnote{In the present Appendix we set the sphere curvature radius in \eqref{Dirac1} to $\ell=1$.}
\be
\slashed{\nabla}_{S^{N}}\Psi^{(a_n,a_{n-1},...a_1)}_{(\pm) L_N L_{N-1}...L_2\ell} = \pm i \left( L_N + \frac{N}{2} \right) \Psi^{(a_n,a_{n-1},...a_1)}_{(\pm)L_N L_{N-1}...L_2\ell} \,\,,
\label{DiracN}
\ee
with quantum numbers taking integer values
$L_N,L_{N-1},...L_2,\ell\in\{0,1,2,...\}$
and satisfying
\be
L_{N}\geq L_{N-1}\geq...\geq L_2\geq\ell\geq0 \,\,.
\label{regu}
\ee
Here $L_N$ is the principal quantum number characterizing the UIR and $L_2,...,L_{N-1}$ label the eigenvalues of the Dirac operator on lower dimensional spheres.
Schematically, 
(cf. \eqref{s3})
\begin{align}
(\hat{\slashed{\nabla}}_{S^{N-1}})^2\Psi^{(a_n,a_{n-1},...a_1)}_{(\pm) L_N L_{N-1}...L_2\ell} &= - \left( L_{N-1} + \frac{N-1}{2} \right)^2 \Psi^{(a_n,a_{n-1},...a_1)}_{(\pm)L_N L_{N-1}...L_2\ell} \,\,,\nn\\
(\hat{\slashed{\nabla}}_{S^{N-2}})^2\Psi^{(a_n,a_{n-1},...a_1)}_{(\pm) L_N L_{N-1}...L_2\ell} &= - \left( L_{N-2} + \frac{N-2}{2} \right)^2 \Psi^{(a_n,a_{n-1},...a_1)}_{(\pm)L_N L_{N-2}...L_2\ell} \,\,,\nn\\
\vdots\nn\\
(\hat{\slashed{\nabla}}_{S^{2}})^2\Psi^{(a_n,a_{n-1},...a_1)}_{(\pm) L_N L_{N-1}...L_2\ell} &= - \left( L_{2} + 1 \right)^2 \Psi^{(a_n,a_{n-1},...a_1)}_{(\pm)L_N L_{N-1}...L_2\ell} \,\,.
\label{casis}
\end{align}
where $\hat{\slashed{\nabla}}_{S^{n}}$ denotes the Dirac operator on a lower dimensional sphere  $S^n$ 
but acting on the $N$-dimensional spinor. More precisely, this operator should be understood as a tensor product  $\hat{\slashed{\nabla}}_{S^{n}}={\mathds 1}_2\otimes...\otimes {\slashed{\nabla}}_{S^{n}}$.  The  notation $\ell$ for the last subindex emphasizes its role as an azimuthal quantum number
\begin{align}
{\cal L}_{\partial_\phi}\Psi^{(...,\uparrow)}_{...\ell}&=+ i\,(\ell+\tfrac12)\Psi^{(...,\uparrow)}_{...\ell} \,\,,\nn\\
{\cal L}_{\partial_\phi}\Psi^{(...,\downarrow)}_{...\ell}&=- i\,(\ell+\tfrac12)\Psi^{(...,\downarrow)}_{...\ell}\,\,,~~~~\ell = 0,1,2,... \,\,.
\label{LieKill}
\end{align}
Here ${\cal L}_{\partial_\phi}$ is the spinor Lie derivative \eqref{SLD} along $\partial_\phi$, which, by virtue of \cite{Unruh,CMcL,Miz}
\be
[{\cal L}_\zeta,\cancel\nabla]=0~~~~~\text{iff}~~~\zeta~\text{ is a Killing vector},
\label{KillDirac}
\ee
can be simultaneously diagonalized with the Dirac operator \cite{Unruh, CMcL, Miz}. The dimension of the vector $\vec a$  is $n=[\tfrac N2]$ with $[..]$ denoting the integer part, and the possible values are $a_i\in\{\uparrow,\downarrow\}$
\footnote{In the recursive construction to be summarized below, the set $\{a_i\}$  denotes the two possible ways  the higher (even) dimensional spinor  can be build out from a lower dimensional  one  (see \eqref{a1}-\eqref{a2}). Up and down arrow account for the possibility of constructing with positive or negative lower dimensional Dirac eigenvalues.}. Finally, the subindex $(\pm)$ denotes  the sign of the Dirac   eigenvalue in \eqref{DiracN}.

\vspace{2mm}

\nin {\bf On transformation properties}  \\
It follows from \eqref{KillDirac} that for any choice of $\{L_{N-1}, \dots, L_2, \ell\}$ and $\{a_i\}$, the set of eigenfunctions with a fixed $(\pm)$ sign and Dirac eigenvalue $L_N$ transform into one another under the action of the sphere Killing vectors, thereby forming a UIR. In other words, the eigen-spaces corresponding to positive and negative Dirac eigenvalues in \eqref{DiracN} do not mix under the action of the symmetries. Concomitantly, each subspace with a fixed Dirac eigenvalue spans a distinct UIR. The sets
$$\{\Psi^{(a_n,a_{n-1},...a_1)}_{(+) L_N L_{N-1}...L_2\ell}\}\,\,,\{\Psi^{(a_n,a_{n-1},...a_1)}_{(-) L_N L_{N-1}...L_2\ell}\}\,\,,~~~L_N=\text{fixed}\,\,,$$
span independent UIRs. In even dimensions, the sets are equivalent UIRs, related by the action of the chiral matrix $\Psi _{(-) L_N ...}=\gamma_*\Psi _{(+) L_N ...}$. Whereas for odd dimensional spheres they are distinct irreps. These statements, from a group theory point of view, are expressed naturally in terms identical or distinct highest-weight vectors (see the discussion below eqn. (2.25) in \cite{vasilirarita}). For completeness, we quote that in all dimensions the labels $\vec a=(\uparrow,\vec a')$ and $\vec a=(\downarrow,\vec a')$ mix under the symmetry group. In even dimensions,  this is seen by performing an analytically continuation of  eqn (5.46) in \cite{lets1/2}. In odd dimensions, the Killing vector mixing the labels lives in the lower dimensional sphere.   

An explicit example of the inequivalence of  positive and negative Dirac eigenmodes on odd dimensional spheres can be seen  in Appendix \ref{casimirsS3}, where we consider the case of $S^3$. There we compute the Spin(4) Casimir operators for the spin-$\frac12$ and spin-$\frac32$ harmonics. The result is that one of the Casimirs coincides with the Dirac squared shifted by a constant (cf. \eqref{C1spin12} and \eqref{C1RS}), whereas the second (quadratic) Casimir reduces to the (first order) Dirac operator on $S^{3}$, hence distinguishing the set $\{ \Psi_{(+)L_N...}^{...} \}$ from the set $\{ \Psi_{(-)L_N...}^{...}\}$ as inequivalent irreps (see \eqref{c2spin12},\eqref{c2spin32}).
\vspace{2mm}

\nin {\bf On complex conjugation}  \\
Since spinor harmonics are unique on the sphere, complex conjugation does not give new or independent UIRs other than those already found by solving \eqref{DiracN} with both signs. 

In technical terms, consider $B_\pm$ matrices satisfying \cite{tanii} (see also \cite{freedman,KT})
$$B_\pm^{-1}\gamma^{\ud a}B_\pm=\pm(\gamma^{\ud a})^* \,\,.$$
These matrices always exist for both $\pm$ signs in even dimensions, whereas only one of them exists in odd dimensions. It is then straightforward to see that given a spinor harmonic $\Upsilon=\Psi_{(+)L_N...}^{...}$ satisfying \eqref{DiracN} the conjugate spinor
$$\Upsilon^{c_\pm}:=B_\pm\Upsilon^* \,\,,$$ 
whenever exists, satisfies
$$\slashed\nabla\Upsilon =  i\left( L_N + \frac{N}{2} \right)\Upsilon ~~\Rightarrow~~\slashed\nabla\Upsilon^{c_\pm} =\mp  i\left( L_N + \frac{N}{2} \right)\Upsilon^{c_\pm} \,\,.$$
Therefore, since the vector spaces associated to each given eigenvalue are unique, charge conjugation of the harmonic basis generates no new independent solution spaces other than those already found for both signs in \eqref{DiracN}. This is not the case in Lorentzian signature see below.

\subsubsection{$N$ even}

The Dirac operator $\slashed{\nabla}:=\gamma^\mu\nabla_\mu$ with our definition of gamma matrices results in
\begin{align}
N\text{\sf ~even}:~~\slashed{\nabla}_{S^{N}}  &=\gamma^{\ud N} \left( \partial_{\theta_N} + \frac{N-1}{2} \cot \theta_N\right)  + \frac{1}{\sin \theta_N}\left(
\begin{array}{cc}
 0 & i \slashed{\nabla}_{S^{N-1}} \\
-i\slashed{\nabla}_{S^{N-1}} & 0 \\ 
\end{array}
\right) \nn\\ 
&=\left(\begin{array}{cc}
 0 & (\partial_{\theta_N} + \tfrac{N-1}{2} \cot \theta_N)\mathds{1}+\tfrac i{\sin\theta_N} \slashed{\nabla}_{S^{N-1}} \\
(\partial_{\theta_N} + \tfrac{N-1}{2} \cot \theta_N)\mathds{1}-\tfrac i{\sin\theta_N} \slashed{\nabla}_{S^{N-1}} & 0 \\ 
\end{array}\right) \,\,.
\label{DiracEven}
\end{align}
Our aim is to construct the eigenfunctions satisfying \eqref{DiracN}. 
Two types of solutions can be constructed out from the $S^{N-1}$ eigen-spinor  
\begin{equation}
  \slashed\nabla_{S^{N-1}}  \bm\chi^{(a_{n-1},...,a_1)}_{(\pm) L_{N-1}...\ell}  = \pm i \left(L_{N-1} + \frac{N-1}{2} \right)\bm\chi^{(a_{n-1},...,a_1)}_{(\pm) L_{N-1}...\ell}\,\,, 
  \label{SN-1}
\end{equation}
they are denoted by an extra supra-index $\uparrow,\downarrow$ appended to the set $(a_n,...,a_1)$
\begin{equation}
\Psi^{(\uparrow,a_{n-1},...,a_1)}_{(\pm) L_N L_{N-1}...\ell}(\theta_N,\Omega_{N-1}) =   \left(
\begin{array}{c}
  i\,f (\theta_N) \,\bm\chi^{(a_{n-1},...,a_1)}_{(+)L_{N-1}...\ell} (\Omega_{N-1}) \\
  \pm g (\theta_N) \,\bm\chi^{(a_{n-1},...,a_1)}_{(+)L_{N-1}...\ell} (\Omega_{N-1}) \\
\end{array}
\right)\,\,,
\label{a1}
\end{equation}
\begin{equation}
\Psi^{(\downarrow,a_{n-1},...,a_1)}_{(\pm) L_N L_{N-1}...\ell}(\theta_N,\Omega_{N-1}) =   \left(
\begin{array}{c}
  g (\theta_N)\, \bm\chi^{(a_{n-1},...,a_1)}_{(-)L_{N-1} ... \ell} (\Omega_{N-1}) \\
  \pm i\,f (\theta_N) \,\bm\chi^{(a_{n-1},...,a_1)}_{(-)L_{N-1}...\ell} (\Omega_{N-1}) \\
\end{array}
\right)\,\,,
\label{a2}
\end{equation}
here $\Omega_{N-1}$ denotes the whole set of  $S^{N-1}$ coordinates, and the quantum number associated to the eigenvalue in \eqref{DiracN} was added as the subindex $L_N$. Notice that the spinor has doubled its size wrt $\bm\chi$. In the following we will simplify the notation by denoting $\vec a=(a_{n},...,a_1)$. Alternatively, the $\uparrow,\downarrow$-indices denote the sign of lower dimensional eigen-spinor.

The functions $f(\theta_N)$ and $g(\theta_N)$ become related when we insert the ansatze \eqref{a1}-\eqref{a2} in \eqref{DiracN}.
One obtains,
\begin{equation}
    \left[\partial_{\theta_N} + \frac{N-1}{2}\cot \theta_N - \frac{1}{\sin \theta_N} \left(L_{N-1} +\frac{N-1}{2} \right) \right] g(\theta_N)=- \left( L_N + \frac{N}{2} \right)  f(\theta_N)\,\,, 
  \label{phipsi}
\end{equation}
\begin{equation}
    \left[\partial_{\theta_N} + \frac{N-1}{2}\cot \theta_N + \frac{1}{\sin \theta_N} \left(L_{N-1} +\frac{N-1}{2} \right) \right] f(\theta_N)=\left( L_N + \frac{N}{2} \right)g(\theta_N)\,\,.
    \label{psiphi}
\end{equation}
Regularity of the solutions in $\theta_N$ implies the eigenvalue must be quantized $L_N=0,1,2...$ with $L_N \geq L_{N-1}$. The final expressions are
\begin{align}
    f(\theta_N)=& C_N\frac{L_N + N/2}{L_{N-1}+N/2 }\cos^{L_{N-1}} \left( \frac{\theta_N}{2} \right)  \sin^{L_{N-1} + 1} \left( \frac{\theta_N}{2} \right) \nn\\&{_2}F_1\left(L_{N}+L_{N-1}+N,-L_{N}+L_{N-1} ;\,L_{N-1}+\tfrac{N}{2} + 1;\sin
   ^2\left(\frac{\theta_N}{2}\right)\right) \,\,,
   \label{psi}
\end{align}
\begin{align}
    g(\theta_N) =&  C_N \cos^{L_{N-1} + 1} \left( \frac{\theta_N}{2} \right)  \sin^{L_{N-1}} \left( \frac{\theta_N}{2} \right) \nn\\
    &{_2}F_1\left(L_{N}+L_{N-1}+N,-L_{N}+L_{N-1} ;\,L_{N-1}+\tfrac{N}{2};\sin
   ^2\left(\frac{\theta_N}{2}\right)\right) \,\,.
   \label{phi}
\end{align}
The quantum numbers satisfy
\begin{equation}
    L_{N} \geq L_{N-1} \,\,, \quad L_{N},L_{N-1}=0,1,2...\,\,.
\label{spin12QN}
\end{equation}
The normalization constant $C_N$ is fixed using the inner product over $S^N$
\begin{equation}
\int_{S^{N}} d\Omega_{N} \, \Psi^{(\vec a)}_{(\pm)L_N,  ...\ell}(\Omega)^{\dagger}  \Psi^{(\vec b)}_{(\pm)L'_{N}...\ell'}(\Omega) = \delta^{\vec a,\vec b}\delta_{L_{N}, L'_{N}}...\,\delta_{\ell,\ell'}\,\,,
    \label{SN-1 product}
\end{equation}
for the ease of notation we have collected the set $\{a_i\}$  in the vector $\vec a=(a_n,...,a_1)$.

It is easy to see that  in  even dimensional spheres the negative eigenvalue modes can be obtained from  positive  modes (and  vice-versa) acting with the chiral matrix  \eqref{5mat}
\begin{equation}
    \gamma_* \Psi^{(\vec a)}_{(+)L_{N}...\ell}= \Psi^{(\vec a)}_{(-)L_{N}...\ell}\,\,.
    \label{plmn}
\end{equation}

\subsubsection{$N$ odd}

The Dirac operator $\slashed{\nabla}:=\gamma^\mu\nabla_\mu$  reads
\begin{align}
    \text{\sf N odd }: \quad \slashed{\nabla}\Psi_{}&= 
\gamma^{\ud N}\left(\partial_{\theta_N} + \frac{N-1}{2} \cot \theta_N \right)  + \frac{1}{\sin \theta_N} \slashed{\nabla}_{S^{N-1}}\,\,.
\end{align}
When compared to \eqref{DiracEven} we see that all block components are non-zero, however its explicit form is not needed for the computations below. Since in odd dimensions the size of the spinor does not change wrt the (even) lower dimensional one, concomitantly no additional $a$-index appears. The positive and negative eigen-spinors of the Dirac operator in $S^N$ are built from the $S^{N-1}$ eigen-spinors $\bm \chi$ satisfying \eqref{SN-1} as \cite{higu+camp}
\begin{equation}
    \Psi^{(\vec a)}_{(\pm)L_N...\ell}(\theta_N, \Omega_{N-1})=  \frac{(1+i\gamma^{\ud N})}{\sqrt{2}} (g(\theta_N) \pm i f(\theta_N)\gamma^{\ud N}) {\bm\chi}^{(\vec a)}_{(-)L_{N-1}...\ell}(\Omega_{N-1})
    \label{1/2Nodd}\,\,.
\end{equation}
Notice that  $ {\bm \chi}$ is an even dimensional sphere eigen-spinor, thus, $\gamma^{\ud N}$ for $N$ odd  is the  chiral matrix in one lower dimension (cf. \eqref{5mat} and \eqref{N odd gamma}), thus, it fulfills the following relation (cf. \eqref{plmn})
\begin{equation}
    \gamma^{\ud N} {\bm \chi}^{(\vec a )}_{(\pm)L_{N-1}...\ell}=  \bm \chi^{(\vec a )}_{(\mp)L_{N-1}...\ell}\,\,.
    \label{oddgammaN}
\end{equation}
This relation allows to write  \eqref{1/2Nodd}  as
\begin{align}
    \Psi^{(\vec a)}_{(\pm) L_N...\ell}(\theta_N, \Omega_{N-1}) = \frac{1}{\sqrt{2}}&\big(g(\theta_N)\mp f(\theta_N)\big)\, {\bm\chi}^{(\vec a)}_{(-)L_{N-1}...\ell}(\Omega_{N-1})\nn\\
    &+ \frac{i}{\sqrt{2}}\big(g(\theta_N) \pm f(\theta_N)\big)\, {\bm\chi}^{(\vec a)}_{(+)L_{N-1}...\ell}(\Omega_{N-1})\,\,.
\end{align}
Inserting this ansatz into  \eqref{DiracN} one finds that the functions $f(\theta_N)$ and $g(\theta_N)$ are again related by equations \eqref{phipsi} \eqref{psiphi}, so that the \emph{smooth} solutions  are given by \eqref{phi} and \eqref{psi}.

In odd $N$ dimensions, negative and  positive eigenvalue modes  are connected through the application of the  Dirac operator in one lower dimension, i.e. 
\begin{equation}
    \gamma^{\ud N} \cancel{\tilde \nabla} \Psi^{(\vec a)}_{(+)L_{N}...\ell} = - \left(L_{N-1 }+ \tfrac{N-1}{2}\right) \Psi^{(\vec a )}_{(-)L_{N}...\ell}\,\,,
    \label{oddgamma5}
\end{equation}
here  $\tilde{\slashed\nabla}=\slashed\nabla_{S^{N-1}}$.

\subsubsection{Compendium of explicit solutions and its construction}
\label{S2 spinor}

\subsection*{. $S^2$}

Line element  
\begin{equation}
    ds^2 = d\theta_2 ^2 + \sin^2 \theta_2 \, d\phi \,\,. 
\end{equation}
Zweibeins  
\begin{equation}
    \bm e^{\ud{2}}= \bm d\theta_2 \,\,, \quad \bm e^{\ud{1}}= \sin \theta_2 \, \bm d\phi \,\,,~~\leadsto~~g_{\mu \nu} = e_{\mu}^{\ud{a}} e_{\nu}^{\ud{b}}\,\delta_{\ud{ab}} \,\,.
\end{equation}
Christoffel symbols 
\begin{equation}
    \Gamma^{\theta_2}_{\phi \phi} = - \cos \theta_2 \sin \theta_2 \,\,, \quad \Gamma^{\phi}_{\phi \,\theta_2} = \Gamma^{\phi}_{\theta_2 \,\phi} = \cot \theta_2 \,\,.
\end{equation}
Spin connection  
$$\bm\omega^{\ud {21}}=-\cos\theta_2\,\bm d\phi \,\,.$$
Covariant derivative $\nabla_\mu=\partial_\mu+\frac14\omega_{\mu\ud{ab}}\gamma^{\ud{ab}}$ is
\begin{equation}
\nabla_{\theta_2} = \partial_{\theta_2} \,\,, \quad \nabla_{\phi} = \partial_\phi + \frac{1}{2} \cos \theta_2 \, \gamma^{\ud{12}}\,\,.
\end{equation}
Flat Gamma matrices (chiral representation)
\begin{equation}
    \gamma^{\ud{2}} = \sigma^{1}
\,\,, \quad \gamma^{\ud{1}} = -\sigma^{2} \,\,,
~~\leadsto~~\{ \gamma^{\underline a},\gamma^{\underline b}\}=2\,\delta^{\ud{ab}}\,\,.
\end{equation}
gives the chiral matrix
$$\gamma_*=\sigma^3 \,\,.$$
Our aim is to solve the spin-$\frac12$ harmonic  equation
\begin{equation}
\slashed{\nabla}_{S^{2}}\Psi_{(\pm)}= \pm i(L_2+1)\Psi_{(\pm)}\,\,,~~~L_2\ge0\,\,,
\label{spHarm}
\end{equation}
the parametrization of the eigenvalue will become clear below. Here $(\pm)$ refer to positive   and negative eigenvalues of the Dirac operator.  
Dirac eigenfunctions $\Psi$ can be also diagonalized wrt the azimuthal Killing vector $\bm\partial_\phi$ (cf. \eqref{LieKill}-\eqref{KillDirac}). We separate positive and negative eigenvalues of ${\cal L}_{\partial_\phi}$ as 
\begin{align}
{\cal L}_{\partial_\phi}\Psi^{(\uparrow)}&=+ i\,(\ell+\tfrac12)\Psi^{(\uparrow)} \,\,, \nn\\
{\cal L}_{\partial_\phi}\Psi^{(\downarrow)}&=- i\,(\ell+\tfrac12)\Psi^{(\downarrow)}\,\,,~~~~\ell = 0,1,2,...\,\,.
\label{azi}
\end{align}
Thus, the explicit equations to be solved are
\be
\eqref{spHarm}~~\leadsto \quad    \gamma^{\ud{2}} \left( \partial_{\theta_2} + \frac{1}{2} \cot \theta_2 \right)\Psi_{(\pm)}^{(a)} + \frac{1}{\sin \theta_2} \gamma^{\ud{1}}\partial_\phi \Psi_{(\pm)}^{(a)} = \pm i (L_2+1) \Psi_{(\pm)}^{(a)}\,\,,~~~~~a=1,2\,.
\label{spH2}
\ee
$$\eqref{azi}~~\leadsto \quad\partial_\phi\Psi^{(a)}_{(\pm)}=s_a i\,(\ell+\tfrac12 )\Psi^{(a)}_{(a)}\,\,,~~~~~~~~s_\uparrow=+\,\,,~s_\downarrow=-\,\,.$$
The appropriate ans\"atze are
\begin{equation}
        \Psi^{(\uparrow)}_{(\pm)} (\theta_2, \phi)= \left(
\begin{array}{c}
i f(\theta_2)   \\
  \pm g(\theta_2)  \\
\end{array}
\right) e^{ i (\ell+ \frac{1}{2})\phi} \,\,, \quad \ell = 0,1,2,...  \, \, ,
\nn
\end{equation}
\begin{equation}
    \Psi^{(\downarrow)}_{(\pm)} (\theta_2, \phi)= \left(
\begin{array}{c}
g(\theta_2)  \\
  \pm i f(\theta_2) \\
\end{array}
\right) e^{ -i (\ell+ \frac{1}{2})\phi} \,\,, \quad \ell = 0,1,2,...  \, \,,
\label{ans1}
\end{equation}
which plugged  into \eqref{spHarm}, result in
\begin{equation}
   \left( \partial_{\theta_2}  + \frac{1}{2}\cot \theta_2  - \frac{\ell+\frac12  }{\sin \theta_2} \right)g(\theta_2) = - (L_2+1)\, f(\theta_2) \,\,,\nn
\end{equation}
\begin{equation}
      \left( \partial_{\theta_2}  + \frac{1}{2}\cot \theta_2  + \frac{\ell+\frac12}{\sin \theta_2} \right)f(\theta_2) =  (L_2+1)\, g(\theta_2)\,\,.
\end{equation}
Imposing regularity of $g,f$ at the north and south pole of the 2-sphere ($\theta_2= 0,\pi$) demands  the eigenvalue $L_2$  to be quantized and the azimuthal quantum number $\ell$ to be bounded by the principal quantum number 
$$ L_2=0,1,2,...~~\text{and}~~  \ell\leq L_2\,\,.$$ 
The solutions are found to be
\begin{align}
    f(\theta_2)&= \frac{L_2+1}{\ell+1}\sin ^{\ell+1}\left(\frac{\theta_2 }{2}\right) \cos ^{\ell}\left(\frac{\theta_2
   }{2}\right) \, _2F_1\left(L_2+\ell+2,\ell-L_2;\ell+2;\sin ^2\left(\frac{\theta_2
   }{2}\right)\right) \,\,,\nn\\
g(\theta_2)&= \sin ^{\ell}\left(\frac{\theta_2 }{2}\right) \cos ^{\ell+1}\left(\frac{\theta_2
   }{2}\right) \, _2F_1\left(L_2+\ell+2,\ell-L_2;\ell+1;\sin ^2\left(\frac{\theta_2
   }{2}\right)\right)\,\,.
   \label{ans2d}
\end{align}
Summarizing,
\begin{align}
\slashed{\nabla}_{S^{2}}\Psi^{(a)}_{(\pm)L_2 \ell} &= \pm i (L_2 +1)\Psi^{(a)}_{(\pm)L_2 \ell}\,\,, ~~~~ L_2,\ell=0,1,2...\,\,,~~a=\uparrow,\downarrow ~~~  \text{ with }~~\ell\leq L_2 \,\,,\nn\\
{\cal L}_{\partial_\phi}\Psi^{(a)}_{(\pm)L_2 \ell}&=s_a i\,(\ell+\tfrac12)\Psi^{(a)}_{(\pm)L_2 \ell}\,\,,~~~~~~~~~s_\uparrow=+1\,\,,~~s_\downarrow=-1\,\,.
\label{s2spin}
\end{align}

\subsection*{. $S^{3}$}
\label{S3 spinor}
The 3-sphere metric reads
\begin{equation}
    ds^2= d\theta_3^2 + \sin^2 \theta_3 \big(d\theta^2_2 +  \sin^2 \theta_2 d\phi^2\big) \,\,.
\label{g3}
\end{equation}
Christoffel symbols
\begin{equation}
\Gamma^{\theta_3}_{\theta_2  \theta_2} = -\cos \theta_3 \sin \theta_3 \,\,, \quad \Gamma^{\theta_3}_{\phi \phi} =-\cos \theta_3 \sin \theta_3 \sin^2 \theta_2 \,\,, \quad \Gamma^{\theta_2}_{\theta_2 \theta_3} = \Gamma^{\theta_2}_{\theta_3 \theta_2}=  \cot \theta_3 \,\,,
\end{equation}
\begin{equation}
    \Gamma^{\theta_2}_{\phi \phi}= - \cos \theta_2 \sin \theta_2  \,\,, \quad \Gamma^{\phi}_{\phi \theta_3} = \Gamma^{\phi}_{ \theta_3 \phi} = \cot \theta_3 \,\,,  \quad \Gamma^{\phi}_{\phi\theta_2}=\Gamma^{\phi}_{\theta_2 \phi}= \cot \theta_2 \,\,.
\end{equation}
Dreibeins,
\begin{equation}
 \bm e^{\ud 3}= \bm d\theta_3 \,\,, \quad \bm e^{\ud 2}= \sin \theta_3\, \bm d\theta_2 \,\,, \quad \bm e^{\ud 1 }= \sin \theta_3 \sin \theta_2\, \bm d\phi \,\,.
\end{equation}
Spin connection
\be
\bm\omega^{\ud{12}}=\cos\theta_2\,\bm d\phi\,\,,~~~~ \bm\omega^{\ud{13}}=\cos\theta_3\sin\theta_2\,\bm d\phi\,\,,~~~~ \bm\omega^{\ud{23}}=\cos\theta_3 \,\bm d\theta_2 \,\,.
\label{scS3}
\ee
Covariant derivatives:
\begin{equation}
    \nabla_{\theta_3} = \partial_{\theta_3}  \,\,,\quad \nabla_{\theta_2}= \partial_{\theta_2} +\frac{1}{2}\cos \theta_3  \gamma^{\ud{23}} \,\,, \quad \nabla_{\phi}= \partial_\phi + \frac{1}{2} \cos \theta_3 \sin \theta_2 \gamma^{\ud{13}} + \frac{1}{2} \cos \theta_2 \, \gamma^{\ud{12}}\,\,. 
\end{equation}
Flat Gamma matrices,
\begin{equation}
\gamma^{\ud 3}=\sigma^{3} \,\,, \quad \gamma^{\ud 2}= \sigma^{1} \,\,, \quad \gamma^{\ud 1}= - \sigma^{2}\,\,.
\label{Gm}
\end{equation}
Spinor harmonics on $S^3$ satisfy
\begin{align}
\slashed{\nabla}_{S^{3}}\Psi^{(a)}_{(\pm)}&=\pm i (L_3+\tfrac32) \Psi^{(a)}_{(\pm)}\,\,,~~~~~~ ~~a=\uparrow,\downarrow\nn\\
{\cal L}_{\partial_\phi}\Psi^{(a)}_{(\pm)}&=s_a i\,(\ell+\tfrac12)\Psi^{(a)}_{(\pm)}\,\,,~~~~\ell = 0,1,2,...\,\,,~~s_\uparrow=+1\,\,,~~~s_\downarrow=-1\,\,.
\end{align}
where  $(\pm)$ denotes the sign of the Dirac operator eigenvalue whereas $a$ denotes the sign of the spinor Lie derivative eigenvalue $s_a$. 
The equation to be solved explicitly reads
\begin{equation}
\left[\gamma^{\ud{3}} \left( \partial_{\theta_3} + \cot \theta_3\right) + \frac{1}{\sin \theta_3} \slashed{\nabla}_{S^{2}}\right]\Psi_{(\pm)}^{(a)} = \pm i (L_3+\tfrac32) \Psi_{(\pm)}^{(a)}\,\,,~~~~\lambda>0\,\,. 
\label{spH3}
\end{equation}
To solve it, following \cite{higu+camp}, we propose the ansatz
\begin{equation}
    \Psi_{(\pm)}^{(a)}(\theta_3, \theta_2, \phi) = (g(\theta_3)\mp f(\theta_3))\bm\chi^{(a)}_{(-)L_2 \ell}(\theta_2,\phi) + i(g(\theta_3) \pm f(\theta_3))\bm\chi^{(a)}_{(+)L_2\ell}(\theta_2, \phi) \,\,,
    \label{ansatz3}
\end{equation}
with $$\slashed{\nabla}_{S^{2}}\bm\chi^{(a)}_{(\pm) L_2 \ell }= \pm i (L_2+1) \bm\chi^{(a)}_{(\pm) L_2 \ell}\,\,,$$ 
the $S^2$ eigenspinor found in \eqref{s2spin}. Plugin \eqref{ansatz3} into \eqref{spH3} one obtains 
\begin{equation}
\left(  \partial_{\theta_3} +\cot \theta_3- \frac{L_2+1}{\sin\theta_3} \right)g(\theta_3)= - (L_3+\tfrac32) \, f(\theta_3) \,\,,
\end{equation}
\begin{equation}
\left(  \partial_{\theta_3} +\cot \theta_3 + \frac{L_2+1}{\sin\theta_3} \right)f(\theta_3)=  (L_3+\tfrac32) \, g(\theta_3) \,\,.
\end{equation}
Regularity at $\theta_3=0,\pi$ quantizes the eigenvalue $L_3$ and imposes 
$$ L_3=0,1,2,...~~\text{and}~~   L_3\geq L_2\geq\ell\,\,.$$
The regular solutions for $g,f$ are (cf. \eqref{ans2d})
\begin{align}
    f(\theta_3) &= \frac{L_3+\frac{3}{2}}{L_2+\frac{3}{2}  } \cos^{L_2} \left( \frac{\theta_3}{2} \right)  \sin^{L_2+1} \left( \frac{\theta_3}{2} \right) {_2}F_1\left(L_3+L_2+3,-L_3+L_2 ;\,L_2+\tfrac{5}{2};\sin
   ^2\left(\frac{\theta_3}{2}\right)\right) \,\,,\nn\\
    g(\theta_3) &= \cos^{L_2+1} \left( \frac{\theta_3}{2} \right)  \sin^{L_2} \left( \frac{\theta_3}{2} \right) {_2}F_1\left(L_3+L_2+3,-L_3 + L_2; \,L_2+\tfrac{3}{2};\sin
   ^2\left(\frac{\theta_3}{2}\right)\right) \,\,.
   \label{sols3}
\end{align}
Summarizing, 
\begin{align}   \slashed{\nabla}_{S^{3}}\Psi^{(a)}_{(\pm) L_3L_2\ell}&=  \pm i\left(L_3 + \tfrac{3}{2}\right) \Psi^{(a)}_{(\pm) L_3 L_2 \ell} \,\,,\nn\\
(\slashed{\nabla}_{S^{2}})^2\Psi^{(a)}_{(\pm) L_3L_2\ell}&=-\left(L_2 + 1\right)^2\Psi^{(a)}_{(\pm) L_3 L_2 \ell} \,\,,\nn\\
{\cal L}_{\partial_\phi}\Psi^{(a)}_{(\pm) L_3L_2\ell}&=s_a i\,(\ell+\tfrac12)\Psi^{(a)}_{(\pm) L_3L_2\ell}\,\,,~~~~s_a=\pm1 \,\,.
\label{s3}
\end{align}
the quantum numbers $L_3,L_2,\ell=0,1,2,....$ satisfy
$$L_3\ge L_2\ge \ell\ge 0~~~~~\text{and}~~~~a=\uparrow,\downarrow \,\,.$$


\subsection*{. $S^{4}$}
\label{S4 spinor}

4-sphere metric 
\begin{equation}
ds^2 = d\theta_4 ^2 +\sin^2 \theta_4 \, \left(   d\theta_3 ^2 + \sin^2 \theta_3 (  d\theta^2_{2} + \ \sin^2 \theta_2 d\phi ^2)\right)\,\,.
\end{equation}
Christoffel symbols:
$$ 
\Gamma^{\theta_4}_{\theta_3 \theta_3}=-\cos \theta_4 \sin \theta_4 \,\,, \quad \Gamma^{\theta_4}_{\theta_2 \theta_2}= - \cos \theta_4 \sin \theta_4 \sin^2 \theta_3 \,\,, \quad \Gamma^{\theta_4}_{\phi\phi}=-\cos \theta_4 \sin\theta_4 \sin^2 \theta_3 \sin^2 \theta_2 \,\,,
$$ $$
\Gamma^{\theta_3}_{\theta_3 \theta_4}= \cot \theta_4 \,\,, \quad \Gamma^{\theta_3}_{\theta_2 \theta_2}=- \cos \theta_3 \sin\theta_3 \,\,, \quad \Gamma^{\theta_3}_{\phi\phi} = -\cos \theta_3 \sin \theta_3 \sin^2 \theta_2 \,\,,
$$ 
$$  
\Gamma^{\theta_2}_{\theta_2 \theta_4}= \cot \theta_4 \,\,, \quad \Gamma^{\theta_2}_{\theta_2 \theta_3}=\cot \theta_3 \,\,, \quad \Gamma^{\theta_2}_{\phi \phi} = - \cos \theta_2 \sin \theta_2 \,\,,
$$  
\begin{equation}
\Gamma^{\phi}_{\phi \theta_4} = \cot \theta_4 \,\,, \quad \Gamma^{\phi}_{\phi \theta_3} = \cot \theta_3  \,\,, \quad \Gamma^{\phi}_{\phi \theta_2}= \cot \theta_2 \,\,.
\end{equation}
which can be succinctly  written as
$$\Gamma^{\theta_4}_{\theta_i \theta_j}=- \cos \theta_4 \sin \theta_4\, \tilde g_{ij} \,\,,$$
with $\tilde g_{ij}$ the  3-sphere metric \eqref{g3}.\\
Vierbeins are 
\begin{equation}
    \bm  e^{\ud{4}}=\bm d\theta_4  \,\,, \quad \bm e^{\ud{3}} = \sin \theta_4 \,\bm d\theta_3\,, \quad \bm e^{\ud{2}} = \sin \theta_4 \sin \theta_3 \,\bm d\theta_2\,, \quad \bm e^{\ud{1}} = \sin \theta_4 \sin \theta_3 \sin \theta_2\,\bm d\phi\,.
\end{equation}
Covariant derivatives (spin connection):
\begin{equation}
    \nabla_{\theta_4}=\partial_{\theta_4}\,\,, \quad \nabla_{\theta_3}= \partial_{\theta_3}+\frac{1}{2}\cos \theta_4 \gamma^{\ud{34}}\,\,, \quad \nabla_{\theta_2}= \partial_{\theta_2} + \frac{1}{2}\cos \theta_4 \sin \theta_3 \gamma^{\ud{24}}+ \frac{1}{2}\cos \theta_3 \gamma^{\ud{23}}\,\,,\nn
\end{equation}
\begin{equation}
    \nabla_{\phi}= \partial_{\phi}+ \frac{1}{2}\cos\theta_4 \sin \theta_3 \sin \theta_2 \gamma^{\ud{14}} + \frac{1}{2}\cos \theta_3 \sin\theta_2 \gamma^{\ud{13} }+ \frac{1}{2}\cos \theta_2 \gamma^{\ud{12}}\,\,.
\end{equation}
Gamma matrices:
\begin{equation}
\gamma^{\ud{4}}=   \left(
\begin{array}{cc}
 0 & \mathds{1} \\
\mathds{1} & 0 \\ 
\end{array}
\right)  \,\,, \quad \gamma^{\ud{i}}=   \left(
\begin{array}{cc}
 0 & i \tilde{\gamma}^i \\
-i\tilde{\gamma}^i & 0 \\ 
\end{array}
\right)~~~\text{then}~~~\gamma_*= \left(
\begin{array}{cc}
  \mathds{1} &0 \\
0&-\mathds{1}   \\ 
\end{array}
\right)  \,\,.
\label{higu}
\end{equation}
where $\tilde\gamma^{i}$ are 3d   gamma matrices \eqref{Gm}.\\
The $\gamma$-representation \eqref{higu} implies the following relations
\begin{equation}
 \nabla_{\theta_4}
 =  \left(
\begin{array}{cc}
 \partial_{\theta_4} & 0 \\
0 & \partial_{\theta_4} \\ 
\end{array}
\right),  \quad \nabla_{\alpha} 
=  \left(
\begin{array}{cc}
 \tilde\nabla_{\alpha} + \frac{i}{2}\cos \theta_4 \tilde \gamma_{\alpha} & 0 \\
0& \tilde\nabla_{\alpha} - \frac{i}{2}\cos \theta_4 \tilde \gamma_{\alpha} \\ 
\end{array}
\right)   , \quad \alpha=\phi,\theta_2,\theta_3   
\label{nablaprops4}
\end{equation}
The spinor harmonics equation to be solved is
$$\slashed{\nabla}_{S^{4}}\Psi _{(\pm)}=\pm i (L_4+2) \Psi _{(\pm)}\,\,, $$  
which explicitly reads
\begin{equation}
    \gamma^{\ud{4}}\left( \partial_{\theta_4} + \frac{3}{2} \cot \theta_4\right) \Psi_{(\pm)}+ \frac{1}{\sin \theta_4} \left(
\begin{array}{cc}
 0 & i \slashed{\nabla}_{S^{3}} \\
-i\slashed{\nabla}_{S^{3}} & 0 \\ 
\end{array}
\right)\Psi_{(\pm)} = \pm i  (L_4+2)  \Psi_{(\pm)} \,\,.
\label{spH4}
\end{equation}
There are two possible ansatze among the 3d spinors \eqref{s3} which we denote by an additional upper index $\uparrow,\downarrow$. Notice this index now refers to the sign of the lower dimensional Dirac eigenvalue. They are
\begin{align}
    \Psi^{(\uparrow,a_1)}_{(\pm)}
    (\theta_4,\Omega_3) =   \left(
\begin{array}{c}
  if (\theta_4)  \\
  \pm g (\theta_4)  \\
\end{array}
\right)\otimes \bm\chi^{(a_1)}_{(+)L_3L_2\ell } (\Omega_3)\,\,, \nn\\
\Psi^{(\downarrow,a_1)}_{(\pm)}
(\theta_4,\Omega_3) =   \left(
\begin{array}{c}
  g (\theta_4)   \\
  \pm if (\theta_4)  \\
\end{array}
\right)\otimes \bm\chi^{(a_1)}_{(-)L_3L_2\ell } (\Omega_3)\,\,.
\label{s4sol}
\end{align}
Plugging into \eqref{spH4} and imposing regularity quantizes $L_4$  and leads to 
$$L_4\ge L_3\ge L_2\ge \ell\ge 0,~~~L_i,\ell=0,1,2,3,...$$ 
The result for $f,g$ are
\begin{equation}
    f(\theta_4)=  \frac{L_4+2}{L_3+2} \sin ^{L_3+1}\left(\frac{\theta_4
   }{2}\right) \cos
   ^{L_3}\left(\frac{\theta_4
   }{2}\right) \,
   _2F_1\left(L_4+L_3 +4,L_3-L_4
   ;L_3+3;\sin
   ^2\left(\frac{\theta_4
   }{2}\right)\right)\,\,,\nn
\end{equation}
\begin{equation}
    g(\theta_4) = \sin ^{L_3}\left(\frac{\theta_4
   }{2}\right) \cos
   ^{L_3+1}\left(\frac{\theta_4
   }{2}\right) \,
   _2F_1\left(L_4+ L_3 +4,L_3-L_4;L_3+2;\sin
   ^2\left(\frac{\theta_4
   }{2}\right)\right)\,\,.
\end{equation}
The final result
\begin{equation}
\slashed{\nabla}_{S^{4}}\Psi^{(a_2,a_1)}_{(\pm) L_4L_3L_2\ell} = \pm i (L_4 +2)\Psi^{(a_2,a_1)}_{(\pm) L_4 L_3L_2 \ell} \,\,, \quad L_4\ge L_3\ge L_2\ge \ell\ge 0\,\,,
\label{S4Es}
\end{equation}
with $a_1,a_2\in\{\uparrow,\downarrow\}$.

\subsection*{. $S^{5}$}
\label{S5 spinor}

The spinor harmonics equation  $\slashed{\nabla} \Psi_{(\pm)}=\pm i (L_5+\frac52)\Psi_{(\pm)}$ takes the explicit form  
\begin{equation}
 \gamma^{\ud{5}}(\partial_{\theta_5} +2 \cot \theta_5)\Psi_{(\pm)} + \frac{1}{\sin \theta_5} \slashed{\nabla}_{S^{4}}\Psi_{(\pm)} = \pm i\left(L_5+\frac52\right) \Psi_{(\pm)}\,\,,
\end{equation}
with $\gamma^{\ud 5}=\gamma_*$ (see \eqref{higu}). Inserting the ansatz  
\begin{equation}
    \Psi_{(\pm)L_5L_4...}^{(a_2,a_1)}(\theta_5, \Omega_4) = (g(\theta_5)\mp f(\theta_5))\bm\chi^{(a_2,a_1)}_{(-)L_4...\ell}(\Omega_4) + i(g(\theta_5) \pm f(\theta_5))\bm\chi^{(a_2,a_1)}_{(+)L_4...\ell}(\Omega_4)\,\,,
\end{equation}
with  $\bm\chi^{(a_2,a_1)}_{(\pm) L_4...\ell}$  an $S^4$ eigenspinor satisfying \eqref{S4Es}, and demanding  regularity at $\theta_5=0,\pi$ quantizes the eigenvalue $L_5$ to integer values and demands
$$ L_5=0,1,2...\,\,,~~~~L_4\leq L_5\,\,.$$
The results for $f,g$ are
\begin{equation}
    f(\theta_5)=  \frac{L_5+\frac{5}{2}}{ L_4+\frac{5}{2}} \cos^{L_4} \left( \frac{\theta_5}{2} \right)  \sin^{L_4+1} \left( \frac{\theta_5}{2} \right) \,{_2}F_1\left(L_5+L_4+5,-L_5+L_4 ;\,L_4+\tfrac{7}{2};\sin
   ^2\left(\frac{\theta_5}{2}\right)\right) \,\,,\nn
\end{equation}
\begin{equation}
    g(\theta_5)=   \cos^{L_4+1} \left( \frac{\theta_5}{2} \right)  \sin^{L_4} \left( \frac{\theta_5}{2} \right)\, {_2}F_1\left(L_5+L_4+5,-L_5+L_4 ;\,L_4+\tfrac{5}{2};\sin
   ^2\left(\frac{\theta_5}{2}\right)\right) \,\,.
\end{equation}
such that
\begin{equation}
\slashed{\nabla}_{S^{5}}\Psi^{(a_2,a_1)}_{(\pm) L_5...\ell} = \pm i \left(L_5 + \frac{5}{2}\right) \Psi^{(a_2,a_1)}_{(\pm) L_5...\ell} \,\,, \quad L_5 \geq L_4\ge L_3\ge L_2\ge \ell\ge 0\,\,.
\end{equation}

\subsection{Spin$-\frac32$ harmonics on $S^{N}$}

This Appendix provides a summary of the TT  spin-$\frac3 2$ harmonics on $S^{N}$  based on  \cite{cho,vasilirarita}. We denote the vector-spinor  field as $\Psi_\mu$ suppressing the spinor index.  Concretely, in $S^{N}$, we denote
\begin{equation}
    \Psi_\mu = (\Psi_{\theta_N}, \Psi_{\theta_{N-1}},...,\Psi_{\theta_1})\,\,.
\end{equation}
The vector-spinor harmonics can be separated into non-TT (longitudinal/gamma-trace)  and TT (transverse and gamma-traceless)  sectors. The non-TT modes  can be constructed  from the spin-$\frac12$ Dirac eigenmodes $\Psi$, given in \eqref{DiracN},  as  linear combinations   $\alpha\nabla_\mu\Psi+\beta\gamma_\mu\Psi$ 
(a comprehensive analysis can be found below eqn (C.11) in \cite{Higu+lets}). We will not discuss them in the present work. The additional modes, which complete the basis for expanding an arbitrary spin-$\frac32$ configuration on $S^N$, are TT, i.e. they satisfy $\nabla^\mu\Psi_\mu = \gamma^\mu\Psi_\mu = 0$. As we now elaborate the TT-modes separate into type I and type II according to whether they are built from lower dimensional vector-spinor TT-modes (type II) or not (type I). 

\vspace{2mm}

\nin {\bf Covariant derivative}\\ 
Our spinor conventions are spelled out in app. \ref{s1/2Con}. Covariant derivatives on vector-spinors $\nabla_\mu$ gets contributions from the spin connection and the Christo\-ffel symbol
\begin{equation}
\nabla_\mu \Psi_\nu = \partial_\mu \Psi_\nu + \frac{1}{4}\omega_\mu{}^{\ud{ab}} \gamma_{\ud{ab}}\Psi_\nu - \Gamma^{\lambda}_{\mu \nu } \Psi_{\lambda}\,\,.
\end{equation}
They explicitly read 
\begin{align}
\nabla_{\theta_N}\Psi_{\theta_i}&=\partial_{\theta_N}\Psi_{
\theta_i}-\cot\theta_N\Psi_{\theta_i}\,\,,~~~~~~~ 
\nabla_{\theta_N}\Psi_{\theta_N}=\partial_{\theta_N}\Psi_{
\theta_N} \,\,,
\end{align}
and
\begin{align}
\nabla_{\theta_i}\Psi_{\theta_j}&=\hat\nabla_{\theta_i}\Psi_{
\theta_j}+\tilde g_{\theta_i\theta_j}\sin\theta_N\cos\theta_N\Psi_{\theta_N}+\frac12\cot\theta_N \,\gamma_{\theta_i}\gamma_{\ud N} \Psi_{\theta_j}\,\,,
\nn\\
\nabla_{\theta_i}\Psi_{\theta_N}&=\hat\nabla_{\theta_i}\Psi_{
\theta_N} -\cot\theta_N\Psi_{\theta_i}+\frac12\cot\theta_N\,\gamma_{\theta_i}\gamma_{\ud N}\Psi_{\theta_N}\,\,.
\end{align}
Here, a hat denotes quantities in one lower dimensional sphere $S^{N-1}$ 
$$\hat\nabla_{\theta_i}\Psi_{
\theta_j}=\partial_{\theta_i}\Psi_{\theta_j}-\tilde\Gamma_{\theta_i\theta_j}^{\theta_k}\Psi_{\theta_k}+\frac14\tilde \omega_{\theta_i}{}^{\ud{kl}}\gamma_{\ud {kl}}\Psi_{\theta_j}\,\,,$$ 
$$\hat\nabla_{\theta_i}\Psi_{
\theta_N}=\partial_{\theta_i}\Psi_{\theta_N}+\frac14\tilde\omega_{\theta_i}{}^{\ud{  ij}}\gamma_{\ud {ij}}\Psi_{\theta_N}\,\,.$$
but with $N$-dimensional flat gammas matrices $\gamma_{\ud i}$.

---

\nin Type I and type II vector-spinor  TT-eigenmodes are spin-$\frac32$ harmonics satisfying
\begin{align}
\text{\sf TT-modes}:~~~\slashed{\nabla}_{S^{N}}\Psi^{(\vec a)}_{\mu\:(\pm) L_N ...}   &= \pm i \left( L_N + \frac{N}{2} \right) \Psi^{(\vec a)}_{\mu\: (\pm)L_N...}\,\,, 
\label{diracrarita}\\
\nabla^{\mu}\Psi^{(\vec a)}_{\mu\:(\pm)L_N...}&=\gamma^{\mu}\Psi^{(\vec a)}_{\mu\:(\pm)L_N...}= 0 \,\,.
\label{TT}
\end{align}
The full set of indices classifying the UIRs states are explained in app.  \ref{s1/2Con}.   To avoid a convoluted notation we will refrain from denoting whether the spinor is  type I or type II. The type should be clear from the context.

\vspace{2mm}

\nin \textbf{On transformation properties of spin-3/2 modes on $S^N$} \\
For a Killing vector  $\bm k$ of the background geometry, the spinor Lie derivative \eqref{RSLD} satisfies 
$${\cal L}_k\gamma^{\ud a}=0,~~{\cal L}_k e_\mu^{\ud a}=0,~~{\cal L}_k\nabla_\mu\Psi_\rho=\nabla_\mu{\cal L}_k\Psi_\rho   \,\,,
$$
here ${\cal L}_k$ is the Kosmann/Lorentz Lie derivative extension discussed in \cite{ortin,kos,FoF,ortin2,letsios3/2,vasilirarita}. These equations imply that the set of TT-modes are closed under the symmetry group action. Moreover, as a consequence of $[{\cal L}_k,{\cal L}_{k'}]={\cal L}_{[k,k']}$ the space of eigenmode solutions provides a representation of  Spin($N+1$).

The type I and type II modes to be constructed below for fixed sign of the Dirac eigenvalue in \eqref{diracrarita}, mix among themselves under the SO($N+1$) symmetry. In other words, the irreducible multiplets are
\begin{equation}
    \{ \Psi^{\,{\sf I}\,  (\vec{a})}_{ \mu\: (+) L_{N}L_{N-1}...\ell} ,  \Psi^{\,{\sf II}\,  (\vec{a})}_{ \mu\:(+)  L_{N}L_{N-1}...\ell} \} ~~\text{and}~~ \{\Psi^{\,{\sf I}\,  (\vec{a})}_{ \mu\: (-) L_{N}L_{N-1}...\ell} ,  \Psi^{\,{\sf II}\,  (\vec{a})}_{ \mu\:(-)  L_{N}L_{N-1}...\ell} \}\,\,,~~~L_N=\text{fixed}\,. 
    \label{irrepSN3/2}
\end{equation}
We refer the reader to eqs (6.10) and (6.15) in \cite{vasilirarita} where the mixing has been explicitly shown. The highest weights are spelled in eqn (4.3) in \cite{vasilirarita}.

~

\nin{\bf On complex conjugation}\\
The situation for spin-$\frac32$ is analogous to that of spin-$\frac12$ fields. Vector-spinor harmonics are unique and their conjugates expand the vector spaces already expanded by the original harmonics.

\subsubsection{N even}

The discussion of spin-$\frac32$ TT-eigenspinors becomes fully non-trivial for $N\ge4$. On $S^2$, the spin-$\frac1 2$ eigenspinors $\Psi$ generate two orthogonal linear combinations of $\gamma_\mu \Psi$ and $\nabla_\mu \Psi$ that diagonalize the Dirac operator. These modes are non-TT and constitute a complete basis for the expansion of arbitrary vector-spinors on $S^2$. In higher dimensions,  these two non-TT modes alone are not sufficient to span the full space of vector-spinor  fields, and additional modes must be introduced which are transverse and traceless (TT). For example, on $S^{3}$, the remaining  vector-spinor eigenmode  is constructed from  the $S^2$ vector-spinors on $S^{2}$ and a non-trivial   $\Psi_{\theta_3}$ which is mandatory to guarantee transverse-tracelessness on $S^3$. This configuration build out solely from spin-$\frac12$ harmonics defines the TT-mode  referred to as type I. On $S^{4}$, the pattern extends naturally. In addition to the aformentioned modes,  we can build $S^4$ TT-modes from $S^3$ TT-modes, these configurations define  type II TT-modes. The complete set of eigenmodes on $S^{4}$ thus includes two non-TT modes, one type I TT-mode I, and one type II TT-mode.

\subsubsection*{Type I modes}

These  TT-modes are built from   lower dimensional spinor eigenmodes. They are characterized by a non-zero $\Psi_{\theta_N}\ne0$ component. Hence, we write 
\begin{equation}
  \text{\sf Type I}:~~~~~  \Psi_\mu^{\,\sf I} = (\Psi_{\theta_N}, \Psi_{\theta_i })\,\,,~~~\Psi_{\theta_N}\ne0 \,\,.
\end{equation}
For the representation \eqref{chir}, the upper and lower  components corresponding to chiral projections will be denoted as $L/R$, 
\begin{equation}
\Psi^{\,{\sf I}}_{\theta_N} = 
\begin{pmatrix}
\Psi_{\theta_N}^{ L } \\ 
\Psi _{\theta_N }^{ R }
\end{pmatrix} \,\,, \quad \Psi_{\theta_i}^{\,{\sf I}} = 
\begin{pmatrix}
\Psi_{\theta_i}^{ L } \\ 
\Psi_{\theta_i}^{ R }
\end{pmatrix} \,\,, \quad i=1,..,N-1\,\,.
\label{chirals}
\end{equation}
The gamma traceless condition relates the $\Psi^{\,{\sf I}}_{\theta_N}$ chiral components to those of $\Psi^{\,{\sf I}}_{\theta_i}$
\begin{equation}
\gamma^{\mu}\Psi_\mu^{\,{\sf I}} =0~~~ \leadsto~~~ \Psi_{\theta_N}^{\alpha} = s_\alpha \frac{i}{\sin \theta_N} \tilde\gamma^{\theta_i}\Psi_{\theta_i}^{\alpha}\,\,,~~~~~s_L=1,s_R=-1\,\,,
\label{gammaT}
\end{equation}
while the transverse condition implies
\begin{equation}
    \nabla^{\mu} \Psi^{\,{\sf I}}_\mu = 0 \leadsto \left( \partial_{\theta_N  } + \left(N-\tfrac{1}{2} \right)\cot \theta_N\right)\Psi_{\theta_N}^{\alpha} + \frac{1}{\sin^2 \theta_N} \tilde \nabla^{\theta_i}\Psi_{\theta_i}^{\alpha}=0\,\,,~~~~~\alpha=L,R\,\,.
    \label{Transv}
\end{equation}
where we used equation \eqref{N even spin}. Inserting these conditions into the $\theta_N$-component of the eigenvalue eqn. \eqref{diracrarita} leads to
\begin{equation}
 \left(\partial_{\theta_N} + \tfrac{N+1}{2} \cot \theta_N\right) \Psi_{\theta_N\,(\pm)}^{\alpha} -s_a \frac{i}{\sin \theta_N} \tilde{\slashed \nabla} \Psi_{\theta_N\,(\pm)}^{\alpha} =\pm i \left( L_N + \tfrac{N}{2} \right)d^{\alpha\beta}\,\Psi_{\theta_N\,(\pm)}^{\beta}\,\,,
\label{difeqN}
\end{equation}
where
$$ d^{\alpha\beta}=\begin{pmatrix}
    0&1\\
    1&0
\end{pmatrix}$$
while the $\theta_i$-components lead to
\begin{equation}
\partial_{\theta_N} \Psi_{\theta_i\,(\pm)}^{\alpha} + \tfrac{N-1}{2} \cot \theta_N \Psi_{\theta_i\,(\pm)}^{\alpha} + \cot\theta_N \tilde \gamma_{\theta_i}{}^{\theta_j}\,\Psi_{\theta_j\,(\pm)}^{\alpha} \mp \tfrac{i}{\sin \theta_N}   \tilde{ \slashed\nabla} \Psi_{\theta_i\,(\pm)}^{\alpha} =\pm i \left( L_N + \tfrac{N}{2} \right) d^{\alpha\beta}\,\Psi_{\theta_i\,(\pm)}^{\beta}\,\,.
\label{difeqI}
\end{equation}
In these expressions  $ \tilde{\slashed\nabla} =\tilde \gamma^{\theta_j} \tilde{  \nabla}_{\theta_j}$ denotes the Dirac operator on $S^{N-1}$ and 
\begin{equation}
\tilde\gamma_{\theta_i}{}^{ \theta_j} := \frac{1}{2}[\tilde\gamma_{\theta_i} , \tilde\gamma^{\theta_j}]\,\,,
\end{equation}
is built with $S^{N-1}$ curved gamma matrices. 

To solve the eigenvalue equation \eqref{difeqN}, we make an ansatz using $S^{N-1}$ spin-$\frac12$ harmonics $\bm \chi$. In even dimensional $S^N$, the eigenmode is denoted with an extra supraindex $\uparrow$ or $\downarrow$ and an extra quantum number $L_N$ as subindex
\begin{equation}
\Psi^{{\sf I}\,(\uparrow,a_{n-1},...,a_1)}_{\theta_N\,(\pm)L_NL_{N-1}...\ell}(\theta_N, \Omega_{N-1}) = 
\begin{pmatrix}
\Psi_{\theta_N}^{ L } \\ 
\Psi_{\theta_N}^{ R }
\end{pmatrix} = 
\begin{pmatrix}
if(\theta_N)\, \bm \chi_{(+)L_{N-1...\ell}}^{(a_{n-1},...,a_1)}(\Omega_{N-1}) \\ 
\pm g(\theta_N)\, \bm  \chi_{(+) L_{N-1}...\ell}^{(a_{n-1},...,a_1)}(\Omega_{N-1})
\end{pmatrix}\,\,,
\end{equation}
\begin{equation}
\Psi^{{\sf I}\,(\downarrow,a_{n-1},...,a_1)}_{\theta_N\,(\pm)L_NL_{N-1}...\ell}(\theta_N, \Omega_{N-1}) = 
\begin{pmatrix}
\Psi_{\theta_N}^{ L } \\ 
\Psi_{\theta_N}^{ R }
\end{pmatrix} = 
\begin{pmatrix}
g(\theta_N)  \, \bm \chi_{(-) L_{N-1}...\ell}^{(a_{n-1},...,a_1)}(\Omega_{N-1}) \\ 
\pm if(\theta_N)\, \bm  \chi_{(-)L_{N-1}...\ell}^{(a_{n-1},...,a_1)}(\Omega_{N-1})
\end{pmatrix} \,\,.
\end{equation}
Inserting into \ref{difeqN} one finds that $f(\theta_N)$ and $g(\theta_N)$ are related by the following equations (cf. \eqref{phipsi}-\eqref{psiphi})
\begin{equation}
  \left[ \partial_{\theta_N } + \frac{N+1}{2} \cot\theta_N + \frac{L_{N-1}+(N-1)/2}{\sin\theta_N} \right]f(\theta_N) = \left( L_N + \tfrac{N}{2} \right)  g(\theta_N)\,\,,
\end{equation}
\begin{equation}
  \left[ \partial_{\theta_N } + \frac{N+1}{2} \cot\theta_N -\frac{L_{N-1}+(N-1)/2}{\sin\theta_N} \right]g(\theta_N) = -\left( L_N + \tfrac{N}{2} \right)  f(\theta_N)\,\,.
  \label{difeqA1}
\end{equation}
Regularity on $\theta_N \in [0,\pi]$ quantizes $L_{N}$ to integer values  and imposes $L_{N-1}\le L_N$. The regular solutions are found to be 
\begin{align}
 f(\theta_N)=& \frac{L_N+N/2}{L_{N-1}+N/2} \cos^{L_{N-1}-1}\left(\frac{\theta_N}{2} \right) \,\sin^{L_{N-1}} \left(\frac{\theta_N}{2}\right)\nn\\
 &
 \cross {_2}F_1\left(L_{N}+L_{N-1}+N,-L_{N}+L_{N-1};L_{N-1} + \tfrac{N}{2}+1 ; \sin^2 \frac{\theta_N}{2} \right) \,\,, 
 \label{A1}\\
g(\theta_N)=&  \cos^{L_{N-1}} \left(\frac{\theta_N}{2} \right)\, \sin^{L_{N-1}-1} \left(\frac{\theta_N}{2}\right)\nn\\
 &\cross {_2}F_1\left(L_N+L_{N-1}+N,-L_N+L_{N-1};L_{N-1} + \tfrac{N}{2} ; \sin^2 \frac{\theta_N}{2} \right)\,\,.
 \label{A2}
\end{align}
The quantum numbers satisfy
\begin{equation}
    L_{N} \geq L_{N-1} \,\,, \quad L_{N},L_{N-1}=1,2,3...\,\,.
    \label{qn3/2I}
\end{equation}
Notice that the principal quantum number $L_N$ is bounded below by 1, whereas for spin-$\frac12$  $L_N=L_{N-1}=0$ were allowed (cf. \eqref{spin12QN}).

The ansatz for $\Psi_{\theta_i}$ involves linear combinations of $S^{N-1}$ vector-spinors $\tilde\nabla_{\theta_i}\bm\chi$ and $\tilde\gamma_{\theta_i} \bm\chi$ with $\theta_N$-dependent functions $C,D$ as
\begin{align}
\Psi^{{\sf I}\,(\uparrow,a_{n-1},...,a_1)}_{\theta_i\,(\pm)L_NL_{N-1}...\ell} 
&= 
\begin{pmatrix}
\Psi_{\theta_i}^{L} \\ 
\Psi_{\theta_i}^{R}
\end{pmatrix} 
\label{32psi}\\
&= 
\begin{pmatrix}
iC^{(1)}(\theta_N)\, \tilde \nabla_{\theta_i} \bm\chi_{(+)L_{N-1}...\ell}^{(a_{n-1},...,a_1)}(\Omega_{N-1} ) - i D^{(1)}(\theta_N)\, \tilde \gamma_{\theta_i} \bm\chi_{(+)L_{N-1}...\ell }^{(a_{n-1},...,a_1)}(\Omega_{N-1})\\ 
\pm C^{(2)}(\theta_N)\, \tilde \nabla_{\theta_i}   \bm\chi^{(a_{n-1},...,a_1)}_{(+)L_{N-1}...\ell}(\Omega_{N-1}) \mp iD^{(2)}(\theta_N ) \,\tilde \gamma_{\theta_i} \bm\chi_{(+)L_{N-1}...\ell}^{(a_{n-1},...,a_1)}(\Omega_{N-1})
\end{pmatrix}\,\,.\nn
\end{align}
\begin{align}
\Psi^{{\sf I}\,(\downarrow,a_{n-1},...,a_1)}_{\theta_i\,(\pm)} &= 
\begin{pmatrix}
\Psi_{\theta_i}^{L} \\ 
\Psi_{\theta_i}^{R}
\end{pmatrix}
\label{32psi2}
\\
&= 
\begin{pmatrix}
C^{(2)}(\theta_N)\, \tilde \nabla_{\theta_i} \bm\chi^{(a_{n-1},...,a_1)}_{(-)L_{N-1}...\ell}(\Omega_{N-1} ) +  D^{(2)}(\theta_N)\, \tilde \gamma_{\theta_i} \bm\chi^{(a_{n-1},...,a_1)}_{(-)L_{N-1}...\ell }(\Omega_{N-1})\\ 
\pm iC^{(1)}(\theta_N)\, \tilde \nabla_{\theta_i} \bm \chi^{(a_{n-1},...,a_1)}_{(-)L_{N-1}...\ell}(\Omega_{N-1}) \mp iD^{(1)}(\theta_N ) \,\tilde \gamma_{\theta_i} \bm\chi^{(a_{n-1},...,a_1)}_{(-)L_{N-1}...\ell}(\Omega_{N-1})
\end{pmatrix}\,\,.
\nn
\end{align}
The solutions for $C,D$ can be found from the TT-constraints. Inserting the ansatz \eqref{32psi}-\eqref{32psi2} in the  gamma traceless condition \eqref{gammaT} one  finds 
\begin{align}
    D^{(1)}(\theta_N)& =-\frac{i}{N-1} \left[ -\left(L_{N-1} + \tfrac{N-1}{2} \right)C^{(1)}(\theta_N) - \sin \theta_N\, f(\theta_N) \right]\,\,,\nn\\
    D^{(2)}(\theta_N) &=-\frac{i}{N-1} \left[ -\left(L_{N-1} + \tfrac{N-1}{2} \right)C^{(2)}(\theta_N) + \sin \theta_N \, g(\theta_N) \right]\,\,,
    \label{D2}
\end{align}
and from the transverse condition \eqref{Transv} in combination with  \eqref{difeqA1} one finds 
\begin{align}
  C^{(1)}(\theta_N) =&  \frac{\sin \theta_N}{L_{N-1}(L_{N-1}+N-1)} \nn\\&~~\times \Bigg[  \left(\tfrac{N-1}{2}\cos \theta_N - L_{N-1} -\tfrac{N-1}{2}  \right) f(\theta_N)  +  \tfrac{N-1}{N-2} (L_{N} + \tfrac{N}{2}) \sin \theta_N \,g(\theta_N)\Bigg]\,\,,
  \nn
\end{align}
\begin{align}
C^{(2)}(\theta_N)=& \frac{\sin \theta_N}{L_{N-1}(L_{N-1}+N-1)} \nn\\
&~~\times \Bigg[  \left(\tfrac{N-1}{2}\cos \theta_N + L_{N-1} +\tfrac{N-1}{2}  \right) g(\theta_N)  -  \tfrac{N-1}{N-2} (L_{N} + \tfrac{N}{2}) \sin \theta_N\, f(\theta_N)\Bigg]\,\,.
\label{C2}
\end{align}

\subsubsection*{Type II modes}
These  are vector-spinor solutions with $\Psi_{\theta_N}=0$. They  are built from lower dimensional TT solutions. Thus, type II vector-spinor components can be written as
\begin{equation}
\text{\sf Type II}:~~~~~ \Psi_\mu^{\,\sf II} = (0, \Psi_{\theta_i})\,\,.
\end{equation}
For $\Psi_{\theta_N}^{\,\sf II}=0$, the gamma traceless condition \eqref{gammaT} reduces to gamma-traceless in one lower dimension,
\begin{equation} \gamma^{\theta_i}\Psi^{\sf II}_{\theta_i}=0 \leadsto \tilde \gamma^{\theta_i} \Psi_{\theta_i}^{\alpha}= 0 \,\,,~~~~\alpha=L,R\,\,.
\label{TTch}
\end{equation}
Here $\alpha=L,R$  denote the chiral components of $\Psi_\mu$ (see \eqref{chirals}). By virtue of \eqref{TTch} the third term in \eqref{difeqI} simplifies to 
\begin{equation}
\tilde \gamma_{\theta_i}{}^{ \theta_j} \Psi_{\theta_j}^{\alpha}=(\tilde \gamma_{\theta_i} \underbrace{\tilde \gamma^{\theta_j} \Psi_{\theta_j}^{\alpha}}_{=0} - \delta_i^{\,\,j} \Psi_{\theta_j}^{\alpha}) \,\,.
\end{equation}
Type II modes rely on lower dimensional spin-$\frac32$ TT-harmonics $\bm\chi_{\theta_i}$ satisfying
\begin{equation}
\tilde{\cancel\nabla}  \bm\chi^{(\vec a')}_{\theta_i\,(\pm)L_{N-1}...\ell}= \pm i\left(L_{N-1} + \frac{N-1}{2} \right) \bm\chi^{(\vec a')}_{\theta_i\,(\pm)L_{N-1}...\ell}\,\,,
    \label{SN-1 spinvec}
\end{equation}
\begin{equation}
    \tilde\gamma^{\theta_i} \bm\chi^{(\vec a')}_{\theta_i\,(\pm)L_{N-1}...\ell} = \tilde \nabla^{\theta_i} \bm\chi^{(\vec a')}_{\theta_i\,(\pm)L_{N-1}...\ell} = 0 \,\,.\nn
\end{equation}
where  $\vec a'=(a_{n-1},...,a_1)$ and $\tilde{\cancel\nabla}=\cancel\nabla_{S^{N-1}}$ is the Dirac operator on $S^{N-1}$. The ansatz to solve \eqref{difeqI} is
\begin{align}
\Psi^{{\sf II}\,(\uparrow,\vec a')}_{\theta_i\,(\pm)L_NL_{N-1}...\ell}(\theta_N, \Omega_{N-1}) &= 
\begin{pmatrix}
\Psi_{\theta_i}^{L} \\ 
\Psi_{\theta_i}^{R}
\end{pmatrix}  
= \begin{pmatrix}
iB^{(1)}(\theta_N)\,\bm\chi^{(\vec a')}_{\theta_i\,(+)L_{N-1}...\ell}(\Omega_{N-1} )\\ 
\pm B^{(2)}(\theta_N) \,\bm\chi^{(\vec a')}_{\theta_i\,(+)L_{N-1}...\ell}(\Omega_{N-1} )
\end{pmatrix}\,\,, \nn \\  \Psi^{{\sf II}\,(\downarrow,\vec a)}_{\theta_i\,(\pm)L_NL_{N-1}...\ell}(\theta_N, \Omega_{N-1}) &= 
\begin{pmatrix}
\Psi_{\theta_i}^{L} \\ 
\Psi_{\theta_i}^{R}
\end{pmatrix}  
=\begin{pmatrix}
B^{(2)}(\theta_N)\,\bm\chi^{(\vec a')}_{\theta_i\,(-)L_{N-1}...\ell}(\Omega_{N-1} )\\ 
\pm iB^{(1)}(\theta_N)\,\bm\chi^{(\vec a')}_{\theta_i\,(-)L_{N-1}...\ell}(\Omega_{N-1} )
\end{pmatrix}\,\,.
\end{align}
Inserting in \eqref{difeqI} gives
\begin{align}
  \left[ \partial_{\theta_N } + \frac{N-3}{2} \cot\theta_N + \frac{L_{N-1}+(N-1)/2}{\sin\theta_N} \right]B^{(1)}(\theta_N) &= \left( L_N + \frac{N}{2} \right) B^{(2)}(\theta_N)\,\,, \nn\\
  \left[ \partial_{\theta_N } + \frac{N-3}{2} \cot\theta_N -\frac{L_{N-1}+(N-1)/2}{\sin\theta_N} \right]B^{(2)}(\theta_N) &= -\left( L_N + \frac{N}{2} \right)B^{(1)}(\theta_N)\,.
  \label{difeqB1}
\end{align}
Regularity on $\theta_N \in [0,\pi]$ quantizes $L_{N}$,  requires $L_N \geq L_{N-1}$ and imposes $L_{N-1}=1,2,...$. The regular solutions result
\begin{align}
 B^{(1)}(\theta_N)=& \frac{L_N+N/2}{L_{N-1}+N/2} \cos^{L_{N-1}+1}\left(\frac{\theta_N}{2} \right) \sin^{L_{N-1}+2} \left(\frac{\theta_N}{2}\right)\nn\\
 &
 \cross {_2}F_1\left(L_{N}+L_{N-1}+N,-L_{N}+L_{N-1};L_{N-1} + \tfrac{N}{2}+1 ; \sin^2 \frac{\theta_N}{2} \right)\,\,,  
 \label{B1}
\end{align}
\begin{align}
 B^{(2)}(\theta_N)=&  \cos^{L_{N-1}+2} \left(\frac{\theta_N}{2} \right) \sin^{L_{N-1}+1} \left(\frac{\theta_N}{2}\right)\nn\\
 &\cross {_2}F_1\left(L_N+L_{N-1}+N,-L_N+L_{N-1};L_{N-1} + \tfrac{N}{2}; \sin^2 \frac{\theta_N}{2} \right)\,\,.
 \label{B2}
\end{align}
The quantum numbers satisfy
\begin{equation}
    L_{N} \geq L_{N-1} \,\,, \quad L_{N},L_{N-1},...=1,2,3...\,\,.
    \label{qn3/2II}
\end{equation}
In summary, Type I and II modes, have identical eigenvalues. A posteriori this is to be expected since they mix under the action of the symmetry group. We conclude by reminding the reader that in even dimensions, positive and negative eigenvalue solutions are connected by the chiral matrix
\begin{equation}
\text{\sf Even dimensions}:~~~~\gamma_{*}\Psi^{(\vec a)}_{ \mu\,(\pm)L_N...} =\Psi^{(\vec a)}_{ \mu\,(\mp)L_N...} \,\,.
\end{equation}


\subsubsection{N odd}


\subsubsection*{Type I}

These modes have non-zero $\Psi_{\theta_N}$ component
\begin{equation}
   \text{\sf Type I}:~~~~~ \Psi_{\mu\,(\pm)}^{\,\sf I} = (\Psi^{\,\sf I}_{ \theta_N\,(\pm)}, \Psi^{\,\sf I}_{\theta_i\,(\pm)})\,\,,~~~\Psi_{\theta_N}\ne0 \,\,. 
\end{equation}
They are built from lower dimensional spin-$\frac12$ eigenspinors $\bm\chi$ as 
\begin{equation}
\Psi^{{\sf I}\,(\vec a)}_{\theta_N\,(\pm)L_NL_{N-1}...\ell}=\frac{(1+i\gamma^{\ud N})}{\sqrt{2}}\left(g(\theta_N) \pm i \,f(\theta_N) \gamma^{\ud N}\right)\bm{\chi}^{(\vec a)}_{(-)L_{N-1}...\ell}(\Omega_{N-1}) \,\,,
\end{equation}
\begin{align}  
\Psi^{{\sf I}\,(\vec a)}_{\theta_i\,(\pm)L_NL_{N-1}...\ell} =&\frac{(1 +i\gamma^{\ud N})}{\sqrt{2}} \Bigg[ \left(C^{(2)}(\theta_N) \pm i  C^{(1)}(\theta_N)\, \gamma^{\ud N}\right)
\tilde\nabla_{\theta_i} \bm{\chi}^{(\vec a)}_{(-)L_{N-1...\ell}}(\Omega_{N-1})\nn\\
& ~~~~~~+\left(D^{(2)}(\theta_N) \pm i D^{(1)}(\theta_N)\, \gamma^{\ud N}\right) \,\tilde\gamma_{\theta_i}\bm{\chi}^{(\vec a)}_{(-)L_{N-1...\ell}}(\Omega_{N-1}) \Bigg]\,\,.
\end{align}
Recall that the lower dimensional spinors live in even dimensional spheres hence satisfy $\gamma^{\ud N}\bm{\chi}_{(-)}= \bm \chi_{(+)}$. The functions $f(\theta_N)$,$g(\theta_N)$ are given in \eqref{A2}, while the functions $C^{(1)}(\theta_N),$ $ C^{(2)}(\theta_N)$ and  $D^{(1)}(\theta_N) , D^{(2)}(\theta_N)$ are given in  \eqref{D2}   and \eqref{C2}.

\subsubsection*{Type II}

These modes are built from lower dimensional  spin-$\frac32$ harmonics $\bm \chi_{\theta_i}$ and have vanishing $\Psi_{\theta_N}$ component. They take the form 
\begin{equation}
    \Psi^{\,\sf II}_{\mu\, (\pm ) } = (0 , \Psi^{\,\sf II}_{ \theta_i(\pm)})\,\,,
\end{equation}
with
\begin{equation}
    \Psi^{\,\sf II}_{\theta_i\,(\pm) } = \frac{(1+i\gamma^{\ud N})}{\sqrt{2}}\left(B^{(2)}(\theta_N) \pm i B^{(1)}(\theta_N) \gamma^{\ud N}\right) \tilde\Psi_{\theta_i\,(-) }(\Omega_{N-1})\,\,.
    \label{IIodd}
\end{equation}
The functions $B^{(2)}(\theta_N)$, $B^{(1)}(\theta_N)$ are given by \eqref{B1} and \eqref{B2}. $\tilde \Psi_{\theta_i\,(-)}$ is a vector-spinor on the even sphere $S^{N-1}$ which satisfies \eqref{SN-1 spinvec}. Notice also, that since $\tilde \Psi_{\theta_i\,(-)}$ live  in an even dimensional sphere, they satisfy $\gamma^{\ud N} \tilde\Psi_{\theta_i\,(-)}=\tilde \gamma_* \tilde\Psi_{\theta_i\,(-)} = \tilde \Psi_{\theta_i\,(+) }$. Using this fact, we can rewrite \eqref{IIodd} as
\begin{equation}
    \Psi^{\,\sf II}_{\theta_i\,(\pm) } = \frac{1}{\sqrt{2}}(B^{(2)}(\theta_N) \mp B^{(1)}(\theta_N) ) \tilde \Psi_{\theta_i\,(-)}(\Omega_{N-1}) + \frac{i}{\sqrt{2}}(B^{(2)}(\theta_N) \pm B^{(1)}(\theta_N))\tilde \Psi_{\theta_i\,(+)}(\Omega_{N-1})\,\,.
\end{equation}
Note the similarity between the odd dimensional Type II modes solutions \eqref{IIodd} and Dirac spinor solutions \eqref{1/2Nodd}.

\section{Fermion solutions in de Sitter space}
\label{spin dS}

In this section we summarize the mode solutions to the (Lorentzian) spin-$\frac12$ and spin-$\frac32$ massive  equations in de Sitter found in \cite{lets1/2} and \cite{vasilirarita,lets1/2}. In those works, the solutions  were obtained by  analytic  continuation of the sphere solutions \eqref{DiracN} and \eqref{diracrarita}. Solutions in other coordinate patches can be found in \cite{other} .

---

\nin {\bf de Sitter parametrization}\\
de Sitter global coordinates are denoted $x^\mu=(t, ..., \theta_2, \theta_1)$ with $t\in\mathbb R$, $\theta_i\in[0,\pi] ~ (i=2,...,N-1)$ and $\theta_1=\phi\in[0,2\pi]$. The  metric can be obtained from the $S^N$ sphere metric  \eqref{SN}. Making
\be
\theta_N \mapsto \frac\pi2 -i t\,\,,
\label{Wick}
\ee
one finds
\begin{equation}
  \text{\sf de Sitter}_N:~~~  ds^2= -dt^2 + \cosh^2 t \, d\Omega_{N-1}^2 \,\,.
  \label{dSN}
\end{equation}
we will denote by $d\tilde s^2=d\Omega_{N-1}^2$ the round metric on $S^{N-1}$. The tilded objects denote quantities on the $(N-1)$-sphere.

\vspace{2mm}

\nin {\bf Christoffel and Spin connections}\\
The non-zero Christoffel symbols of dS$_N$ are
\begin{equation}
\Gamma^{t}_{\theta_i \theta_j} =  \cosh t \sinh t\, \tilde g_{\theta_i \theta_j} \,\,, \quad \Gamma^{\theta_j}_{\theta_i t} = \tanh t   \, \delta_i^j  \,\,, \quad \Gamma^{\theta_k}_{\theta_i \theta_j} = \tilde\Gamma^{\theta_k}_{\theta_i \theta_j} \,\,,\,~~~~~  i,j,k=1,...,N-1\,\,.
\end{equation}
$N-$beins are chosen as   
\begin{equation}
    \bm{e}^{\ud 0} = \bm dt \,\,, \quad \bm{e}^{\ud i} = {\cosh t}\, \bm{\tilde e}^{\ud i} \,\,,~~~ \quad i=1,...,N-1\,\,.
\end{equation}
with $\bm{\tilde e}^{\ud i}$  the $n-$beins of  $S^{N-1}$. The solution to \eqref{torsion} for the spin connection components  reads 
\begin{equation}
    \bm\omega ^{ \ud{ij}}=   \tilde {\bm \omega} ^{\ud{ij}} \,\,, \quad \bm\omega ^ {\ud{iN}}=  \sinh t \,\tilde {\bm e}^ {\ud i } \,\,, ~~~~\quad i,j =1,...,N-1\,\,.
    \label{sPConL}
\end{equation}

\vspace{2mm}

\nin {\bf Spinor conventions}\\
Dirac spinors are $2^{[\frac N 2]}$ dimensional column vectors, where $[...]$ denotes integer part. Flat gamma matrices signature convention is
\begin{equation}
    \{ \gamma^{\ud a},\gamma^{\ud b} \}=2 \eta^{\ud{ab}} \,\,, \quad \eta^{\ud{ab}}=(-1,1,...,1)\,\,.
\end{equation}
We can obtain them from the Euclidean ones by inserting an $`i$' as $\gamma^{\ud 0}:=i\gamma^{\ud N}$

\vspace{1mm} 

\nin {\sf $N$ even}: We write the set of $\gamma$-matrices as $\gamma^{\ud a} = (\gamma^{\ud 0} , \gamma^{\ud i}) $, and construct them from gamma matrices $\tilde\gamma^{\ud i}$ in one-lower dimension  as
\begin{equation}
\text{\sf Chiral representation}:~~~\gamma^{\ud{0}}=  i \left(
\begin{array}{cc}
 0 & \mathds{1} \\
\mathds{1} & 0 \\ 
\end{array}
\right)  \,\,, \quad \gamma^{\ud{i}}=   \left(
\begin{array}{cc}
 0 & i \tilde{\gamma}^{\ud i} \\
-i\tilde{\gamma}^{\ud i} & 0 \\ 
\end{array}
\right) \,\,, \quad i=1,...,N-1\,\,.
\label{chirL}
\end{equation}
The chiral matrix is defined in \eqref{5mat} irrespective  of the signature.

\nin{\sf  $N$ odd}: as we increase the dimension from even to odd, the gamma matrices keep their size. We obtain the odd-dimensional set of gamma matrices appending the  `chiral matrix' to the even-dimensional set\footnote{We need to properly adjust a phase for the new matrix to square to one.}. Writing $\gamma^{\ud a} = (\gamma^{\ud 0},\gamma^{\ud{N-1}},\gamma^{\ud i})$ we have
\begin{equation}
\gamma^{\ud 0}=i\left(
\begin{array}{cc}
 \mathds{1} & 0 \\
0 & -\mathds{1} \\ 
\end{array}
\right)   \,\,, \quad 
\gamma^{\ud{N-1}}=   \left(
\begin{array}{cc}
 0 & \mathds{1} \\
\mathds{1} & 0 \\ 
\end{array}
\right)  \,\,, \quad 
\gamma^{\ud i}=   \left(
\begin{array}{cc}
 0 & i \tilde{\tilde{\gamma}}^{\ud i} \\
-i\tilde{\tilde{\gamma}}^{\ud i} & 0 \\ 
\end{array}
\right) \,\,, \quad i=1,...,N-2\,\,.
\label{N odd gammaL}
\end{equation}
here $\tilde{\tilde{\gamma}}$ denote gamma matrices in $N-2$ dimensions.

\vspace{2mm}

\nin{\bf Covariant derivatives}\\
We denote by $\tilde\nabla_\mu=\partial_\mu+\frac14\tilde\omega_\mu{}^{\ud{ab}}\tilde\gamma_{\ud{ab}},~\mu\in(\theta_1,...\theta_{N-1})$   the covariant derivative on $S^{N-1}$ and   $\tilde \gamma_{\theta_i} = \tilde e_{\theta_i}^{\ud i}\tilde\gamma_{\ud i}$ is the curved one-lower dimension gamma matrix with $\tilde e^{\ud i}_{ \theta_i}$   the $n$-beins of $S^{N-1}$.

\vspace{1mm}

\nin {\sf $N$ even}: from \eqref{sPCon}, we have 
\begin{equation}
\nabla_{t}= \partial_{t} \,\,, \quad \nabla_{\theta_i} =  \left(
\begin{array}{cc}
 \tilde\nabla_{\theta_i} -\frac{1}{2}\sinh t \, \tilde \gamma_{ \theta_i} & 0 \\
0& \tilde\nabla_{\theta_i} + \frac{1}{2}\sinh t\, \tilde \gamma_{ \theta_i} \\ 
\end{array}
\right)   \,\,.
\label{N even spinL}
\end{equation}

\vspace{1mm}

\nin {\sf $N$ odd}:  the spinor covariant derivative now reads
\begin{equation}
    \nabla_{t}= \partial_{t}\,\,,~~~~~~~\nabla_{\theta_i}=\tilde\nabla_{\theta_i}- \frac{1}{2}\sinh t \, \gamma^{\ud{0}}\,  {\tilde \gamma_{ \theta_i}}\,\,.
\end{equation}

\subsection{Spin-$\frac{1}{2}$: Dirac fermions }

\label{ferds1/2}

The massive Dirac equation in de Sitter is\footnote{Performing the Wick rotation \eqref{Wick} and replacing the eigenvalue by $L_N+\frac N2 \mapsto-iM $ in \eqref{DiracN} one finds \eqref{LorDir}.}  
\be
\slashed{\nabla}_{\text{\sf dS}_{N}} \Psi_{M...}^{(\vec a)}  = M \Psi_{M...}^{(\vec a)} \,\,.
\label{LorDir}
\ee 
The values $M\in\mathbb R$ give UIRs. In even dimensional de Sitter, positive and negative $M$ solutions give equivalent UIRs, as they are related by the action of the chiral matrix. In odd dimensions, $M$ and $-M$ define two distinct, inequivalent UIRs of the connected group Spin$(N,1)$. This follows from the highest-order Casimir operator which turns out to change its sign under $M\to -M$ (see \eqref{C2D31/2} for the result in dS$_3$).\footnote{In embedding space notation, the top Casimir for odd dimensional de Sitter  is written as   ${\cal C}^{top}=\epsilon_{A_1..A_{2k+2}}M^{A_1A_2}...M^{A_{2k+1}A_{2k+2}}$. Hence, the solution spaces for $M$ and $-M$ are mapped into each other by   (spatial) parity transformation.} As we now discuss, to each given mass $M$, two sets of solutions can be constructed.

\vspace{2mm}

\nin {\bf On complex conjugation} \\
In Lorentzian signature,  the lack of a global timelike Killing vector invalidates the traditional flat-space separation between \emph{positive-} and \emph{negative-energy} modes. Instead, in de Sitter space, the mode separation is dictated by the short-wavelength limit  as exhibited in \eqref{CompCh} \cite{ChTag,BD}. In physics parlance, we abuse language and call the solutions to \eqref{CompCh}  ``\emph{positive-energy}''.  Thus, we have  two distinct bases of modes since we can alternatively construct modes with a positive sign on the right-hand side of \eqref{CompCh}. Of course, the positive sign modes can be found by taking the complex conjugates of those  satisfying \eqref{CompCh}  (more precisely the charge conjugate spinor). They are customarily referred to in physics parlance as the ``\emph{negative-energy}'' modes. As demonstrated in \cite{lets1/2}, the  conjugated modes $\{(\Psi_{ML_{N-1}\dots\ell}^{(\vec{a})})^*\}$ are orthogonal to the original set $\{\Psi_{ML_{N-1}\dots\ell}^{(\vec{a})}\}$ with respect to the canonical Dirac inner product \eqref{DIP} (see eqn. (6.5) in \cite{lets1/2}). Consequently, the subspaces 
$$\text{span}\{\Psi_{ML_{N-1}\dots\ell}^{(\vec{a})}\}~\text{  and }~ \text{span}\{(\Psi_{ML_{N-1}\dots\ell}^{(\vec{a})})^*\}\,\,,$$ 
are orthogonal with respect to the Dirac inner product,  and carry independent, mutually conjugate UIRs of the de Sitter group. At the quantum level, these distinct sectors  encode the particle and antiparticle degrees of freedom (see discussion in \cite{dio+silva}).

\vspace{2mm}

\nin \textbf{Spin-$\frac12$ modes transformation properties}

\nin The transformation properties of the solutions $\{\Psi^{(\vec a)}_{ M L_{N-1}\dots\ell}\}$ were analyzed in \cite{lets1/2}. The symmetry acts on these modes via the Lie derivative \eqref{SLD} along the de Sitter Killing vectors. 

\vspace{1mm}

\nin{\sf Even $N$}: the relevant Killing vector governing the mixing of the modes corresponds to an $\text{so}(1,1) \subset \text{so}(N,1)$ subalgebra. Without loss of generality it can be chosen to be
\begin{equation}
\text{\sf de Sitter boost}:~~~~~\bm{D} = \cos \theta_{N-1} \,\partial_t - \tanh t \sin \theta_{N-1} \, \partial_{\theta_{N-1}} \,\,.
\label{boost2}
\end{equation}
This specific choice yields a simple and important result: in even dimensional de Sitter space $\bm D$ mixes the $\uparrow$- and $\downarrow$-components in \eqref{PsidsUP}. Indeed (cf. eqn.(5.45) in \cite{lets1/2}), 
\be
\mathcal{L}_{D}\Psi_{ML_{N-1}\dots\ell}^{(\uparrow, \vec{a}')} = R_+ \, \Psi^{(\uparrow , \vec{a}')}_{M\, L_{N-1}+1 \dots\ell } + R_-\, \Psi^{(\uparrow , \vec{a}')}_{M\, L_{N-1}-1 \dots\ell } + A \, \Psi^{(\downarrow , \vec{a}')}_{M\, L_{N-1}\dots\ell }\,\,.
\label{Lie1/2}
\ee
Here, the terms proportional to $R_\pm$ correspond to modes wherein the $\uparrow$ state is preserved, whereas the highest angular quantum number, $L_{N-1}$, is shifted up or down by one. On the other hand the last term, proportional to $A$, shows a flip $\uparrow\:\leadsto\:\downarrow$.

The coefficients $R_\pm,A$ can be found in \cite{lets1/2}. They are functions of the dimension $N$, the mass $M$ and the quantum numbers $ L_{N-1}$ and $L_{N-2}$. For a massive Dirac field, all these coefficients are non-zero. Therefore the set $\{\Psi_{M \dots }^{(\uparrow, \vec{a}')}, \Psi_{M\dots }^{(\downarrow, \vec{a}')} \}$ furnishes a UIR,
\begin{equation}
   M\in\mathbb R\neq 0. \text{\sf ~Principal series UIR}:~~~~~ \{\Psi_{ML_{N-1}\dots\ell}^{(\uparrow, \vec{a}')}, \Psi_{ML_{N-1}\dots\ell}^{(\downarrow, \vec{a}')} \}\,\,.
\end{equation}
The massless case ($M=0$) is special since $\uparrow$ and $\downarrow$ cease to be connected, i.e. in even dimensions $A=0$  if $M=0$. Therefore, in even dimensional de Sitter one finds two separate sets (cf. \eqref{transf})
\begin{equation}
     M=0.\text{\sf ~Discrete series UIR}:~~~~\{\Psi_{ML_{N-1}...\ell}^{(\uparrow,\vec{a}')} \}\,\,, \ \{\Psi_{ML_{N-1}...\ell}^{(\downarrow,\vec{a}')} \}\,\,.
     \label{DiscreteDirac}
\end{equation}
The decoupling of modes is the hallmark of Discrete/Exceptional series representations. This feature will also appear for spin-$\frac{3}{2}$ modes in dS$_N$ (see below).  

 \vspace{1mm}

\nin{\sf Odd $N$}: the $\uparrow$- and $\downarrow$-components in the massless case do mix, but the Killing vector responsible for this is one of the $S^{N-1}$ Killing vectors.

\vspace{2mm}

\subsubsection{Derivation of the mode solutions} 

Using  
\begin{equation}
    e^{\theta_i}_{\,\,\,\ud i } = \frac{1}{\cosh t} \tilde e^{\theta_i}_{\, \,  \ud i} \leadsto  \gamma^{\theta_i}= e^{\theta_i}_{\,\, \ud i}\gamma^{\ud i }= \frac{1}{\cosh t} \tilde e^{\theta_i}_{\, \, \ud i} \, \gamma^{\ud i}\,\,,
\end{equation} 
with $\tilde e^\theta$ the $(N-1)$ dimensional vielbeins on $S^{N-1}$, 
along with the fact that  $\gamma^{t}= \gamma^{\ud 0}$ and $\nabla_t = \partial_t$,   the Dirac equation reads
 \begin{equation}
     \gamma^{\ud 0} \partial_t \Psi +  \frac{1}{\cosh t} \tilde e^{\theta_i}_{\, \, \ud i } \gamma^{\ud i} \nabla_{\theta_i} \Psi = M \Psi \,\,.
 \end{equation}
Now we use our representation of the gamma matrices for even dimensions, also with \ref{N even spinL} . At some point the contraction $\tilde \gamma^{\theta_i} \tilde \gamma_{\theta_i}= \tilde \gamma_{\ud i} \tilde \gamma^{\ud i }$ appears, that is the only dimension dependent contraction, which will give $N-1$. The Dirac equation gives
\begin{equation}
    \gamma^{\ud 0}\left(\partial_t + \frac{N-1}{2} \tanh t \right)\Psi + \frac{1}{\cosh t } \left(
\begin{array}{cc}
 0 & i \slashed{\nabla}_{S^{N-1}} \\
-i\slashed{\nabla}_{S^{N-1}} & 0 \\ 
\end{array}
\right) \Psi = M \Psi\,\,,
\end{equation}
\begin{equation}
    \left(\begin{array}{cc}
 0 & i(\partial_{t} + \tfrac{N-1}{2} \tanh t )\mathds{1}+\tfrac i{\cosh t } \slashed{\nabla}_{S^{N-1}} \\
i(\partial_{t} + \tfrac{N-1}{2} \tanh t )\mathds{1}-\tfrac i{\cosh t } \slashed{\nabla}_{S^{N-1}} & 0 \\ 
\end{array}\right) \Psi = M \Psi\,\,.
\end{equation}

We now proceed to write the explicit form of the solutions for even dimensional de Sitter. The Dirac equation is solved   using the  (odd dimensional) $S^{N-1}$  spinor harmonics $\bm\chi_{(\pm)L_{N-1}...}^{...}$. The ansatze are 
\begin{align}
\text{\sf N even}:~~~~\Psi^{(\uparrow,a_{n-1},...,a_1)}_{ M L_{N-1}...\ell}(t,\Omega_{N-1}) &=   \left(
\begin{array}{c}
  if  (t) \,\bm\chi^{(a_{n-1},...,a_1)}_{(+)L_{N-1}...\ell} (\Omega_{N-1}) \\
  g  (t) \,\bm\chi^{(a_{n-1},...,a_1)}_{(+)L_{N-1}...\ell} (\Omega_{N-1}) \\
\end{array}
\right)\,\,,\nn\\ 
\Psi^{(\downarrow,a_{n-1},...,a_1)}_{ M L_{N-1}...\ell}(t,\Omega_{N-1}) &=   \left(
\begin{array}{c}
  g (t)\, \bm\chi^{(a_{n-1},...,a_1)}_{(-)L_{N-1} ... \ell} (\Omega_{N-1}) \\
  if  (t) \,\bm\chi^{(a_{n-1},...,a_1)}_{(-)L_{N-1}...\ell} (\Omega_{N-1}) \\
\end{array}
\right)\,\,.
\label{PsidsUP}
\end{align}
Inserted in the Dirac equation \eqref{LorDir} gives the following differential equations for $f,g$
\begin{align}
\left[ \frac{d}{dt}+\frac{N-1}{2} \tanh t  + \frac{i}{\cosh t } \left(L_{N-1 } + \frac{N-1}{2} \right)\right]g (t)&=M f(t)\,\,,\nn\\
\left[ \frac{d}{dt}+\frac{N-1}{2} \tanh t  - \frac{i}{\cosh t } \left(L_{N-1 } + \frac{N-1}{2} \right)\right]f (t)&=-M g(t)\,\,.
\end{align}
The solutions are
\begin{align}
    f(t)&=  C_{N}\frac{-i M}{N/2+L_{N-1}} \left(\cos \frac{\pi/2 -it}{2}\right)^{L_{N-1}} \left(\sin \frac{\pi/2 -it}{2}\right)^{L_{N-1}+1} \nn\\&~~~{_2}F_1\left(\frac{N}{2}+i M+L_{N-1}, \frac{N}{2}-i M + L_{N-1}, \frac{N}{2}+L_{N-1}+1;
   \frac{1-i \sinh t}{2}\right)\,\,,\\
  g(t) &=  C_N  \left(\cos \frac{\pi/2 -it}{2}\right)^{L_{N-1}+1} \left(\sin \frac{\pi/2 -it}{2}\right)^{L_{N-1}} \nn\\
    &~~~{_2}F_1\left(\frac{N}{2}+ i M+L_{N-1},\frac{N}{2}-iM+L_{N-1}, \frac{N}{2}+L_{N-1} ;\frac{1-i\sinh t }{2}\right)\,\,, 
\end{align}
with
\begin{equation}
\cos \frac{\pi/2-it}{2} = \frac{\sqrt{2}}{2} \left( \cosh \frac{t}{2} +i \sinh \frac{t}{2}\right)\,\,,
\end{equation}
\begin{equation}
\sin \frac{\pi/2-it}{2} = \frac{\sqrt{2}}{2} \left( \cosh \frac{t}{2} -i \sinh \frac{t}{2}\right)\,\,.
\end{equation}
We can simplify the arguments of the Hypergeometric function using the property 
$$F(a,b,c,z)= (1-z)^{c-a-b}F(c-a,c-b,c,z)\,\,.$$ 
The result is 
\begin{align}
    f(t)&=  C_{N}\frac{-i M}{N/2+L_{N-1}} \left(\cos \frac{\pi/2 -it}{2}\right)^{-N-L_{N-1}+2} \left(\sin \frac{\pi/2 -it}{2}\right)^{L_{N-1}+1} \nn\\&~~~{_2}F_1\left(i M+1, -i M + 1, \frac{N}{2}+L_{N-1}+1;
   \frac{1-i \sinh t}{2}\right)\,\,,\nn\\
  g(t) &=  C_N  \left(\cos \frac{\pi/2 -it}{2}\right)^{-N-L_{N-1}+1} \left(\sin \frac{\pi/2 -it}{2}\right)^{L_{N-1}} \nn\\
    &~~~{_2}F_1\left( i M,-iM, \frac{N}{2}+L_{N-1} ;\frac{1-i\sinh t }{2}\right)\,\,.
    \label{fg1/2dS}
\end{align}
These solutions are intrinsically complex, a feature that becomes evident by noting that in the short wavelength limit ($L_{N-1}\gg1$) they satisfy \cite{lets1/2}
\begin{align}
\frac{d}{dt}f(t) &\sim -i \frac{L_{N-1}}{\cosh t} f(t)\,\,,\nn \\[1.5ex]
\frac{d}{dt}g(t) &\sim -i \frac{L_{N-1}}{\cosh t} g(t)\,\,.
\label{CompCh}
\end{align}
The negative sign on the right-hand side should be interpreted as signaling the \emph{positive-energy} character of the mode. At the quantum level, these are modes that define the Bunch-Davies/Euclidean/Hartle-Hawking vacuum \cite{lets1/2,dio+silva}.

The normalization factor $C_N$  is adjusted using the (Lorentzian) Dirac inner product
\be
(\Psi,\Psi')=i\int_{\Sigma} d\Sigma\,  n_\mu\bar\Psi\gamma^\mu\Psi'=\int_{S^{N-1}}d\Omega_{N-1  }\cosh^{N-1}t\,\Psi^\dagger\Psi' \,\,.
\label{DIP}
\ee
Here $\bar\Psi= \Psi^\dagger\beta=i\Psi^\dagger\gamma^{\ud 0}$ is the Dirac adjoint\footnote{The $\beta$ matrix satisfies $\beta\gamma^{\ud a}\beta^{-1}=-(\gamma^{\ud a})^\dagger$. The uniqueness of the gamma matrices representation in even dimensions ensure that $\beta$ always exists.} and   $\Sigma$ is a spacelike  hipersurface with $n_\mu$ unit normal.  As mentioned, functions $f(t)$ and $g(t)$ can be obtained by analytically continuing  \eqref{psi}-\eqref{phi} using \eqref{Wick}.

In the massless case ($M=0$) the function $f(t)=0$. Since  we are considering even dimensional de Sitter spacetimes, the solutions are naturally decomposed  in their chiral components
$$\text{\sf Massless fermion }M=0:~~~~~~\gamma_*\Psi^{(a,...)}_{ M=0\, L_{N-1}... }  =s_a\Psi^{(a,...)}_{ M=0\, L_{N-1}... }\,\,, ~~~~~~~s_\uparrow=+1\,\,,~~ s_\downarrow=-1\,\,.$$
Their explicit form is
\begin{equation}
\Psi^{(\uparrow,a_{n-1},...,a_1)}_{ M\, L_{N-1}...\ell}{}_{\rfloor_{M=0}} =   \left(
\begin{array}{c}
  0\\
  g (t)_{\rfloor_{M=0}} \,\bm\chi^{(a_{n-1},...,a_1)}_{(+)L_{N-1}...\ell} (\Omega_{N-1}) \\
\end{array}
\right)\,\,,\nn
\end{equation}
\begin{equation}
\Psi^{(\downarrow,a_{n-1},...,a_1)}_{ M\, L_{N-1}...\ell}{}_{\rfloor_{M=0}} =   \left(
\begin{array}{c}
  g  (t)_{\rfloor_{M=0}}\, \bm\chi^{(a_{n-1},...,a_1)}_{(-)L_{N-1} ... \ell} (\Omega_{N-1}) \\
  0 \\
\end{array}
\right)\,\,.
\label{masslesspa}
\end{equation}

\vspace{2mm}

\subsection{Spin-$\frac32$: Rarita-Schwinger fermions}
\label{discretes}

It was recognized long ago that masslessness, gauge invariance, and lightcone propagation are not synonymous in curved spacetime \cite{DeserNepo}. The classification of spinning fields with $s\ge1$ as \emph{massive, partially massless, or strictly massless} reflects a striking phenomenon unique to de Sitter and Anti-de Sitter backgrounds for spin $s\ge1$: the emergence of gauge invariances that project out specific lower-helicity states. This results in a physical spectrum with missing intermediate helicities, named as partial masslessness. In the present paper we will refer to the so called spin-$\frac32$  \emph{strictly massless}  fermionic field as \emph{fermionic gauge fields}. We prefer this terminology because the fermion gauge invariant field equations display a mass-like term in its field equation.

---

\nin To understand why gauge invariance requires a specific non-zero value of the mass parameter, we consider the Rarita-Schwinger action in curved $N$-dimensional de Sitter space (see also \cite{DWnull}) 
\begin{equation}
    S= \int d^{N} x \sqrt{g} \, \Bar{\psi}_\mu \left(\gamma^{\mu \nu \rho} \nabla_\nu - M \gamma^{\mu \rho}\right) \psi_\rho\, \,.
\end{equation}
The equation of motion reads
\begin{equation}
\text{\sf Rarita-Schwinger}:~~~~\big(\gamma^{\mu \nu \rho} \nabla_\nu - M \gamma^{\mu \rho}\big) \psi_\rho = 0 \,\,,
\label{RSC}
\end{equation}
here $\nabla_\mu$ denotes the spin-$\frac32$ covariant derivative (see Appendix \ref{B} for conventions). We now determine the  values of  $M$ and $\beta$  that guarantee the invariance of \eqref{RSC} under the gauge transformation
\begin{equation}
  \text{\sf Gauge transformation}:~~~  \psi_\mu \rightarrow \psi_\mu + (\nabla_\mu + \beta \gamma_\mu) \epsilon \,\,,
\end{equation}
with $\epsilon=\epsilon(x)$ an arbitrary fermionic gauge parameter. Inserting this transformation in \eqref{RSC} yields
\begin{equation}
\frac{1}{2}\gamma^{\mu \nu \rho } [\nabla_\nu , \nabla_\rho] \epsilon    + \big(\beta(N-2) - M \big)\gamma^{\mu \rho} \nabla_\rho \epsilon - \beta M (N-1)\gamma^{\mu} \epsilon = 0  \,\,,
\label{Gi}
\end{equation}
where we have used 
 $\gamma^{\mu \nu \rho} \gamma_\rho = (N-2)\gamma^{\mu \nu}    , \quad \gamma^{\mu \nu}\gamma_\nu = (N-1)\gamma^{\mu}$\,\,.
Since de Sitter spacetime is maximally symmetric, the  first term of \eqref{Gi} can be further reduced to give
\begin{align}
 \frac{1}{2}\gamma^{\mu \nu \rho } [\nabla_\nu , \nabla_\rho] \epsilon  =   - \frac{1}{4\ell^2} (N-2)(N-1) \gamma^{\mu}\epsilon\,\,.
\end{align}
Inserting this result back into \eqref{Gi}, one finds
\begin{equation}
   \big(\beta(N-2)-M\big)\gamma^{\mu \rho} \nabla_\rho \epsilon - \left( \frac{1}{4\ell^2}(N-2)(N-1) + \beta M (N-1) \right) \gamma^{\mu} \epsilon  = 0 \,\,,
\end{equation}
Demanding that this expression vanishes for an arbitrary spinor $\epsilon(x)$ requires   
\begin{equation}
    \beta(N-2)- M = 0, \quad
    \frac{1}{4\ell^2}(N-2)+ \beta M = 0 ~~\Rightarrow~~\beta^2=-\frac1{4\ell^2} \,\,.
\end{equation}
This yields two solutions for the   pair $(\beta, M)$ 
$$(\beta,M)=\pm i\left(\frac 1{2\ell}, \frac{N-2}{2\ell}\right)\,\,.$$
In conclusion, the Rarita-Schwinger field equation \eqref{RSC} with mass $M = \pm i \frac{(N-2)}{2\ell}$ is invariant under the gauge transformations 
\be
\text{\sf spin-$\tfrac32$ gauge invariance}: ~~~~ \delta\psi_\mu= \big(\nabla_\mu \pm \frac{i}{2\ell } \gamma_\mu\big) \epsilon~~\Leftrightarrow~~M =\pm i\tfrac{N-2}{2\ell}
\label{Ginv}
\ee
At this point it is important to recognize that precisely for $M=\pm i \frac{(N-2)}{2\ell}$ the Type I modes to be discussed below take the form  \cite{vasilirarita,letsios3/2}:
\begin{equation}
  \text{\sf Type I. }M=\pm i\tfrac{N-2}{2\ell}:~~~~  \Psi_{\pm \mu}(t,\Omega_{N-1})= \left( \nabla_\mu \pm \frac{i}{2} \gamma_\mu\right)\epsilon_{\pm}(t,\Omega_{N-1}),
  \label{PG}
\end{equation}
with the spinor gauge function $\epsilon_{\pm}$ satisfying
\begin{equation}
    \slashed{\nabla} \epsilon_{\pm} = \mp  i \frac{N}{2} \epsilon_{\pm }.
\end{equation}
This means that for $M=\pm i \frac{(N-2)}{2\ell}$, Type I modes are pure gauge mode solutions.
 
\vspace{2mm}

\nin {\sf Comment on higher spin fields}: it is well known that fermionic  spinning fields with $s=\frac12 + r$, $r\in\mathbb N$ and mass parameter $M$ are described in dS$_N$ by a Dirac-type equation together with gamma-traceless and  divergence-free conditions (a.k.a. TT-constraints) 
\begin{equation}\slashed\nabla\Psi_{\mu_1 ...\mu_r}= M \Psi_{\mu_1 ...\mu_r} ~~
\label{diracrarita2}
 \text{such that}~~   \gamma^{\nu} \Psi_{\nu \mu_2...\mu_r} = \nabla^{\nu}\Psi_{\nu \mu_2...\mu_r}=0\,.
\end{equation}
Partially massless fields require the equation of motion to be invariant under gauge symmetries. These show up at masses \cite{DWnull,HinterFerm,vasilirarita}
\begin{equation}
     \ell^2M^2=-\left(r-\tau  + \frac{N-2}{2}\right)^2 , \quad \tau=1,...,r.
    \label{mass partially}
\end{equation}
which are pure imaginary in de Sitter spacetime.  The parameter $\tau$ is known as the depth of the partially massless field with $\tau=1$ corresponding to the fermionic gauge field (aka strictly massless) possessing 2 propagating degrees of freedom (pdof). The values $\tau=2,...,r$ lead to partially massless fermionic fields having $ 2\tau$ propagating helicities $h=\pm r,\pm (r-1),...,\pm (r-(\tau-1))$. 

\subsubsection{Mode solutions in even dimensional de Sitter}

In the bulk of the paper we discuss the $N=4$ spin-$\frac32$ field ($r=1$) which only has UIRs for  $M\in\mathbb R$ (massive/principal series)\footnote{The zero Dirac mass $M=0$ case however is reducible (see \eqref{Mzero32irrep}.).}, and $M=\pm i$ (fermionic gauge field/discrete series) which corresponds to $\tau=1$ in \eqref{mass partially}.  Nevertheless we sketch below the mode solutions for even $N$.

\vspace{2mm}

\nin {\sf Principal series. Massive fermions}: we consider $M\in\mathbb R$ and start by denoting the vector-spinor components as
\begin{equation}
    \Psi_{\mu}= (\Psi_t, \Psi_{\theta_i}).
\end{equation}
Using TT-conditions in the $t$-component of equation \eqref{diracrarita2} yields
\begin{equation}
  \left( \partial_t +\frac{N+1}{2}\tanh t \right)\gamma^{t}\Psi_t + \frac{1}{\cosh t} \left(
\begin{array}{cc}
 0 & i \cancel{\tilde\nabla} \\
-i \cancel{\tilde\nabla} & 0 \\ 
\end{array}
\right) \Psi_t = \ M \Psi_t,
\label{tcompo}
\end{equation}
and the $\theta_i$-components   read
\begin{equation}
  \left(  \partial_t + \frac{N-3}{2} \tanh t \right) \gamma^{t} \Psi_{\theta_j} + \frac{1}{\cosh t}\left(
\begin{array}{cc}
 0 & i \cancel{\tilde\nabla} \\
-i \cancel{\tilde\nabla} & 0 \\ 
\end{array}
\right) \Psi_{\theta_j}-\tanh t  \, \gamma_{\theta_j} \Psi_t =  M \Psi_{\theta_j}.
\label{thetacomp}
\end{equation}
As for the case of $S^N$, the strategy to solve these equations is the following:

\vspace{1mm}

\nin{\sf Type I}: we solve the  equation \eqref{tcompo} for the $t$-component. The $\theta_i$-components are then obtained by substituting the appropriately Wick-rotated ansatz~\eqref{32psi}--\eqref{32psi2} with unknown parameters $C^{(1,2)}$ and $D^{(1,2)}$ into the transverse and traceless constraints.\\
{\sf Type II}: we solve  \eqref{thetacomp}, setting $\Psi_t=0$,  by inserting an ansatz involving a spin-vector on the sphere (see \eqref{b} below).\\
The solutions are complex valued. Two sets are found colloquially referred to as \emph{positive/ negative-energy}.\footnote{See the discussion on complex conjugation in sect. \ref{ferds1/2}) and around eqn. \eqref{CompCh}} 

\vspace{2mm}

\nin . \textbf{Type I modes}: the ansatz for the temporal component involves a $S^{N-1}$  spin-$\frac12$   harmonic $\bm\chi$
\begin{equation}
 {\sf Type \,I}:\quad   \Psi_{ t\,ML_{N-1 ...\ell}}^{{\sf I}\,(\uparrow,\vec{a}')} (t,\Omega_{N-1})=   \left(
\begin{array}{c}
  i\, f (t) \,\bm \chi^{(\vec{a}')}_{(+)L_{N-1}...\ell} (\Omega_{N-1}) \\
   g (t) \,\bm\chi^{(\vec{a}')}_{(+)L_{N-1}...\ell} (\Omega_{N-1}) 
  \label{aa}
\end{array}
\right)  .
\end{equation}
We can also build $\Psi^{{\sf I}\, (\downarrow, \vec{a}')}$ by using $\bm \chi^{(\vec{a}')}_{(-)L_{N-1}...\ell}$  as in \eqref{PsidsUP}. Substituting into the Dirac equation we find  
\begin{align}
\left[ \frac{d}{dt}+\frac{N+1}{2} \tanh t  + \frac{i}{\cosh t } \left(L_{N-1 } + \frac{N-1}{2} \right)\right]g (t)&=M f(t),\nn\\
\left[ \frac{d}{dt}+\frac{N+1}{2} \tanh t  - \frac{i}{\cosh t } \left(L_{N-1 } + \frac{N-1}{2} \right)\right]f (t)&=-M g(t).
\end{align}
The \emph{positive-energy} solutions satisfying \eqref{CompCh} are
\begin{align}
    f(t)=&\frac{-iM}{L_{N-1}+N/2} \left(\cos \frac{\pi/2-it}{2} \right)^{L_{N-1}-1} \left(\sin \frac{\pi/2-it}{2} \right)^{L_{N-1}} \\
    &\cross F \left(iM + \frac{N}{2}+L_{N-1} , -iM + \frac{N}{2}+L_{N-1}, L_{N-1} + \frac{N+2}{2}, \sin^2 \frac{\pi/2-it}{2} \right).\nn 
\end{align}
\begin{align}
    g(t)=& \left(\cos \frac{\pi/2-it}{2} \right)^{L_{N-1}} \left(\sin \frac{\pi/2-it}{2} \right)^{L_{N-1}-1} \\
    &\cross F \left(i M + \frac{N}{2}+L_{N-1} , -iM + \frac{N}{2}+L_{N-1}, L_{N-1} + \frac{N}{2}, \sin^2 \frac{\pi/2-it}{2} \right). \nn
\end{align}
The components $\Psi^{{\sf I}\,(\uparrow,\vec{a}')}_{\theta_i \, ML_{N-1}...\ell}(t,\Omega_{N-1})$ can be obtained from \eqref{aa} by imposing the transverse-traceless conditions in \eqref{diracrarita2}. We do not present their explicit expressions, as they are rather lengthy and are not needed for our analysis. Their structure is analogous to that of the Euclidean solutions in \eqref{32psi}. As mentioned before,  the  Lorentzian mode solutions can be found by analytic continuation from the Euclidean ones.

\vspace{2mm}

\nin . \textbf{Type II modes}: the temporal component vanishes and they involve $S^{N-1}$ spin-$\frac32$ harmonics $\bm\chi_{\theta_i}$ in the ansatz
\begin{equation}
{\sf Type \,II}: \quad  \Psi_{ t \, M L_{N-1}...\ell}^{{\sf II}}=0, \quad \Psi_{ \theta_i \, ML_{N-1}...\ell}^{{\sf II}\,(\uparrow,\vec{a}')} (t,\Omega_{N-1})=   \left(
\begin{array}{c}
  i\, f (t) \,\bm \chi^{(\vec{a}')}_{\theta_i(+)L_{N-1}...\ell} (\Omega_{N-1}) \\
   g (t) \,\bm \chi^{(\vec{a}')}_{\theta_i(+)L_{N-1}...\ell} (\Omega_{N-1}) \\ 
\end{array}
\right)  .
\label{b}
\end{equation}
The $\Psi^{{\sf II\,} (\downarrow,\vec{a}')}$ solution is built using the $S^{N-1}$ spin-vectors $\bm \chi_{(-)\theta_i}$. Inserting  \eqref{b} into \eqref{thetacomp} one finds 
\begin{align}
\left[ \frac{d}{dt}+\frac{N-3}{2} \tanh t  + \frac{i}{\cosh t } \left(L_{N-1 } + \frac{N-1}{2} \right)\right]g (t)&=M f(t),\nn\\
\left[ \frac{d}{dt}+\frac{N-3}{2} \tanh t  - \frac{i}{\cosh t } \left(L_{N-1 } + \frac{N-1}{2} \right)\right]f (t)&=-M g(t).
\end{align}
The solutions satisfying \eqref{CompCh} read 
\begin{align}
    f^{}(t)&= \frac{-i M}{L_{N-1}+N/2} \left(\cos \frac{\pi/2-it}{2} \right)^{L_{N-1}+1} \left(\sin \frac{\pi/2-it}{2} \right)^{L_{N-1}+2}\nn \\
    &~~\cross F \left(i M + \frac{N}{2}+L_{N-1} , -iM + \frac{N}{2}+L_{N-1}, L_{N-1} + \frac{N+2}{2}, \sin^2 \frac{\pi/2-it}{2} \right), \nn\\
    g(t)&=  \left(\cos \frac{\pi/2-it}{2} \right)^{L_{N-1}+2} \left(\sin \frac{\pi/2-it}{2} \right)^{L_{N-1}+1} \nn\\
    &~~\cross F \left(i M + \frac{N}{2}+L_{N-1} , -i M + \frac{N}{2}+L_{N-1}, L_{N-1} + \frac{N}{2}, \sin^2 \frac{\pi/2-it}{2} \right).
\end{align}
As expected, in the $M=0$ (zero Dirac mass limit) the  equations for $f,g$ decouple and we get  chiral solutions  similar  to those in \eqref{masslesspa}.

\vspace{2mm}

\nin {\sf Discrete series. Fermionic gauge fields}: we now consider $M=i \frac{(N-2)}{2}$ and write\footnote{The modes for the opposite sign for the mass can be easily obtained using the chiral matrix.}
\begin{equation}
\slashed{\nabla}\Psi_{ \mu}=   i \tilde M \Psi_{ \mu} ,\quad \tilde M= \tfrac{N-2}{2}
\end{equation}
Type I and type II solutions can be succinctly described in terms of the ansatze \eqref{aa} and \eqref{b}, which only differ by either involving a $S^{N-1}$ spinor or a $S^{N-1}$ vector-spinor harmonic respectively. The solutions are  
\begin{align}
    f^{(b)}(t)&= \frac{\tilde M}{L_{N-1}+N/2} \left(\cos \frac{\pi/2-it}{2} \right)^{L_{N-1}-b} \left(\sin \frac{\pi/2-it}{2} \right)^{L_{N-1}+1-b}\nn \\
    &~~\cross F \left(-\tilde M + \frac{N}{2}+L_{N-1} , \tilde M + \frac{N}{2}+L_{N-1}, L_{N-1} + \frac{N+2}{2}, \sin^2 \frac{\pi/2-it}{2} \right), \nn\\
    g^{(b)}(t)&=  \left(\cos \frac{\pi/2-it}{2} \right)^{L_{N-1}+1-b} \left(\sin \frac{\pi/2-it}{2} \right)^{L_{N-1}-b} \nn\\
    &~~\cross F \left(-\tilde M + \frac{N}{2}+L_{N-1} , \tilde M + \frac{N}{2}+L_{N-1}, L_{N-1} + \frac{N}{2}, \sin^2 \frac{\pi/2-it}{2} \right),
\end{align}
where $b=1$ for Type I and $b=-1$ for Type II modes. 

Some comments are in order:\\
i. In $N=4$ the RS equation displays gauge invariance  for $M=i$ (see \eqref{Ginv}) and the solution space becomes indecomposable.\\
ii.  It is important to realize that type I modes  are ``pure gauge'' (see comments  in \eqref{PG}).\\
iii. It is only in $N=4$ that the Type II are unitary (with an invariant positive inner product \cite{letsios3/2,vasilirarita}).

\vspace{2mm}

\nin \textbf{Type I and Type II  modes  transformation properties} \\
The relevant isometry for the study of the mixing between type I and II modes is the  boost \eqref{boost2}. In even dimensions Type I modes transform as 
\begin{align}
\mathcal{L}_{D}\Psi_{\mu\,M L_{N-1}...\ell}^{{\sf I}\,(\uparrow,\vec{a}')} &=  R_{+}^{({\sf I})}\Psi_{\mu\,M L_{N-1}+1\,...\ell}^{{\sf I}\,(\uparrow,\vec{a}')} + R_{-}^{({\sf I})}\Psi_{\mu\,M L_{N-1}-1\,...\ell}^{{\sf I}\,(\uparrow,\vec{a}')}+A^{({\sf I})}\Psi_{\mu\,M  L_{N-1}...\ell}^{{\sf I}\,(\downarrow,\vec{a}')}\nn\\
&~~~+ B^{({\sf I}\rightarrow {\sf II})} \Psi_{\mu\,M L_{N-1}...\ell}^{{\sf II}\,(\uparrow,\vec{a}')}\,\,,\nn\\
\mathcal{L}_{D}\Psi_{\mu\,M L_{N-1}...\ell}^{{\sf II}\,(\uparrow,\vec{a}')} &=  R^{({\sf II})}_{+}\Psi_{\mu\,M L_{N-1}+1\,...\ell}^{{\sf II}\,(\uparrow,\vec{a}')} + R^{({\sf II})}_{-}\Psi_{\mu\,M L_{N-1}-1\,...\ell}^{{\sf II}\,(\uparrow,\vec{a}')}+A^{({\sf II})}\Psi_{\mu\,M  L_{N-1}...\ell}^{{\sf II}\,(\downarrow,\vec{a}')}\nn\\
&+ B^{({\sf II}\rightarrow {\sf I})} \Psi_{\mu\,M L_{N-1}...\ell}^{{\sf I}\,(\uparrow,\vec{a}')}\,\,.
\label{mixing}
\end{align}
For the   Killing vector \eqref{boost2}, the coefficients $R_{\pm},A$ and $B$  depend solely on $M,L_{N-1}$  and the spacetime dimension $N$. They can be found in section VIII.A of \cite{vasilirarita}. Notice that $R_{\pm}$ account for a change in the principal quantum number of the sphere, meaning $L_{N-1}\to L_{N-1}\pm1$. The coefficient $A$ implies a mixing between $(\uparrow,...)-$ $(\downarrow,...)$-modes, and the $B$-coefficient shows a mixing between type I and type II modes.   Their computation shows  that for the principal series ($M\in\mathbb R$) all the coefficients are non-vanishing. In turn, this implies the UIR multiplet in even dimensions is
\begin{equation}
{\sf\,\, Principal \,\,series}.~M  \in \mathbb{R} :  \quad    \{  \Psi_{\mu \,M L_{N-1}...\ell}^{{\sf I}\,(\vec{a})} , \Psi_{\mu\, M L_{N-1}...\ell}^{{\sf II}\,(\vec{a})}  \} \,\,.
\end{equation}
In the zero Dirac mass case $M=0$, the coefficients $A$  vanish. Consequently, the $(\vec a)=(\uparrow,...)$ and $(\vec a)=(\downarrow,...)$ sectors are no longer connected under the symmetry. This leads  to a splitting of the multiplet,
\begin{equation}
{\sf  Discrete \,\,series}.~ M  = 0: \quad \{ \Psi_{\mu,\, ML_{N-1}...\ell}^{{\sf I}\,(\uparrow,\vec{a}')},  \ \Psi_{\mu,\, ML_{N-1}...\ell}^{{\sf II}\,(\uparrow,\vec{a}')} \} \,\,, \, \{ \Psi_{\mu,\, ML_{N-1}...\ell}^{{\sf I}\,(\downarrow,\vec{a}')},  \ \Psi_{\mu,\, ML_{N-1}...\ell}^{{\sf II}\,(\downarrow,\vec{a}')} \} \,\,.
\label{Mzero32irrep}
\end{equation}
A similar halving was found for the spin-$\frac{1}{2}$ massless fields \eqref{DiscreteDirac}.

As demonstrated in \eqref{Ginv}, at   $M= \pm i (N-2)/2$  the Rarita-Schwinger field equation \eqref{RSC} acquires gauge invariance. Precisely at this same value, the coefficient $B^{({\sf I} \rightarrow {\sf II})} $ vanishes, thus  Type I (pure gauge) modes close among themselves under the action of the symmetry.
As $B^{({\sf II} \rightarrow {\sf I})} $ remains non-vanishing, the multiplet becomes indecomposable. Moreover, one can show that the Type I modes  have vanishing norm, implying that the Type II modes, with non-zero positive norm in $N=4$, constitute the physical sector  \cite{letsios3/2,vasilirarita}.

\section{Casimir operators and their relation to  Dirac  and Rarita-Schwinger fields on $S^{3}$ and dS$_3$}
\label{casimirsS3}

In this appendix we give a realization of the Casimir operators of SO(4) and SO(3,1) on spinors and vector-spinors.

\subsection{Round 3-sphere}

The metric for the round 3-metric is given in \eqref{g3}. Its isometry group is Spin(4) $\sim$ SU(2)$_L$ $\times$ SU(2)$_{R}$ which is realized in terms of two sets of mutually commuting non-vanishing Killing vectors $\{\bm L_i\}$ and $\{\bm R_i\}~(i=1,2,3)$ satisfying
\begin{equation}
    [\bm L_i, \bm R_j]= 0 \,\,, \quad [\bm L_i , \bm L_j] = -2\epsilon_{ijk}\bm L_k \,\,, \quad [\bm R_i , \bm R_j ]= 2 \epsilon_{ijk}\bm R_k\,\,.
\end{equation}
with so(3)$_L$ generators
\begin{align}
\bm L_1&= \sin \theta_2 \cos \phi \, \bm \partial_{\theta_3} + (\cot \theta_3 \cos \theta_2 \cos \phi - \sin \phi) \bm \partial_{\theta_2} - (\cot \theta_3 \csc \theta_2 \sin \phi + \cot \theta_2 \cos \phi) \bm \partial_{\phi} \nn\\ 
\bm L_2&= \sin \theta_2 \sin \phi \, \bm \partial_{\theta_3} + (\cot \theta_3 \cos \theta_2 \sin \phi + \cos \phi) \bm\partial_{\theta_2} + (\cot \theta_3 \csc \theta_2 \cos \phi - \cot \theta_2 \sin \phi) \bm\partial_{\phi} \nn\\ 
\bm  L_3 &=  \cos \theta_2 \, \bm\partial_{\theta_3} - \cot \theta_3 \sin \theta_2 \, \bm\partial_{\theta_2} + \bm\partial_\phi\,\,,
\end{align}
with Casimir
$${\cal C}_L= (\bm L_i)^2=-4j_L(j_L+1)\,\,,$$
and  so(3)$_R$ generators 
\begin{align}
\bm R_1 &= \sin \theta_2 \cos \phi \, \bm\partial_{\theta_3} + (\cot \theta_3 \cos \theta_2 \cos \phi + \sin \phi) \bm\partial_{\theta_2} + (\cot \theta_2 \cos \phi - \cot \theta_3 \csc \theta_2 \sin \phi) \bm\partial_\phi \nn\\ 
\bm R_2 &= \sin \theta_2 \sin \phi \,\bm\partial_{\theta_3} + (\cot \theta_3 \cos \theta_2 \sin \phi - \cos \phi) \bm\partial_{\theta_2} + (\cot \theta_3 \csc \theta_2 \cos \phi + \cot \theta_2 \sin \phi) \bm\partial_\phi \nn\\ 
\bm R_3 &= \cos \theta_2 \, \bm\partial_{\theta_3} - \cot \theta_3 \sin \theta_2 \,\bm\partial_{\theta_2} - \bm\partial_\phi\,\,,
\end{align}
with Casimir
$${\cal C}_R= (\bm R_i)^2=-4j_R(j_R+1)\,\,,$$
The change of basis 
\begin{equation}
\bm J_i := \frac{1}{2} (\bm R_i - \bm L_i) \,\,, \quad \bm P_i := \frac{1}{2}(\bm R_i +\bm L_i)\,\,,
\end{equation}
expresses the so(4) algebra as 
\begin{equation}
[\bm J_i, \bm J_j]= \epsilon_{ijk}J_k \,\,, \quad    [\bm P_i, \bm P_j]= \epsilon_{ijk}\bm J_k \,\,,  \quad [\bm P_i , \bm J_j]=\epsilon_{ijk}\bm P_k\,\,.
\label{leftright}
\end{equation}
Now,
\be 
\bm J_1 =\sin \phi \, \bm \partial_{\theta_2} + \cot \theta_2 \cos \phi \, \bm\partial_\phi\,\,, ~~
\bm J_2 = -\cos \phi \, \bm \partial_{\theta_2} + \cot \theta_2 \sin \phi \, \bm\partial_{\phi}\,\,,~~
\bm J_3  = -\bm\partial_\phi\,\,.\nn
\ee
and
\begin{align}
\bm P_1 &=
\sin \theta_2 \cos \phi \, \bm \partial_{\theta_3} + \cot \theta_3 \cos \theta_2 \cos \phi \, \bm\partial_{\theta_2} - \cot \theta_3 \csc \theta_2 \sin \phi \, \bm\partial_\phi\nn
\\ 
\bm P_2 &=
\sin \theta_2 \sin \phi \, \bm\partial_{\theta_3} + \cot \theta_3 \cos \theta_2 \sin \phi \, \bm\partial_{\theta_2} + \cot \theta_3 \csc \theta_2 \cos \phi \, \bm\partial_\phi\nn
\\ 
\bm P_3 &=
\cos \theta_2 \, \bm\partial_{\theta_3} - \cot \theta_3 \sin \theta_2 \, \bm\partial_{\theta_2}\,\,.
\label{Pset}  
\end{align}
In   the  $\{ \bm J_i,\bm P_i\}$ basis, the two so(4) Casimirs are 
\begin{alignat}{3}
    &\mathcal{C}_1 &&=  (\bm J_i)^2 +(\bm P_i)^2 &&=-2\left(j_R(j_R+1)+j_L(j_L+1)\right) \,\,,\nn\\
    &\mathcal{C}_2 &&=~\, - \bm J_i \bm P_i&&=j_R(j_R+1)-j_L(j_L+1)\,\,.
    \label{C1y2}
\end{alignat}
Here the values $j_L,j_R=0,\frac12,1,\frac32,..$ give UIRs.

In the next section we turn to the connection between Casimir operators and field equations. The general idea is that the abstract Casimir operator is realized on a particular tensor field $\Psi$ by the replacement of the  $\bm T_a\to\bm k_a\to{\cal L}_{ k_a}$, where ${\cal L}_{k_a}$ is the Lie derivative along the Killing vector $\bm k_a$ realizing the abstract Lie algebra generator $\bm T_a$. Thus, our aim is to compute 
\begin{align}
\mathcal{C}_1 \Psi &= ( \mathcal{L}_{J_1}\mathcal{L}_{J_1} + \mathcal{L}_{J_2}\mathcal{L}_{J_2} + \mathcal{L}_{J_3}\mathcal{L}_{J_3}+\mathcal{L}_{P_1}\mathcal{L}_{P_1}+ \mathcal{L}_{P_2}\mathcal{L}_{P_2}+\mathcal{L}_{P_3}\mathcal{L}_{P_3}) \Psi \,\,, \nn\\ 
 \mathcal{C}_2 \Psi&= -( \mathcal{L}_{J_1}\mathcal{L}_{P_1} +\mathcal{L}_{J_2} \mathcal{L}_{P_2} +\mathcal{L}_{J_3} \mathcal{L}_{P_3} ) \Psi \,\,.
 \label{LieCas}
\end{align}
Since we will be working with spinor fields, the Lie derivative is that defined by Kosmann long ago  given in \eqref{SLD} for spin-$\frac12$ and in \eqref{RSLD} for spin-$\frac32$ (see \cite{ortin,kos,FoF,ortin2}).

\subsubsection{Dirac spinor}

Spinor Lie derivatives along the Killing vectors provide a realization of the symmetry algebra on spinor fields. Their explicit expression for the six $so(4)$ Killing vectors are
\begin{align}
\mathcal{L}_{J_1} \Psi = J_1^{\mu}  \partial_\mu \Psi- \frac{1}{2} \gamma^{\ud{21}}\cos \phi \csc \theta_2 \Psi \,\,,\quad \mathcal{L}_{J_2} \Psi =  J_2^{\mu}  \partial_\mu \Psi - \frac{1}{2}\gamma^{\ud{21}} \sin \phi \csc \theta_2 \Psi \,\,,\nn
\end{align}
$$        \mathcal{L}_{J_3} \Psi= -  \partial_\phi \Psi \,\,,
$$
\begin{alignat}{3}  \mathcal{L}_{P_1}\Psi &=   P_1^{\mu}  \partial_\mu \Psi  - \frac{1}{2}\gamma^{\ud{32}} \cos\phi \cos \theta_2 \csc \theta_3 \Psi+ \frac{1}{2} \gamma^{\ud{31}} \sin\phi \csc \theta_3 \Psi  + \frac{1}{2} \gamma^{\ud{21}} \cot \theta_3 \cot \theta_2 \sin \phi \Psi\nn \\ \mathcal{L}_{P_2}\Psi&=   P_2^{\mu}  \partial_\mu \Psi  -\frac{1}{2}\gamma^{\ud{32}}\sin \phi \cos \theta_2 \csc \theta_3 \Psi - \frac{1}{2} \gamma^{\ud{31}}\cos \phi \csc \theta_3 \Psi  - \frac{1}{2} \gamma^{\ud{21}} \cot \theta_3 \cot \theta_2 \cos \phi \Psi \nn \\
   \mathcal{L}_{P_3}\Psi &= P_3^{\mu} \partial_\mu \Psi + \frac{1}{2} \gamma^{\ud{32}} \sin \theta_2 \csc \theta_3 \Psi \,\,.\nn
\end{alignat}
One verifies that the Lie derivative provides a  representation of the algebra \eqref{leftright},
\begin{equation}
[\mathcal{L}_{J_i}, \mathcal{L}_{J_j}]\Psi= \epsilon_{ijk} \mathcal{L}_{J_k}\Psi\,\,, \quad    [\mathcal{L}_{P_i}, \mathcal{L}_{P_j}]\Psi= \epsilon_{ijk}\mathcal{L}_{J_k}\Psi \,\,,  \quad [\mathcal{L}_{P_i} , \mathcal{L}_{J_j}]\Psi=\epsilon_{ijk}\mathcal{L}_{P_k}\Psi\,\,.\nn
\end{equation}
On $S^{3}$ we find  the relation
\begin{equation}
  ( \mathcal{C}_1 - \nabla^2) \Psi= -\frac{3}{4}\Psi \,\,.
  \label{lapla}
\end{equation}
Here $\nabla^2=g^{\mu\nu}\nabla_\mu\nabla_\nu$ is the spinor Laplacian, with $\nabla_\mu=\partial_\mu+\omega_\mu$ the covariant derivative involving the spin connection (see app.\ref{B} for details). Equation \eqref{lapla} expresses the well known relation between   the Laplacian and the Casimir on maximally symmetric spaces. We can now   relate this last expression to the Dirac operator $\slashed\nabla= \gamma^\mu\nabla_\mu$. Using the relation 
\begin{equation}
    \slashed{\nabla}^2 \Psi = \nabla^2 \Psi - \frac{R}{4} \Psi \,\,, 
  \label{mass2}
\end{equation}
first obtained by Schr\"odinger \cite{schr} (see also \cite{GK}), and using  the fact that the Ricci scalar of the unit $S^{3}$ is $R=N(N-1)=6$, we conclude that
\begin{equation}
{\cal C}_1 \Psi= \left(\slashed \nabla ^2+ \frac34 \right)\Psi\,\,.
\label{C1spin12}
\end{equation}
The relation for the second Casimir  is
\begin{equation}
    \mathcal{C}_2 \Psi = -\frac{i}{2}\slashed{\nabla} \Psi\,\,.
    \label{c2spin12}
\end{equation}
It is amusing to see that even though ${\cal C}_2$ involves second order derivatives, the relation leads to only first order ones.

Consider now the $S^3$ spinor harmonics obeying 
\begin{equation}
\slashed\nabla\Psi_{(\pm)L_3L_2\ell}^{(\vec{a})} = \pm i(L_3+ \tfrac3 2)\Psi_{(\pm) L_3 L_2 \ell}^{(\vec{a})}\,\,.
\end{equation}
We then conclude that 
$$\Psi_{(\pm)L_3L_2\ell}^{(\vec a)}~~\leadsto~~\left\{
\begin{array}{l}
{\cal C}_1= \tfrac{3}{4} - \left(L_3+\tfrac{3}{2} \right)^2 \\[2mm]
{\cal C}_2=\pm  \tfrac{1}{2} \left( L_3+\tfrac{3}{2}\right)
\end{array}.\right.$$
Comparing to \eqref{C1y2} we conclude that each spinor eigenspace constitutes a UIR of Spin(4) where
\begin{align}
\Psi_{(+)L_3L_2\ell}^{(\vec a)}:~~~~&(j_L,j_R)=\left(\frac{L_3}2,\frac{L_3}2+\frac12\right)\,\,,\nn\\
\Psi_{(-)L_3L_2\ell}^{(\vec a)}:~~~~&(j_L,j_R)=\left(\frac{L_3}2+\frac12,\frac{L_3}2\right)\,\,.\nn
\end{align}

\subsubsection{Rarita-Schwinger spinor}

The definitions of the covariant and Lie derivatives on vector-spinors can be found in appendix \ref{B} and equation \eqref{RSLD}. The spin connection components can be found in \eqref{scS3}. To simplify the notation we only denote the vector components   
\begin{equation}
    \Psi_\mu = (\Psi_{\theta_3}, \Psi_{\theta_2}, \Psi_{\phi}) \,\,.
\end{equation}
The explicit expressions for the  covariant derivatives are
$$ \nabla_{\theta_3} \Psi_{\theta_3} = \partial_{\theta_3} \Psi_{\theta_3}\,\,,~~  \nabla_{\theta_3}\Psi_{\theta_2} =\partial_{\theta_3} \Psi_{\theta_2}-\cot \theta_3 \Psi_{\theta_2}\,\,,~~ \nabla_{\theta_3}\Psi_{\phi} = \partial_{\theta_3} \Psi_{\phi} - \cot \theta_3 \Psi_{\phi}\,\,,$$   
\begin{align}
 \nabla_{\theta_2} \Psi_{\theta_3} &= \partial_{\theta_2}\Psi_{\theta_3} - \frac{1}{2}\cos\theta_3 \gamma^{\ud{32}} \Psi_{\theta_3} - \cot \theta_3 \Psi_{\theta_2}\nn \\ 
 \nabla_{\theta_2}\Psi_{\theta_2}&= \partial_{\theta_2}\Psi_{\theta_2} - \frac{1}{2}\cos\theta_3 \gamma^{\ud{32}} \Psi_{\theta_2} + \cos \theta_3 \sin\theta_3\Psi_{\theta_3}\nn \\ 
  \nabla_{\theta_2}\Psi_{\phi} &= \partial_{\theta_2}\Psi_{\phi} - \frac{1}{2}\cos\theta_3 \gamma^{\ud{32}} \Psi_{\phi} - \cot \theta_2 \Psi_{\phi}\nn
\end{align} 
\begin{align}
 \nabla_{\phi}\Psi_{\theta_3} &= \partial_\phi \Psi_{\theta_3}  - \frac{1}{2}\cos\theta_2 \gamma^{\ud{21}} \Psi_{\theta_3} - \frac{1}{2}\cos\theta_3\sin\theta_2 \gamma^{\ud{31}}\Psi_{\theta_3} - \cot \theta_3 \Psi_{\phi}\nn \\
 \nabla_{\phi}\Psi_{\theta_2}&= \partial_\phi \Psi_{\theta_2}  - \frac{1}{2}\cos\theta_2 \gamma^{\ud{21}} \Psi_{\theta_2} - \frac{1}{2}\cos\theta_3\sin\theta_2 \gamma^{\ud{31}}\Psi_{\theta_2} - \cot \theta_2 \Psi_{\phi} \nn\\ 
   \nabla_{\phi}\Psi_{\phi} &=  \partial_\phi \Psi_{\phi}  - \frac{1}{2}\cos\theta_2 \gamma^{\ud{21}} \Psi_{\phi} - \frac{1}{2}\cos\theta_3\sin\theta_2 \gamma^{\ud{31}}\Psi_{\phi} +  \cos \theta_2 \sin\theta_2 \Psi_{\theta_2}\nn \\
   &~\quad~+ \cos \theta_3 \sin \theta_3 \sin^2 \theta_2 \Psi_{\theta_3} \nn
\end{align}
Whereas for the Lie derivatives one obtains
$$ \mathcal{L}_{J_3} \Psi_{\theta_3}= -\partial_\phi \Psi_{\theta_3}\,\,,~~~ \mathcal{L}_{J_3}\Psi_{\theta_2}= - \partial_{\phi} \Psi_{\theta_2} \,\,,~~~ \mathcal{L}_{J_3}\Psi_{\phi}=-\partial_{\phi} \Psi_{\phi}\,\,,$$ 
\begin{align} \mathcal{L}_{J_2}\Psi_{\theta_3}&= J_2^{\mu} \partial_{\mu} \Psi_{\theta_3}-\frac{1}{2}\sin \phi \csc \theta_2  \gamma^{\ud{21}}\Psi_{\theta_3} \nn\\ 
 \mathcal{L}_{J_2}\Psi_{\theta_2} &= J_2^{\mu} \partial_\mu \Psi_{\theta_2} -\frac{1}{2} \sin \phi \csc \theta_2 \gamma^{\ud{21}} \Psi_{\theta_2 } - \sin \phi \csc^2 \theta_2 \Psi_{\phi} \nn\\ \mathcal{L}_{J_2} \Psi_{\phi}& = \xi^{\mu}_{J_2}\partial_\mu  \Psi_\phi  +(\cos \phi \cot \theta_2 - \frac{1}{2} \gamma^{\ud{21}} \csc \theta_2 \sin \phi)\Psi_{\phi}+\sin\phi\,\Psi_{\theta_2}\nn
 \end{align}
 \begin{align}
    & \mathcal{L}_{J_1} \Psi_{\theta_3} = J_1^{\mu}  \partial_\mu \Psi_{\theta_3}  -\frac{1}{2} \cos \phi\csc \theta_2 \gamma^{\ud{21}}\Psi_{\theta_3}\nn \\ 
    &  \mathcal{L}_{J_1} \Psi_{\theta_2} = J_1 ^{\mu} \partial_\mu  \Psi_{\theta_2} -\frac{1}{2}\cos \phi\csc \theta_2 \gamma^{\ud{21} }\Psi_{\theta_2}- \cos \phi \csc^2 \theta_2 \Psi_{\phi} \nn\\ 
    &   \mathcal{L}_{J_1}\Psi_\phi = J_1^{\mu} \partial_\mu \Psi_{\phi}  -\left( \frac{1}{2}\cos \phi \csc \theta_2  \gamma^{\ud{21}} + \cot \theta_2 \sin \phi\right)\Psi_\phi + \cos \phi \,\Psi_{\theta_2} \nn
\end{align}
\begin{align}
&\mathcal{L}_{P_3}\Psi_{\theta_3}={P_3}^{\mu}\partial_\mu \Psi_{\theta_3} + \frac{1}{2} \csc \theta_3 \sin \theta_2 \gamma^{\ud{32}} \Psi_{\theta_3} +\csc ^2\theta_3 \sin \theta_2 \Psi_{\theta_2} \nn\\ 
&  \mathcal{L}_{P_3}\Psi_{\theta_2}= {P_3}^{\mu}\partial_\mu \Psi_{\theta_2}  + \left( \frac{1}{2} \csc \theta_3 \sin \theta_2 \gamma^{\ud{32}} -\cos \theta_2 \cot \theta_3\right) \Psi_{\theta_2}-\sin \theta_2 \Psi_{\theta_3}\nn \\ 
&  \mathcal{L}_{P_3}\Psi_{\phi} =  {P_3}^{\mu}\partial_\mu \Psi_\phi + \frac{1}{2} \csc \theta_3 \sin \theta_2 \gamma^{\ud{32}} \Psi_\phi \nn
\end{align}
\begin{align}
    \mathcal{L}_{P_2} \Psi_{\theta_3} &=   {P_2}^{\mu}\partial_\mu \Psi_{\theta_3} -\frac{1}{2}\left(\cos \phi\csc \theta_3 \gamma^{\ud{31}}+  \cos \phi \cot \theta_2 \cot \theta_3 \gamma^{\ud{21}}+ \cos \theta_2 \sin \phi \gamma^{\ud{32}}\right)\Psi_{\theta_3} \nn\\
    &~~-\csc^2\theta_3 \cos\theta_2 \sin \phi \, \Psi_{\theta_2} -\csc^2\theta_3 \csc\theta_2 \cos\phi \,\Psi_\phi\nn\\
    \mathcal{L}_{P_2} \Psi_{\theta_2} &=   {P_2}^{\mu}\partial_\mu \Psi_{\theta_2} -\frac{1}{2} \left(\cos \phi \csc \theta_3 \gamma^{\ud{31}} +\cos \phi \cot \theta_2\cot\theta_3 \gamma^{\ud{21}}+\sin\phi \csc\theta_3\cos\theta_2 \gamma^{\ud{32}}\right)\Psi_{\theta_2} \nn\\
    &~~ +\cos \theta_2 \sin \phi \,\Psi_{\theta_3}-\cot \theta_3 \sin\theta_2 \sin\phi\, \Psi_{\theta_2}-\cot \theta_3 \csc \theta_2 \cot \theta_2 \cos\phi \,\Psi_\phi\nn\\
    \mathcal{L}_{P_2} \Psi_{\phi} &=   {P_2}^\mu \partial_\mu \Psi_\phi -\frac{1}{2}\left( \cos \phi \csc \theta_3 \gamma^{\ud{31}} +\cos \phi \cot \theta_2\cot\theta_3 \gamma^{\ud{21}}+\sin\phi \csc\theta_3\cos\theta_2 \gamma^{\ud{32}}\right)\Psi_{\phi}\nn\\
    &~~+\sin \theta_2 \cos \phi \, \Psi_{\theta_3}+\cot \theta_3 \cos \theta_2 \cos\phi\,\Psi_{\theta_2 }-\cot \theta_3 \csc \theta_2 \sin \phi \, \Psi_\phi\nn
\end{align}
\begin{align}
    \mathcal{L}_{P_1} \Psi_{\theta_3} &=   P_1^{\mu}\partial_\mu \Psi_{\theta_3} +\frac{1}{2} \left(\csc \theta_3 \sin \phi\,\gamma^{\ud{31}}+ \cot \theta_3 \cot \theta_2 \sin \phi \, \gamma^{\ud{21}} -\csc \theta_3 \cos \theta_2 \cos \phi \, \gamma^{\ud{32}}\right)\Psi_{\theta_3}\nn\\
    &~~-\csc^2 \theta_3 \cos \theta_2 \cos\phi \,\Psi_{\theta_2} +\csc ^2 \theta_3 \csc \theta_2 \sin \phi \, \Psi_\phi\nn\\
    \mathcal{L}_{P_1} \Psi_{\theta_2}& =  P_1^{\mu}\partial_\mu \Psi_{\theta_2}+\frac{1}{2}\left(\csc \theta_3 \sin \phi \, \gamma^{\ud{31}}+ \cot \theta_3 \cot \theta_2 \sin \phi \, \gamma^{\ud{21}} -\csc \theta_3 \cos \theta_2 \cos \phi \, \gamma^{\ud{32}}\right)\Psi_{\theta_2} \nn\\
    &~~+ \cos \theta_2 \cos \phi \, \Psi_{\theta_3}- \cot \theta_3 \sin \theta_2 \cos \phi \, \Psi_{\theta_2}+ \cot \theta_3 \csc\theta_2 \cot \theta_2 \sin \phi \, \Psi_{\phi}\nn\\
    \mathcal{L}_{P_1} \Psi_{\phi} 
    &=  P_1^{\mu}\partial_\mu  \Psi_{\phi}  + \frac{1}{2}\left(\csc \theta_3 \sin \phi \, \gamma^{\ud{31}}+ \cot \theta_3 \cot \theta_2 \sin \phi \, \gamma^{\ud{21}} -\csc \theta_3 \cos \theta_2 \cos \phi \, \gamma^{\ud{32}}\right)\Psi_{\phi} \nn\\
    &~~+ \sin \theta_2 \sin\phi \, \Psi_{\theta_3} -\cot \theta_3 \cos \theta_2 \sin \phi \, \Psi_{\theta_2}- \cot \theta_3 \csc \theta_2 \cos \phi \, \Psi_{\phi}\nn
\end{align}
It can be checked that this set of Lie derivatives provides a realization of \eqref{leftright}.  

Inserting the Lie derivatives in the Casimir ${\cal C}_1$ given in \eqref{LieCas} and subtracting the RS-Laplacian one finds 
\begin{equation}
\mathcal{C}_1\Psi_{\mu}- \nabla^2 \Psi_{\mu} = -\frac{15}{4} \Psi_{\mu}+ \gamma_{\mu} (\gamma \cdot \Psi)\,\,.
\end{equation} 
We will consider below the action of this Casimir operator on TT spin-$\frac32$ eigenmodes. Thus, since they are $\gamma$-traceless the last term will vanish. To connect ${\cal C}_1$ to the Dirac operator we use  Weitzenb\"ock's generalization of \eqref{mass2} to TT vector-spinor fields. The relation is (cf. below eqn (5.5) in \cite{letsios3/2}) 
$$ \nabla^{\mu}\nabla_{\mu} = \slashed\nabla^2 + \frac52\,\,.$$
The final result for the Casimir  on TT spin-$\frac32$ fields is
\begin{equation}
\mathcal{C}_1\Psi_\mu =\left(\slashed\nabla^2 -\frac54 \right)\Psi_\mu\,\,.
\label{C1RS}
\end{equation}
For the second casimir, we find the following identity
\begin{equation}
\mathcal{C}_2 \Psi_\sigma - \frac{1}{2i} \slashed{\nabla}\Psi_\sigma -i g_{\sigma\mu } \gamma^{\mu \nu \rho} \nabla_\nu \Psi_\rho = 0 \,\,.
\label{c2S3/2}
\end{equation}
Using the identity
$      \gamma^{\mu \nu \rho}= \frac{1}{2} (\gamma^{\nu }\gamma^{\rho}\gamma^{\mu} - \gamma^{\mu}\gamma^{\rho}\gamma^{\nu})
$ we can rewrite the last term as
\begin{equation}
 \gamma^{\mu \nu \rho} \nabla_\nu \Psi_\rho =    \slashed{\nabla}\Psi^\mu -  \nabla^\mu (\gamma \cdot\Psi) -   \gamma^{\mu} (\nabla\cdot\Psi) +  \gamma^{\mu} \slashed\nabla  (\gamma \cdot \Psi)\,\, .
\label{idnRS}
\end{equation} 
Using this identity in \eqref{c2S3/2} and considering TT spin-$\frac32$ fields yields
\be
\mathcal{C}_2 \Psi_\mu =-\frac {3i}2   \slashed{\nabla} \Psi_\mu\,\,.  
\label{c2spin32}
\ee
Consider $S^3$ TT vector-spinor harmonics 
\begin{align}
 \slashed{\nabla} \Psi^{(\vec a)}_{\mu\:(\pm) L_3L_2\ell}   = \pm i \left( L_3 + \frac{3}{2} \right) \Psi^{(\vec a)}_{\mu\: (\pm)L_3L_2\ell} 
 ~\oplus~~
\nabla^{\mu}\Psi^{(\vec a)}_{\mu\:(\pm)L_3L_2\ell}=\gamma^{\mu}\Psi^{(\vec a)}_{\mu\:(\pm)L_3L_2\ell}= 0 \,\,.\nn
\end{align}
$$\Psi_{\mu\,(\pm)L_3L_2\ell}^{(\vec a)}~~\leadsto~~\left\{
\begin{array}{l}
{\cal C}_1= -\big(\tfrac{5}{4} + \left(L_3+\tfrac{3}{2} \right)^2\big) \\[2mm]
{\cal C}_2=\pm  \tfrac{3}{2} \left( L_3+\tfrac{3}{2}\right)
\end{array}\right..$$
The corresponding eigenvalues of the su(2)$_L$ $\times$ su(2)$_R$ quantum numbers using \eqref{C1y2} are
$$\Psi_{\mu\,(+)L_3L_2\ell}^{(\vec a)}:~~~~(j_L,j_R)=( \tfrac {L_3}2-\tfrac12,\tfrac {L_3}2+1)\,\,,$$
$$\Psi_{\mu\,(-)L_3L_2\ell}^{(\vec a)}:~~~~(j_L,j_R)=( \tfrac {L_3}2+1,\tfrac {L_3}2-\tfrac12)\,\,.$$
Notice the quantum numbers are non-negative which is consistent with the result obtained in   \eqref{qn3/2I},\eqref{qn3/2II}, i.e. $L_3=1,2,...$\,\,.

\subsection{ 3d de Sitter spacetime}

Metric  in global coordinates $x^\mu=(t,\theta_2,\phi)$: 
\begin{equation}
ds^2 = - dt^2 + \cosh^2 t  (d\theta_2^2 + \sin^2 \theta_2  \, d\phi^2)\,\,.
    \label{ds3}
\end{equation}
Dreibeins:
\begin{equation}
\bm e^{\ud 0}= \bm dt \,\,, \quad \bm e^{\ud 2}=\cosh t\,\bm d\theta_2 \,\,, \quad \bm e^{\ud 1} = \cosh t \sin \theta_2\,\bm d\phi \,\,.
\end{equation}
Christoffel symbols:
\begin{equation}
\Gamma^{t}_{\theta_2 \theta_2}= \cosh t \sinh t , \quad  \Gamma^{t}_{\phi \phi} = \cosh t \sinh t \sin^2 \theta_2, \quad \Gamma^{\theta_2}_{t \theta_2}=\Gamma^{\theta_2}_{\theta_2 t }= \tanh t 
\end{equation}
\begin{equation}
\Gamma^{\phi}_{t\phi}=\Gamma^{\phi}_{\phi t} = \tanh t  ,\quad   \Gamma^{\theta_2}_{\phi \phi}=-\sin \theta_2 \cos \theta_2 , \quad \Gamma^{\phi}_{\theta_2 \phi}= \Gamma^{\phi}_{\phi \theta_2} = \cot \theta_2
\end{equation}
Spin connection:
\begin{equation}
\bm \omega^{ \ud{02}}=\sinh t\,\bm d\theta_2, \quad \bm\omega^{  \ud{01}}=\sinh t \sin \theta_2\,\bm d\phi, \quad \bm\omega^{\ud{21}}=-\cos \theta_2\,\bm d\phi
\label{SCds3}
\end{equation}

---

\nin The isometry group of \eqref{ds3} is SO(3,1) which is realized in terms of six   Killing vectors $\{\bm J_i,\bm K_i\}~(i=1,2,3)$
\begin{align}
\bm J_1 =\sin \phi \, \bm\partial_{\theta_2} + \cot \theta_2 \cos \phi \, \bm\partial_\phi,~~~
\bm J_2 = -\cos \phi \, \bm \partial_{\theta_2} + \cot \theta_2 \sin \phi \, \bm\partial_{\phi},~~~ \bm J_3  = -\bm\partial_\phi\nn
\end{align}
\begin{align}
\bm K_1 &=
\sin\theta_2\cos\phi\,\bm \partial_t
+\tanh t\,\cos\theta_2\cos\phi\,\bm \partial_{\theta_2}
-\tanh t\,\csc\theta_2\sin\phi\,\bm \partial_\phi,\nn
\\ 
\bm K_2 &=
\sin\theta_2\sin\phi\,\bm \partial_t
+\tanh t\,\cos\theta_2\sin\phi\,\bm \partial_{\theta_2}
+\tanh t\,\csc\theta_2\cos\phi\,\bm \partial_\phi,\nn
\\ 
\bm K_3 &=
\cos\theta_2\,\bm \partial_t
-\tanh t\,\sin\theta_2\,\bm \partial_{\theta_2}.\nn
\end{align}
which fulfill the algebra 
\begin{equation}
[\bm J_i, \bm J_j]= \epsilon_{ijk}\bm J_k, \quad    [\bm K_i, \bm K_j]=- \epsilon_{ijk}\bm J_k ,  \quad [\bm K_i , \bm J_j]=\epsilon_{ijk}\bm K_k,
\end{equation}
notice the sign change in the second commutator when compared to \eqref{leftright}.  The $\{\bm K_i\}$   Killing vectors can be obtained by analytically continuing the $\{\bm P_i\}$ set \eqref{Pset}, using \eqref{Wick}.

The Casimir operators for so(3,1) are both quadratic\footnote{These are the traditional quadratic Casimirs  $$\text{\sf Scalar}:~{\cal C}_1 = \frac{1}{2} M_{\mu\nu} M^{\mu\nu},\qquad\text{\sf   Pseudoscalar}:~ {\cal  C}_2 = \frac{1}{8} \epsilon_{\mu\nu\rho\sigma} M^{\mu\nu} M^{\rho\sigma}.$$}
\begin{equation}
\mathcal{C}_1 =(\bm J_i)^2-(\bm K_i)^2 , \quad \mathcal{C}_2 =\bm K\cdot  \bm J 
\end{equation}
Parametrizing the Casimir eigenvalues as \footnote{Notice that the expression for ${\cal C}_1$ matches the universal CFT formula for the so$(d,1)$ quadratic Casimir $-C^{(2)} = \Delta(\Delta - d) + s(s + d - 2)$ evaluated at $d=2$. }
$$\mathcal{C}_1 =-\left(\Delta(\Delta-2)+s^2\right) , \quad \mathcal{C}_2 =-is(\Delta-1) $$
The allowed values for $(\Delta,s)$ giving UIRs are: 

. {\sf Principal series. Heavy fields}: $s=0,\frac12,1,\frac32,...$ and $\Delta=1+i\nu,~\nu\in\mathbb R$.

. {\sf Complementary series. Light fields}: $s=0$ and $\Delta\in(0,2)$\\
We remind the reader  the unitary equivalence of $\Delta$ and $\Delta^s=2-\Delta$: the Harish-Chandra character for a representation with scaling dimension $\Delta$ coincides with the character for $\Delta^s=2 - \Delta$. Therefore, the UIRs labeled by $\Delta \in (0,1)$ are unitarily equivalent to the UIRs labeled by $\Delta \in (1,2)$. Analogously for $\Delta=1+i\nu\leftrightarrow \Delta^s=1-i\nu$ (see \cite{dob,Boers}).

Our aim below  is to compute the action of the Casimir operators as Lie derivative 
\begin{align}
&    \mathcal{C}_1 \Psi = \left( \mathcal{L}_{J_1}\mathcal{L}_{J_1} + \mathcal{L}_{J_2}\mathcal{L}_{J_2} + \mathcal{L}_{J_3}\mathcal{L}_{J_3}-\mathcal{L}_{K_1}\mathcal{L}_{K_1} - \mathcal{L}_{K_2}\mathcal{L}_{K_2}-\mathcal{L}_{K_3}\mathcal{L}_{K_3} \right) \Psi \label{C1ds3}\\
&   \mathcal{C}_2 \Psi= \left( \mathcal{L}_{J_1}\mathcal{L}_{K_1} +\mathcal{L}_{J_2}\mathcal{L}_{K_2} +\mathcal{L}_{J_3}\mathcal{L}_{K_3} \right) \Psi
\label{cas2lor}
\end{align}

\subsubsection{Spin-$\frac{1}{2}$}
Covariant derivatives:
\begin{eqnarray}  
&\nabla_{t}\Psi= \partial_t \Psi, \quad \nabla_{\theta_2}\Psi =\partial_{\theta_2} \Psi - \frac{1}{2} \sinh t \, \gamma^{\ud{02}}\Psi \nn\\
   &\nabla_\phi \Psi = \partial_\phi \Psi - \frac{1}2  \sinh t \sin \theta_2 \gamma^{\ud{01}} \Psi - \frac{1}{2}\cos \theta_2 \gamma^{\ud{21}}\Psi\nn
\end{eqnarray}
Lie derivatives:
\begin{eqnarray}
    &\mathcal{L}_{J_1} \Psi = {J_1}^{\mu} \partial_\mu \Psi - \frac{1}{2} \gamma^{\ud{21}}\cos \phi \csc \theta_2  \Psi,~~\mathcal{L}_{J_2} \Psi =  {J_2}^{\mu} \partial_\mu \Psi  - \frac{1}{2}\gamma^{\ud{21}} \sin \phi \csc \theta_2  \Psi\nn\\
     &\mathcal{L}_{J_3} \Psi= -  \partial_\phi \Psi \nn
\end{eqnarray}
\begin{align}
\mathcal{L}_{K_1} \Psi&= {K_1}^{\mu} \partial_\mu \Psi + \frac{1}{2}\gamma^{\ud{02}} \sech t  \cos \theta_2 \cos\phi \,\Psi- \frac{1}{2}        \gamma^{\ud{01}}  \sech t \sin\phi     \,\Psi \\
&~~+ \frac{1}{2} \gamma^{\ud{21}} \tanh t  \cot \theta_2 \sin \phi \,\Psi\nn \\
\mathcal{L}_{K_2}\Psi&={K_2}^{\mu} \partial_\mu \Psi +\frac{1}{2}\gamma^{\ud{02}} \sech t \cos \theta_2  \sin \phi \, \Psi + \frac{1}{2} \gamma^{\ud{01}} \sech t \cos \phi \, \Psi \nn \\
   &~~ - \frac{1}{2} \gamma^{\ud{21}} \tanh t \cot \theta_2 \cos \phi \, \Psi \nn
 \\
\mathcal{L}_{K_3} \Psi &= K_3^{\mu} \partial_\mu \Psi - \frac{1}{2} \gamma^{\ud{02}}\sech t  \sin \theta_2 \Psi    \nn
\end{align}
Inserting these expressions in \eqref{C1ds3} and subtracting the spinor Laplacian $\nabla^2=g^{\mu\nu}\nabla_\mu\nabla_\nu$  (with $\nabla_\mu=\partial_\mu+\omega_\mu$) we find
\begin{equation}
   \left( \mathcal{C}_1 - \nabla^2 \right) \Psi = -\frac{3}{4} \Psi
\end{equation}
We can relate $\nabla^2$ with $\slashed{\nabla}^2$ using the Schr\"odinger-Lichnerowicz relation \eqref{mass2} \cite{schr,GK,liche}. Since the dS$_3$ Ricci scalar is $R=6$, we obtain
\begin{equation}
    \mathcal{C}_1 \Psi = \left(\slashed{\nabla}^2 + \frac{3}{4} \right)\Psi
\end{equation}
For the second Casimir \eqref{cas2lor}  we find
\begin{equation}
    \mathcal{C}_2 \Psi= \frac{1}{2} \slashed{\nabla} \Psi 
    \label{C2D31/2}
\end{equation}
Consider now the Dirac equation in dS$_3$,
\be
\slashed\nabla\Psi=M\Psi\,,
\label{Dirds3}
\ee
we then conclude that the mass fixes the UIR,
$$ \Psi_ M ~\leadsto~(\Delta,s)=\left(1+iM,\tfrac12\right)$$
More precisely, the Dirac fermion carries two UIRs in its solution space. For additional discussion see the section on complex conjugation in app.\ref{ferds1/2}

\subsubsection{Spin-$\frac{3}{2}$}

We denote the spin-vector components in  dS$_3$ as
\begin{equation}
    \Psi_\mu=(\Psi_t, \Psi_{\theta_2},\Psi_\phi ).
\end{equation}
The covariant derivatives are
$$  \nabla_t \Psi_t= \partial_t \Psi_t,~~ \nabla_t \Psi_{\theta_2}=  \partial_t \Psi_{\theta_2} - \tanh t \, \Psi_{\theta_2},~~ \nabla_t \Psi_\phi= \partial_t \Psi_{\phi}- \tanh t \, \Psi_\phi $$  
\begin{align}
\nabla_{\theta_2} \Psi_{t}\,&=  \partial_{\theta_2} \Psi_t - \tfrac{1}{2} \sinh t \, \gamma^{\ud{02}} \Psi_t - \tanh t \, \Psi_{\theta_2}\nn\\ 
\nabla_{\theta_2} \Psi_{\theta_2}&= \partial_{\theta_2} \Psi_{\theta_2} - \tfrac{1}{2} \sinh t \, \gamma^{\ud{02}} \Psi_{\theta_2}- \cosh t \sinh t \, \Psi_t \nn\\ \nabla_{\theta_2} \Psi_{\phi}&=  \partial_{\theta_2} \Psi_\phi - \tfrac{1}{2} \sinh t \, \gamma^{\ud{02}} \Psi_\phi - \cot \theta_2 \, \Psi_\phi\nn
\end{align}
\begin{align} 
\nabla_\phi \Psi_t&= \partial_\phi \Psi_t - \tfrac{1}2  \left(\sinh t \sin \theta_2 \gamma^{\ud{01}} +\cos \theta_2 \gamma^{\ud{21}}\right)\Psi_t - \tanh t \, \Psi_\phi\nn \\ 
 \nabla_\phi \Psi_{\theta_2} &=  \partial_\phi \Psi_{\theta_2} - \tfrac{1}2 \left( \sinh t \sin \theta_2 \gamma^{\ud{01}} +\cos \theta_2 \gamma^{\ud{21}}\right)\Psi_{\theta_2}  - \cot \theta_2 \, \Psi_\phi \nn\\      \nabla_\phi \Psi_{\phi}&=  \partial_\phi \Psi_\phi - \tfrac{1}2 \left( \sinh t \sin \theta_2 \gamma^{\ud{01}} +\cos \theta_2 \gamma^{\ud{21}}\right)\Psi_\phi - \cosh t \sinh t \sin^2 \theta_2 \, \Psi_t \nn\\
  &~~+ \sin \theta_2 \cos \theta_2 \, \Psi_{\theta_2}\nn
\end{align}
The Lie derivatives are
\begin{align}
\mathcal{L}_{J_3} \Psi_{t}=  - \partial_\phi \Psi_t,~~  \mathcal{L}_{J_3} \Psi_{\theta_2}= - \partial_\phi \Psi_{\theta_2}, ~~\mathcal{L}_{J_3} \Psi_{\phi}= - \partial_\phi \Psi_{\phi} \nn
\end{align}
\begin{align}
\mathcal{L}_{J_2} \Psi_{t}&=  {J_2}^{\mu}\partial_\mu \Psi_t - \tfrac{1}{2}\csc \theta_2 \sin \phi \, \gamma^{\ud{21}} \,\Psi_t \nn\\
\mathcal{L}_{J_2} \Psi_{\theta_2}&=  {J_2}^{\mu}\partial_\mu \Psi_{\theta_2}- \tfrac{1}{2}\csc \theta_2 \sin \phi \, \gamma^{\ud {21}} \,\Psi_{\theta_2} -\csc^2 \theta_2 \sin \phi \, \Psi_\phi \nn\\
\mathcal{L}_{J_2} \Psi_{\phi}&=  {J_2}^{\mu}  \partial_\mu \Psi_\phi + \left(\cot \theta_2 \cos \phi - \tfrac{1}{2} \csc \theta_2 \sin \phi \, \gamma^{\ud {21}} \right)  \Psi_{\phi} + \sin \phi \, \Psi_{\theta_2}  \nn
\end{align}
\begin{align}
 \mathcal{L}_{J_1} \Psi_{t}&=  {J_1}^\mu \partial_\mu \Psi_t - \tfrac{1}{2}\csc \theta_2 \cos \phi \, \gamma^{\ud{21}}\Psi_t \nn \\  \mathcal{L}_{J_1} \Psi_{\theta_2}&=   {J_1}^\mu \partial_\mu \Psi_{\theta_2} - \tfrac{1}{2}\csc \theta_2 \cos \phi \, \gamma^{\ud{21}}\Psi_{\theta_2} -\csc^2 \theta_2 \cos \phi \,\Psi_\phi \nn\\ 
 \mathcal{L}_{J_1} \Psi_{\phi}&=   {J_1}^\mu \partial_\mu \Psi_\phi - \left(\cot \theta_2 \sin \phi+ \tfrac{1}{2}\csc \theta_2 \cos \phi \, \gamma^{\ud{21}} \right)\Psi_\phi   + \cos \phi \, \Psi_{\theta_2} \nn
\end{align}
\begin{align}
\mathcal{L}_{K_3} \Psi_{t}&=  {K_3}^{\mu}\partial_\mu \Psi_t - \tfrac{1}{2}\sech t \sin \theta_2 \, \gamma^{\ud{02}}\,\Psi_t - \sech^2 t \sin \theta_2 \, \Psi_{\theta_2}\nn \\
\mathcal{L}_{K_3} \Psi_{\theta_2}&=  {K_3}^{\mu} \partial_\mu \Psi_{\theta_2} - \left(\tfrac{1}{2} \sech t \sin \theta_2 \, \gamma^{\ud{02}}+ \tanh t \cos \theta_2 \right)\Psi_{\theta_2}-\sin \theta_2 \,\Psi_t  \nn \\ 
 \mathcal{L}_{K_3} \Psi_{\phi}&=    {K_3}^{\mu} \partial_\mu \Psi_{\phi}- \tfrac{1}{2}\sech t \sin \theta_2 \, \gamma^{\ud{02}}\, \Psi_\phi\nn
\end{align}
\begin{align}
\mathcal{L}_{K_2} \Psi_{t}&=    {K_2}^\mu\partial_\mu \Psi_t + \tfrac{1}{2}\left(\sech t \cos \theta_2 \sin \phi \, \gamma^{\ud{02} }+\sech t \cos \phi \, \gamma^{\ud{01}} -\tanh t \cot \theta_2 \cos \phi \, \gamma^{\ud{21}}\right) \Psi_t \nn\\
&~~+\sech^2 t  \cos \theta_2 \sin \phi \, \Psi_{\theta_2} + \sech^2 t \csc \theta_2 \cos \phi \, \Psi_\phi\nn\\[1mm]
\mathcal{L}_{K_2}\Psi_{\theta_2} &=  {K_2}^\mu\partial_\mu \Psi_{\theta_2}+ \tfrac{1}{2}\left(\sech t \cos \theta_2 \sin \phi \, \gamma^{\ud{02}}+\sech t \cos \phi \, \gamma^{\ud{01}}-2\tanh t \cot \theta_2 \cos \phi \, \gamma^{\ud{21}}\right) \Psi_{\theta_2 }\nn\\[1mm]
&~~- \tanh t \sin \theta_2 \sin \phi \, \Psi_{\theta_2 } + \cos \theta_2 \sin \phi \, \Psi_t  -\tanh t \cot \theta_2 \csc \theta_2 \cos \phi \, \Psi_{\phi}\nn\\
\mathcal{L}_{K_2}\Psi_{\phi}&= {K_2}^\mu\partial_\mu \Psi_\phi + \tfrac{1}{2} \left(\sech t \cos \theta_2 \sin \phi \, \gamma^{\ud{02}}+\sech t \cos \phi \, \gamma^{\ud{01} }-  \tanh t \cot \theta_2 \cos \phi \, \gamma^{\ud{21}}\right) \Psi_\phi\nn\\
&~~- \tanh t \csc \theta_2 \sin \phi \, \Psi_\phi + \sin \theta_2 \cos \phi \, \Psi_t  + \tanh t \cos \theta_2 \cos \phi \, \Psi_{\theta_2}\nn
\end{align}
\begin{align}
\mathcal{L}_{K_1}\Psi_t&=    {K_1}^\mu\partial_\mu \Psi_t + \tfrac{1}{2}\left(\sech t \cos \theta_2 \cos \phi\, \gamma^{\ud{02}}-\sech t \sin \phi \, \gamma^{\ud{01}}+\tanh t \cot \theta_2 \sin \phi \, \gamma^{\ud{21}}\right) \Psi_t \nn\\
&~~+ \sech^2 t \cos \theta_2 \cos \phi \, \Psi_{\theta_2}   - \sech^2 t \csc \theta_2 \sin \phi \, \Psi_\phi\nn\\
\mathcal{L}_{K_1}\Psi_{\theta_2}&=  {K_1}^\mu \partial_\mu \Psi_{\theta_2} + \tfrac{1}{2}\left(\sech t \cos \theta_2 \cos \phi \, \gamma^{\ud{02}}-\sech t \sin \phi \, \gamma^{\ud{01}}+2\tanh t \cot \theta_2 \sin \phi \, \gamma^{\ud{21}}\right) \Psi_{\theta_2}\nn\\
&~~- \tanh t \sin \theta_2 \cos \phi \, \Psi_{\theta_2} +\cos \theta_2 \cos \phi \, \Psi_t  + \tanh t \csc \theta_2 \cot \theta_2 \sin \phi \, \Psi_\phi\nn\\
\mathcal{L}_{K_1} \Psi_{\phi}&=  {K_1}^\mu \partial_\mu  \, \Psi_\phi + \tfrac{1}{2}\left(\sech t \cos \theta_2 \cos \phi \, \gamma^{\ud{02}}  -\sech t \sin \phi \, \gamma^{\ud{01}}+\tanh t \cot \theta_2 \sin \phi \, \gamma^{\ud{21}}\right) \Psi_\phi\nn\\
&~~- \tanh t \csc \theta_2 \cos \phi \, \Psi_\phi - \sin \theta_2 \sin \phi \, \Psi_t - \tanh t \cos \theta_2 \sin \phi \, \Psi_{\theta_2}\nn
\end{align}
Inserting these expressions in \eqref{C1ds3} and subtracting the vector-spinor Laplacian we obtain 
\begin{equation}
    \mathcal{C}_1 \Psi_\mu - \nabla^2 \Psi_\mu = - \frac{15}{4} \Psi_\mu+ \gamma_\mu  \, (\gamma\cdot \Psi).
    \label{cas1loren}
\end{equation}
The generalized Schr\"odinger-Lichnerowicz-Weitzenb\"ock relation for a rank-$r$ TT tensor-spinor $\Psi_{\mu_1...\mu_r}$ is
\begin{equation}
    \nabla^2 = \slashed \nabla ^2 +\frac{N(N-1)}{4} + r.
\end{equation}
Inserting this relation for $r=1$ in \eqref{cas1loren} and assuming $\Psi_\mu$ satisfies TT conditions, we arrive to
\begin{equation}
    \mathcal{C}_1\Psi_\mu = \left( \slashed \nabla^2 - \frac{5}{4}\right)\Psi_\mu .
\end{equation}
For the second Casimir we obtain
\begin{equation}
    \mathcal{C}_2 \Psi_\sigma =\frac{1}{2} \slashed{\nabla} \Psi_\sigma + g_{\sigma\mu} \gamma^{\mu \nu \rho} \nabla_\nu \Psi_\rho  .
\end{equation}
Using the identity \eqref{idnRS} and assuming $\Psi_\mu$ satisfies TT conditions
\begin{equation}
    \gamma \cdot \Psi = \nabla \cdot \Psi = 0 ,
\end{equation}
we find the relation
\begin{equation}
\mathcal{C}_2 \Psi_\sigma = \frac{3}{2} \slashed \nabla \Psi_\sigma .
\end{equation}
Consider now a Rarita-Schwinger field satisfying \eqref{RSC} which is known to be equivalently described, for $M\ne0$, by \eqref{diracrarita2} with $r=1$. Then, 
$$\Psi_{\mu\,M} ~~\leadsto~~\left\{
\begin{array}{l}
{\cal C}_1= M^2-\tfrac{5}{4}  \\[2mm]
{\cal C}_2=  \tfrac{3}{2} M
\end{array}\right.~\leadsto~(\Delta,s)=\left(1+iM,\tfrac32\right)$$

\section{Dirac and Rarita-Schwinger spinors on dS$_4$ and their relation to  SO(4,1) UIRs }
\label{apendixE}

In this section we  parametrize dS$_4$ with  
\begin{equation}
    ds^2 = \frac{- dT^2 + d\Omega_3^2}{\sin^2 T} ,
\end{equation}
here $d\Omega_3^2$ is the round 3-sphere metric \eqref{g3}.
 
The vielbeins are
\begin{equation}
    \bm e^{\ud 0}  = \frac{1}{\sin T}\bm dT , \quad \bm e^{\ud 3}  = \frac{1}{\sin T}\bm d\theta_3 , \quad \bm e^{\ud 2} = \frac{\sin \theta_3}{\sin T}\bm d\theta_2, \quad \bm e^{\ud 1} = \frac{\sin \theta_3 \sin \theta_2}{ \sin T }\bm d\phi
\end{equation}
Christoffel symbols:
\begin{equation*}
    \Gamma^{T}_{TT}= \Gamma^{T}_{\theta_3\theta_3}=\Gamma^{\theta_3}_{T \theta_3}=\Gamma^{\theta_2}_{T \theta_2}= \Gamma^{\phi}_{T\phi}= - \cot T , \, \quad \Gamma^{T}_{\theta_2 \theta_2}=-\cot T \sin^2 \theta_3, 
\end{equation*}
\begin{equation*}
    \ \Gamma^{T}_{\phi\phi}=-\cot T \sin^2 \theta_2 \sin^2 \theta_3, \quad \Gamma^{\theta_3}_{\theta_2 \theta_2}= - \cos \theta_3 \sin \theta_3 , \quad \Gamma^{\theta_3}_{\phi \phi}= - \cos \theta_3 \sin \theta_3 \sin ^2 \theta_2
\end{equation*}
\begin{equation*}
    \Gamma^{\theta_2}_{\theta_2 \theta_3}=\Gamma^{\phi}_{\phi \theta_3}= \cot \theta_3 , \quad \Gamma^{\theta_2}_{\phi\phi}=-\cos \theta_2 \sin \theta_2 , \quad \Gamma^{\phi}_{\phi \theta_2}=\cot \theta_2. 
\end{equation*}
Spin connection:
\begin{equation}
    \bm\omega_{\ud{03}}= \cot T\, \bm d\theta_3,  \quad  \bm \omega_{ \ud{02}}= \cot T \sin \theta_3\,\bm d\theta_2 , \quad \bm \omega_{ \ud{32}}=-\cos \theta_3\,\bm d\theta_2
\end{equation}
\begin{equation}
   \bm \omega_{ \ud{01}}=\cot T \sin \theta_2 \sin \theta_3\, \bm d\phi , \quad \bm\omega_{ \ud{31}}=- \cos \theta_3 \sin \theta_2\,\bm d\phi , \quad \bm\omega_{ \ud{21}}=-\cos \theta_2 \bm d\phi
\end{equation}
The isometry group  of dS$_4$ is $SO(1,4)$. 
The Killing vectors  take the form (app. C in \cite{AGS})\footnote{The coordinates  are adapted so that in the $T\to0$ we obtain the CKV of $S^3$.}
\begin{equation}
    \bm D= \sin T  \cos \theta_3 \,\bm \partial_T + \cos T \sin \theta_3 \, \bm\partial_{\theta_3}
\end{equation}
\begin{align}
\bm J_1 &=\sin \phi \, \bm\partial_{\theta_2} + \cot \theta_2 \cos \phi \, \bm\partial_\phi \nn\\ 
\bm J_2 &= -\cos \phi \, \bm\partial_{\theta_2} + \cot \theta_2 \sin \phi \, \bm\partial_{\phi} \nn\\ 
\bm J_3 &= -\bm \partial_\phi
\end{align}
\begin{align}
    & \bm P_1= \sin\theta_2 \cos \phi \, \bm \partial_{\theta_3} + \cot \theta_3 \cos \theta_2 \cos \phi \, \bm \partial_{\theta_2}-\cot \theta_3 \csc \theta_2 \sin \phi \, \bm \partial_\phi \nn\\ 
    & \bm P_2= \sin \theta_2 \sin \phi \, \bm \partial_{\theta_3}+ \cot \theta_3 \cos \theta_2 \sin \phi \, \bm \partial_{\theta_2} + \cot \theta_3 \csc \theta_2 \cos \phi \, \bm \partial_\phi \nn\\ 
    & \bm P_3=  \cos \theta_2 \, \bm \partial_{\theta_3} - \cot \theta_3 \sin \theta_2 \, \bm \partial_{\theta_2}
\end{align}
\begin{align}
\bm K_{1}&= -\sin T\,\sin\theta_{3}\,\sin\theta_{2}\,\cos\phi\, \bm \partial_{T}
+\cos T\,\cos\theta_{3}\,\sin\theta_{2}\,\cos\phi\, \bm \partial_{\theta_{3}}
\nonumber\\
&~~+\cos T\,\csc\theta_{3}\,\cos\theta_{2}\,\cos\phi\, \bm \partial_{\theta_{2}}
-\cos T\,\csc\theta_{3}\,\csc\theta_{2}\,\sin\phi\, \bm \partial_{\phi},
\nn\\ 
\bm K_{2}&= -\sin T\,\sin\theta_{3}\,\sin\theta_{2}\,\sin\phi\, \bm \partial_{T}
+\cos T\,\cos\theta_{3}\,\sin\theta_{2}\,\sin\phi\, \bm \partial_{\theta_{3}}
\nonumber\\
&+\cos T\,\csc\theta_{3}\,\cos\theta_{2}\,\sin\phi\, \bm \partial_{\theta_{2}}
+\cos T\,\csc\theta_{3}\,\csc\theta_{2}\,\cos\phi\, \bm \partial_{\phi},
\nn\\ 
\bm K_{3}&= -\sin T\,\sin\theta_{3}\,\cos\theta_{2}\, \bm \partial_{T}
+\cos T\,\cos\theta_{3}\,\cos\theta_{2}\,\partial_{\theta_{3}}
-\cos T\, \csc\theta_{3}\,\sin\theta_{2}\, \bm \partial_{\theta_{2}}.
\end{align}
The algebra in the basis $\{\bm D,\bm J_i, \bm P_i,\bm  K_i \}$ is
\begin{align}
[\bm J_i,\bm J_j] &= \epsilon_{ijk}\bm J_k,
& [\bm J_i,\bm P_j] &= \epsilon_{ijk}\bm P_k,
& [\bm J_i,\bm K_j] &= \epsilon_{ijk}\bm K_k,\nn 
\\ 
[\bm P_i,\bm P_j] &= \epsilon_{ijk}\bm J_k,
&
[\bm K_i,\bm K_j] &= -\epsilon_{ijk}\bm J_k,
\\ 
[\bm D,\bm P_i] &= -\bm K_i,
&
[\bm D,\bm K_i] &= -\bm P_i,
&
[\bm P_i,\bm K_j] &= -\bm D\,\delta_{ij}.\nn
\end{align}
In this basis the Killing-Cartan metric is diagonal and separates neatly compact and non-compact generators. The quadratic Casimir takes the form \cite{dS4Cas}
\begin{equation}
    \mathcal{C}^{(2)} = -(\bm D)^2-(\bm K_i)^2+(\bm P_i)^2 + (\bm J_i)^2
    \label{C2ds4Cas}
\end{equation}
The second Casimir operator of so(4,1) is quartic in the generators and obtained by squaring the Pauli-Luba\'nski  vector 
$$W_A:=\frac18\epsilon_{ABCDE}M^{BC}M^{DE}$$ 
as\footnote{In eqn \eqref{C4}, $A=0,1,2,3,4$ is the index of the embedding space  Minkowski coordinates $X^A$. For our adapted basis the 5 components of $W$ decompose into two scalars and a 3-vector
$$   
W_0  = \bm P \cdot \bm J,~~W_4 = \bm K \cdot \bm J,~~ \bm W_i = -\bm D\, \bm J_i - (\bm P \times \bm K)_i$$
}$^,$\footnote{We thank M. Mizrahi for reminding us that the Casimir \eqref{C4} can also be obtained using the invariant tensor $g_{a_1...a_4}={\rm STr}[T_{a_1}...T_{a_4}]$ where STr is the symmetrized trace.}
\be
{\cal C}^{(4)}=W_AW^A
\label{C4}
\ee
which in our basis can be written as
\begin{align}
 {\mathcal{C}}^{(4)} &=  -(\bm P\cdot \bm J)^2+(\bm K\cdot \bm J)^2 + (-\bm D\,\bm J_3-\bm P_1\bm K_2 + \bm P_2 \bm K_1)^2 +(\bm D\,\bm J_2 - \bm P_1\bm K_3+\bm P_3\bm K_1)^2 \nn\\
    &~~+ (-\bm D\,\bm J_1 -\bm P_2\bm K_3+\bm P_3\bm K_2)^2
\end{align}
We parametrize the Casimir eigenvalues as \cite{Boers}\footnote{Notice these eigenvalues are invariant if we replace $\Delta \to \Delta^s=3-\Delta$ (cf. \eqref{shadow}), reflecting the fact that $\Delta$ and $\Delta^s$ describe the same underlying representation}
$$\mathcal{C}^{(2)} =-\left( \Delta(\Delta - 3) + s(s + 1)\right)$$
$$\mathcal{C}^{(4)} = -s(s + 1)(\Delta - 1)(\Delta - 2)$$
Unitary representations (positive-definite inner products in the representation space), restrict the parameters $(\Delta,s)$ to the following three series \cite{dS4meta,otto,sch,Bonifacio}: \\
. {\sf  Principal Series.  Heavy fields}:   $s \in \{0, \frac{1}{2}, 1, \frac{3}{2}, 2, \dots\}$ and $\Delta = \frac{3}{2} + i\rho$, $\rho \in \mathbb{R}$.\\
. {\sf  Complementary Series. Light fields}: only exists for boson fields

Scalars: $s=0$ and  $\Delta \in (0, 3)$  

Rank-$s$ tensors: $s=1,2,...$ and $\Delta \in (1, 2)$.\\
. {\sf Discrete Series. Massless and Partially Massless fields}: there appear gauge redundancies which elliminate intermediate helicity states for rank-$s$ tensors. They require specific integer or half-integer real scaling dimension $\Delta$. 

Scalars: $s=0$ and $\Delta=3+n$ with $n\in\mathbb N$

Massless  (2 pdof): exist for $s \ge \frac{1}{2}$ when $\Delta = s + 1$.

Partially Massless Fields ($2t+2$ pdof): exist for $s=2,\frac52,3,...$  at   $\Delta=s+1-t$ 
with 

$t=1,2,...,[s-1]$.\\
The states at $\Delta = 2$ for rank-$s$ tensors correspond to the Partially Massless state of maximal depth ($t = s-1$). In the de Sitter literature, this exact point is the bottom of the Higuchi bound \cite{HiguPhD}
\subsection{Spin-$\frac{1}{2}$}
\label{dS4Lies1/2}
 Covariant derivatives:
\begin{eqnarray*}
  &\nabla_T \Psi= \partial_T \Psi , \quad \nabla_{\theta_3}\Psi= \partial_{\theta_3} \Psi + \frac{1}{2} \cot T \, \gamma^{\ud{03}} \Psi\\
    &\nabla_{\theta_2} \Psi = \partial_{\theta_2} \Psi +\frac{1}{2}\cot T \sin \theta_3 \, \gamma^{\ud {02}}\,\Psi- \frac{1}{2}\cos \theta_3 \, \gamma^{\ud{32}} \,\Psi  \\
    &\nabla_{\phi} \Psi = \partial_\phi \Psi + \frac{1}{2}\cot T \sin \theta_2 \sin \theta_3 \, \gamma^{\ud{01}}\Psi - \frac{1}{2}\cos \theta_3 \sin \theta_2  \, \gamma^{\ud{31}}\Psi - \frac{1}{2}\cos \theta_2 \,\gamma^{\ud{21}}\Psi
\end{eqnarray*}
\nin Spinor Lie derivatives give a representation of the SO(4,1) algebra 
\begin{eqnarray*}
 &\mathcal{L}_{J_1} \Psi=  {J_1}^\mu\partial_\mu \Psi - \frac{1}{2} \csc \theta_2 \cos \phi \, \gamma^{\ud{21}}\,\Psi,~~\mathcal{L}_{J_2} \Psi=  {J_2}^\mu\partial_\mu \Psi - \frac{1}{2} \csc \theta_2 \sin \phi \, \gamma^{\ud{21}}\,\Psi  \\
 &\mathcal{L}_{J_3}\Psi = - \partial_\phi \Psi 
\end{eqnarray*}
\begin{equation*}
    \mathcal{L}_{ D} \Psi =  {D}^\mu\partial_\mu \Psi - \tfrac{1}{2}  \sin T \sin \theta_3 \gamma^{\ud{03}}\,\Psi
\end{equation*}
\begin{align*}
 \mathcal{L}_{P_1}\Psi &=   {P_1}^\mu \partial_\mu \Psi  - \tfrac{1}{2} \csc \theta_3   \cos \theta_2 \cos\phi \, \gamma^{\ud{32}} \,\Psi+ \tfrac{1}{2} \csc \theta_3 \sin\phi  \, \gamma^{\ud{31}} \, \Psi \nn\\
&~~+ \tfrac{1}{2}  \cot \theta_3 \cot \theta_2 \sin \phi \, \gamma^{\ud{21}} \,\Psi \\
\mathcal{L}_{P_2}\Psi&= {P_2}^\mu \partial_\mu \Psi  -\tfrac{1}{2} \csc \theta_3 \cos \theta_2\sin \phi  \, \gamma^{\ud{32}} \, \Psi - \tfrac{1}{2} \csc \theta_3\cos \phi   \, \gamma^{\ud{31}} \,\Psi \nn\\ 
    &~~- \tfrac{1}{2} \cot \theta_3 \cot \theta_2 \cos \phi \,\gamma^{\ud{21}} \,\Psi  \\
 \mathcal{L}_{P_3}\Psi &=  {P_3}^\mu \partial_\mu \Psi + \tfrac{1}{2} \csc \theta_3  \sin \theta_2 \, \gamma^{\ud{32}} \, \Psi
\end{align*}
\begin{align*}
 \mathcal{L}_{K_3}\Psi& =  {K_3}^\mu \partial_\mu \Psi - \tfrac{1}{2} \sin T \cos \theta_3 \cos \theta_2 \, \gamma^{\ud{03}}\, \Psi +\tfrac{1}{2}\sin T \sin \theta_2 \, \gamma^{\ud{02}}\, \Psi \nn\\
   &~~+ \tfrac{1}{2}\cos T \cot \theta_3 \sin \theta_2 \, \gamma^{\ud{32}}\, \Psi\\
\mathcal{L}_{K_2}\Psi &=K_2^{\mu}\partial_\mu \Psi - \tfrac{1}{2}\sin T \cos \theta_3 \sin \theta_2 \sin \phi \, \gamma^{\ud{03}}\,\Psi - \tfrac{1}{2}\sin T \cos \theta_2 \sin \phi \, \gamma^{\ud{02}} \Psi \nonumber \\
&~~+\tfrac{1}{2}\sin T \cos \phi \, \gamma^{\ud{01}}\Psi - \tfrac{1}{2}\cos T \cot \theta_3 \cos \theta_2 \sin \phi \, \gamma^{\ud{32}}\, \Psi \nonumber\\
&~~- \tfrac{1}{2}\cos T \cot \theta_3 \cos \phi \, \gamma^{\ud{31}}\, \Psi-\frac{1}{2}\cos T \csc \theta_3 \cot \theta_2 \cos \phi \, \gamma^{\ud{21}} \, \Psi\\
\mathcal{L}_{K_1}\Psi&=K_1^{\mu}\partial_\mu \Psi -\tfrac{1}{2}\sin T \cos \theta_3 \sin \theta_2  \cos \phi \, \gamma^{\ud{03}} \, \Psi - \tfrac{1}{2}\sin T \cos \theta_2 \cos \phi \, \gamma^{\ud{02}}\Psi \nn\\
&~~+ \tfrac{1}{2} \sin T \sin \phi \, \gamma^{\ud{01}}\Psi - \tfrac{1}{2}\cos T \cot \theta_3 \cos \theta_2 \cos \phi \, \gamma^{\ud{32}}\, \Psi \\
&~~ + \tfrac{1}{2} \cos T \cot \theta_3 \sin \phi \, \gamma^{\ud{31}}\Psi + \tfrac{1}{2} \cos T \csc \theta_3 \cot \theta_2 \sin \phi \, \gamma^{\ud{21}} \Psi
\end{align*}
Even though ${\cal C}^{(4)}$ is quartic in the generators,  we find that both Casimirs are linear functions of the spinor Laplacian 
\begin{equation*}
   \left( \mathcal{C}^{(2)} - \nabla^2 \right)\Psi=- \frac{3}{2}\Psi
\end{equation*}
\begin{equation*}
   \left( \mathcal{C}^{(4)} - \frac{3}{4}\nabla^2 \right) \Psi=-\frac{33}{16}\Psi
\end{equation*}
Using the relation between the Dirac square and the Laplacian given in \eqref{mass2}, and plugging  $R=N(N-1)=12$ for dS$_4$ we find
\begin{equation}
    \mathcal{C}^{(2)} \Psi= \left(\frac{3}{2}+ \slashed \nabla^2 \right)\Psi 
\end{equation}
\begin{equation}
     \mathcal{C}^{(4)} \Psi= \left(\frac{3}{16}+ \frac{3}{4}\slashed \nabla^2 \right)\Psi 
\end{equation}
A massive Dirac fermion
$$\slashed\nabla\Psi=M\Psi$$
gives 
$$\Psi_{M} ~~\leadsto~~\left\{
\begin{array}{l}
{\cal C}^{(2)}= M^2+\tfrac{3}{2}  \\[2mm]
{\cal C}^{(4)}=\tfrac34  M^2+\tfrac{3}{16} 
\end{array}\right.~\leadsto~(\Delta,s)=\left(\tfrac32+iM,\tfrac12\right)$$
Contrary to what happened in (odd) 3d, in (even) 4d  the ${\cal C}^{(4)}$ is blind to the sign of the mass. This is expected since $M$ and $-M$ are equivalent irreps being related by the chiral matrix.

\subsection{Spin-$\frac{3}{2}$}

We write the Rarita-Schwinger field components as
\begin{eqnarray}
    \Psi_\mu= (\Psi_T, \Psi_{\theta_3} , \Psi_{\theta_2}, \Psi_\phi).
\end{eqnarray}
The covariant derivatives are
$$  
    \nabla_T\Psi_\nu=  \partial_T \Psi_\nu+ \cot T \, \Psi_\nu
$$ 
\begin{align*}
    \nabla_{\theta_3}\Psi_T&= \partial_{\theta_3} \Psi_T + \tfrac{1}{2} \cot T \, \gamma^{\ud{03}} \, \Psi_T+ \cot T \, \Psi_{\theta_3}\\
    \nabla_{\theta_3}\Psi_{\theta_3}&= \partial_{\theta_3} \Psi_{\theta_3} + \tfrac{1}{2} \cot T \, \gamma^{\ud{03}} \, \Psi_{\theta_3}+ \cot T \, \Psi_{T}\\
    \nabla_{\theta_3}\Psi_{\theta_2}&= \partial_{\theta_3} \Psi_{\theta_2} + \tfrac{1}{2} \cot T \, \gamma^{\ud{03}} \, \Psi_{\theta_2} - \cot \theta_3 \, \Psi_{\theta_2}\\
    \nabla_{\theta_3}\Psi_{\phi}&= \partial_{\theta_3} \Psi_{\phi} + \tfrac{1}{2} \cot T \, \gamma^{\ud{03}} \, \Psi_{\phi} - \cot \theta_3 \, \Psi_{\phi}
\end{align*}
\begin{align*}
    \nabla_{\theta_2} \Psi_t&= \partial_{\theta_2} \Psi_T + \tfrac{1}{2} \cot T \sin \theta_3 \, \gamma^{\ud{02}} \, \Psi_T - \tfrac{1}{2} \cos \theta_3 \, \gamma^{\ud{32}} \, \Psi_T + \cot T \, \Psi_{\theta_2}\\
    \nabla_{\theta_2} \Psi_{\theta_3}& = \partial_{\theta_2} \Psi_{\theta_3} + \tfrac{1}{2} \cot T \sin \theta_3 \, \gamma^{\ud{02}} \, \Psi_{\theta_3} - \tfrac{1}{2} \cos \theta_3 \, \gamma^{\ud{32}} \, \Psi_{\theta_3} - \cot \theta_3 \, \Psi_{\theta_2} \\
    \nabla_{\theta_2} \Psi_{\theta_2}&=  \partial_{\theta_2} \Psi_{\theta_2} + \tfrac{1}{2} \cot T \sin \theta_3 \, \gamma^{\ud{02}} \, \Psi_{\theta_2} - \tfrac{1}{2} \cos \theta_3 \, \gamma^{\ud{32}} \, \Psi_{\theta_2} + \cot T \sin^2 \theta_3 \, \Psi_T   \\
    &~~+ \sin \theta_3 \cos \theta_3 \, \Psi_{\theta_3}\\
\nabla_{\theta_2} \Psi_{\phi}&= \partial_{\theta_2} \Psi_\phi + \tfrac{1}{2} \cot T \sin \theta_3 \, \gamma^{\ud{02}} \, \Psi_\phi - \tfrac{1}{2} \cos \theta_3 \, \gamma^{\ud{32}} \, \Psi_\phi - \cot \theta_2 \, \Psi_\phi
\end{align*}
\begin{align*}
    \nabla_\phi \Psi_{T}& = \partial_\phi \Psi_T + \tfrac{1}{2}\cot T \sin \theta_3 \sin \theta_2 \, \gamma^{\ud{01}}\, \Psi_T - \tfrac{1}{2}\cos \theta_3 \sin \theta_2 \, \gamma^{\ud{31}}\, \Psi_T - \tfrac{1}{2}\cos \theta_2 \, \gamma^{\ud{21}}\, \Psi_T \\
    &~~+ \cot T \, \Psi_\phi\\
   \nabla_\phi \Psi_{\theta_3}&=  \partial_\phi \Psi_{\theta_3} + \tfrac{1}{2}\cot T \sin \theta_3 \sin \theta_2 \, \gamma^{\ud{01}}\, \Psi_{\theta_3} - \tfrac{1}{2}\cos \theta_3 \sin \theta_2 \, \gamma^{\ud{31}}\, \Psi_{\theta_3} - \tfrac{1}{2}\cos \theta_2 \, \gamma^{\ud{21}}\, \Psi_{\theta_3} \\
    &~~- \cot \theta_3 \, \Psi_\phi\\
\nabla_\phi \Psi_{\theta_2}&= \partial_\phi \Psi_{\theta_2} + \tfrac{1}{2}\cot T \sin \theta_3 \sin \theta_2 \, \gamma^{\ud{01}}\, \Psi_{\theta_2} - \tfrac{1}{2}\cos \theta_3 \sin \theta_2 \, \gamma^{\ud{31}}\, \Psi_{\theta_2} - \tfrac{1}{2}\cos \theta_2 \, \gamma^{\ud{21}}\, \Psi_{\theta_2} \\
&~~- \cot \theta_2 \, \Psi_\phi\\
\nabla_\phi \Psi_{\phi}&=\partial_\phi \Psi_{\phi} + \tfrac{1}{2}\cot T \sin \theta_3 \sin \theta_2 \, \gamma^{\ud{01}}\, \Psi_{\phi} - \frac{1}{2}\cos \theta_3 \sin \theta_2 \, \gamma^{\ud{31}}\, \Psi_{\phi} - \tfrac{1}{2}\cos \theta_2 \, \gamma^{\ud{21}}\, \Psi_{\phi} \\
    &~~+\cot T \sin^2 \theta_3 \sin^2 \theta_2 \, \Psi_T + \sin\theta_3 \cos \theta_3 \sin^2 \theta_2 \, \Psi_{\theta_3} + \sin \theta_2 \cos \theta_2 \, \Psi_{\theta_2}
\end{align*}
By means of simplyfing expressions, we will denote $\mathcal{ L}^{(1/2)}$ as the spin-$\frac{1}{2}$ Lie Derivative. These were calculated in \ref{dS4Lies1/2}. With this notation in mind, the Lie derivatives on the spin-$\frac{3}{2}$ field are
\begin{align*}
& \mathcal{L}_{J_1} \Psi_T=  \mathcal{L}^{(1/2)}_{J_1} \Psi_T \\
& \mathcal{L}_{J_1} \Psi_{\theta_3} = \mathcal{L}^{(1/2)}_{J_1} \Psi_{\theta_3} \\
& \mathcal{L}_{J_1} \Psi_{\theta_2} = \mathcal{L}^{(1/2)}_{J_1} \Psi_{\theta_2}-\csc^2 \theta_2 \cos \phi \, \Psi_\phi \\
& \mathcal{L}_{J_1} \Psi_\phi = \mathcal{L}^{(1/2)}_{J_1} \Psi_\phi + \cos \phi \, \Psi_{\theta_2} - \cot \theta_2 \sin \phi \, \Psi_\phi
\end{align*}
\begin{align*}
&   \mathcal{L}_{J_2} \Psi_T =  \mathcal{L}_{J_2}^{(1/2)} \Psi_T \\
&\mathcal{L}_{J_2}  \Psi_{\theta_3} =   \mathcal{L}_{J_2}^{(1/2)} \Psi_{\theta_3} \\
&  \mathcal{L}_{J_2} \Psi_{\theta_2}= \mathcal{L}^{(1/2)}_{J_2}\Psi_{\theta_2}-\csc^2 \theta_2 \sin \phi \, \Psi_\phi  \\
&   \mathcal{L}_{J_2}\Psi_\phi = \mathcal{L}^{(1/2)}_{J_2}\Psi_\phi + \sin \phi \, \Psi_{\theta_2} + \cot \theta_2 \cos \phi \, \Psi_\phi
\end{align*}
\begin{equation}
    \mathcal{L}_{J_3} \Psi_\mu= -\partial_\phi \Psi_\mu = \mathcal{ L}^{(1/2)}_{J_3}\Psi_\mu
\end{equation}
\begin{align*}
&\mathcal{L}_{D} \Psi_T = \mathcal{L}^{(1/2)} \Psi_T + \cos T \cos \theta_3 \, \Psi_T - \sin T \sin \theta_3 \, \Psi_{\theta_3} \\
& \mathcal{L}_D \Psi_{\theta_3} = \mathcal{L}^{(1/2)}  \Psi_{\theta_3} - \sin T \sin \theta_3 \, \Psi_T + \cos T \cos \theta_3 \, \Psi_{\theta_3}\\
& \mathcal{L}_{D}\Psi_{\theta_2} = \mathcal{L}^{(1/2)} \Psi_{\theta_2} \\
&\mathcal{L}_D \Psi_\phi = \mathcal{L}^{(1/2)} \Psi_\phi
\end{align*}
\begin{align*}
 & \mathcal{L}_{P_3} \Psi_T = \mathcal{L}_{P_3}^{(1/2)}\Psi_T \\
 & \mathcal{L}_{P_3} \Psi_{\theta_3}= \mathcal{L}_{P_3}^{(1/2)} \Psi_{\theta_3} + \csc^2 \theta_3 \sin \theta_2 \, \Psi_{\theta_2}\\
 & \mathcal{L}_{P_3} \Psi_{\theta_2} = \mathcal{L}_{P_3}^{(1/2)} \Psi_{ \theta_2} -\sin \theta_2 \, \Psi_{\theta_3}-\cot \theta_3 \cos \theta_2 \, \Psi_{\theta_2} \\
 & \mathcal{L}_{P_3} \Psi_\phi =  \mathcal{L}_{P_3}^{(1/2)}\Psi_\phi 
\end{align*}
\begin{align*}
 &\mathcal{L}_{P_2} \Psi_T=  \mathcal{L}_{P_2}^{(1/2)}\Psi_T \\
 & \mathcal{L}_{P_2} \Psi_{\theta_3} =   \mathcal{L}_{P_2}^{(1/2)} \Psi_{\theta_3}- \csc^2 \theta_3 \cos \theta_2 \sin \phi \, \Psi_{\theta_2}-\csc^2 \theta_3 \csc \theta_2 \cos \phi \, \Psi_\phi \\
 & \mathcal{L}_{P_2} \Psi_{\theta_2}=  \mathcal{L}_{P_2}^{(1/2)} \Psi_{\theta_2}+\cos \theta_2 \sin \phi \, \Psi_{\theta_3} - \cot \theta_3 \sin \theta_2 \sin \phi \, \Psi_{\theta_2} \\
 & \mathcal{L}_{P_2} \Psi_\phi = \mathcal{L}_{P_2}^{(1/2)} \Psi_\phi + \sin \theta_2 \cos \phi \, \Psi_{\theta_3} + \cot \theta_3 \cos \theta_2 \cos \phi \, \Psi_{\theta_2}- \cot \theta_3 \csc \theta_2 \sin \phi \, \Psi_\phi
\end{align*}
\begin{align*}
&\mathcal{L}_{P_1}\Psi_T = \mathcal{L}_{P_1}^{(1/2)}\Psi_T \\
& \mathcal{L}_{P_1} \Psi_{\theta_3} = \mathcal{L}_{P_1}^{(1/2)}   \Psi_{\theta_3}- \csc^2 \theta_3 \cos \theta_2 \cos \phi \, \Psi_{\theta_2} + \csc^2 \theta_3 \csc \theta_2 \sin \phi \, \Psi_\phi \\
& \mathcal{L}_{P_1} \Psi_{\theta_2} = \mathcal{L}_{P_1}^{(1/2)} \Psi_{\theta_2} + \cos \theta_2 \cos \phi \, \Psi_{\theta_3}- \cot \theta_3 \sin \theta_2 \cos \phi \, \Psi_{\theta_2} + \cot \theta_3 \cot \theta_2 \csc \theta_2 \sin \phi \, \Psi_\phi \\
& \mathcal{L}_{P_1} \Psi_{ \phi} = \mathcal{L}_{P_1}^{(1/2)} \Psi_\phi - \sin \theta_2 \sin \phi \, \Psi_{\theta_3}- \cot \theta_3 \cos \theta_2 \sin \phi \, \Psi_{\theta_2}- \cot \theta_3 \csc \theta_2 \cos \phi \, \Psi_\phi
\end{align*}
\begin{align*}
 & \mathcal{L}_{K_3} \Psi_T=  \mathcal{L}_{K_3}^{(1/2)} \Psi_T - \cos T \sin \theta_3 \cos \theta_2 \Psi_T - \sin T \cos \theta_3 \cos \theta_2 \, \Psi_{\theta_3} + \sin T \csc \theta_3 \sin \theta_2 \Psi_{\theta_2}   \\
 &  \mathcal{L}_{K_3} \Psi_{\theta_3} =  \mathcal{L}_{K_3}^{(1/2)} \Psi_{\theta_3}- \sin T \cos \theta_3 \cos \theta_2 \Psi_T - \cos T \sin \theta_3 \cos \theta_2 \, \Psi_{\theta_3} + \cos T \cot \theta_3 \csc \theta_3 \sin \theta_2 \, \Psi_{\theta_2} \\
 & \mathcal{L}_{K_3} \Psi_{\theta_2}= \mathcal{L}^{(1/2)} \Psi_{\theta_2} + \sin T \sin \theta_3 \sin \theta_2 \, \Psi_T - \cos T \cos \theta_3 \sin \theta_2 \, \Psi_{\theta_3} - \cos T \csc\theta_3 \cos \theta_2 \, \Psi_{\theta_2} \\
 & \mathcal{L}_{K_3} \Psi_{\phi} = \mathcal{L}_{K_3}^{(1/2)} \Psi_\phi   
\end{align*}
\begin{align*}
\mathcal{L}_{K_2} \Psi_T&=  \mathcal{L}_{K_2}^{(1/2)}\Psi_T-\cos T \sin \theta_3 \sin \theta_2 \sin \phi \Psi_T - \sin T \cos \theta_3 \sin \theta_2 \sin \phi \, \Psi_{\theta_3}\\
&~~ - \sin T \csc \theta_3 \cos \theta_2 \sin \phi \Psi_{\theta_2} - \sin T \csc \theta_3 \csc \theta_2 \cos \phi \, \Psi_\phi \\
\mathcal{L}_{K_2} \Psi_{\theta_3} &= \mathcal{L}_{K_2}^{(1/2)}\Psi_{\theta_3}-\sin T \cos \theta_3 \sin \theta_2 \sin \phi \, \Psi_T - \cos T \sin \theta_3 \sin \theta_2 \sin \phi \, \Psi_{\theta_3} \\
&~~ - \cos T \cot \theta_3 \csc \theta_3 \cos \theta_2 \sin \phi \, \Psi_{\theta_2}- \cos T \cot \theta_3 \csc \theta_3 \csc \theta_2 \cos \phi \, \Psi_\phi \\
\mathcal{L}_{K_2} \Psi_{\theta_2} &= \mathcal{L}_{K_2}^{(1/2)} \Psi_{\theta_2}- \sin T \sin \theta_3 \cos \theta_2 \sin \phi \, \Psi_T + \cos T \cos \theta_3 \cos \theta_2 \sin \phi \, \Psi_{\theta_3} \\
&~~- \cos T \csc \theta_3 \sin \theta_2 \sin \phi \, \Psi_{\theta_2} - \cos T \csc \theta_3 \csc \theta_2 \cot \theta_2 \cos \phi \, \Psi_\phi\\
\mathcal{L}_{K_2} \Psi_\phi &= \mathcal{L}_{K_2}^{(1/2)}\Psi_\phi - \sin T \sin \theta_3 \sin \theta_2 \cos \phi \, \Psi_T + \cos T \cos \theta_3 \sin \theta_2 \cos \phi \, \Psi_{\theta_3 }\\
&~~+ \cos T \csc \theta_3 \cos \theta_2 \cos \phi \, \Psi_{\theta_2} - \cos T \csc \theta_3 \csc \theta_2 \sin \phi \, \Psi_\phi
\end{align*}
\begin{align*}
    \mathcal{L}_{K_1} \Psi_{T}&= \mathcal{L}_{K_1}^{(1/2)}\Psi_T  - \cos T \sin \theta_3 \sin \theta_2 \cos \phi \, \Psi_T - \sin T \cos \theta_3 \sin \theta_2 \cos \phi \, \Psi_{\theta_3} \\
&~~ - \sin T \csc \theta_3 \cos \theta_2 \cos \phi \, \Psi_{\theta_2} + \sin T \csc \theta_3 \csc \theta_2 \sin \phi \, \Psi_\phi \\
\mathcal{L}_{K_1} \Psi_{\theta_3} &= \mathcal{L}^{(1/2)}_{K_1} \Psi_{\theta_3} - \sin T \cos \theta_3 \sin \theta_2 \cos \phi \, \Psi_T - \cos T \sin \theta_3 \sin \theta_2 \cos \phi \, \Psi_{\theta_3} \\
&~~-\cos T \cot \theta_3 \csc \theta_3 \cos \theta_2 \cos \phi \, \Psi_{\theta_2 }+ \cos T \cot \theta_3 \csc \theta_3 \csc \theta_2 \sin \phi \, \Psi_\phi \\
\mathcal{L}_{K_1}\Psi_{\theta_2} &= \mathcal{L}^{(1/2)}_{K_1}\Psi_{\theta_2} - \sin T \sin \theta_3 \cos \theta_2 \cos \phi \, \Psi_T + \cos T \cos \theta_3 \cos \theta_2 \cos \phi \, \Psi_{\theta_3} \\
&~~- \cos T \csc \theta_3 \sin \theta_2 \cos \phi \, \Psi_{\theta_2} + \cos T \csc \theta_3 \csc \theta_2 \cot \theta_2 \sin \phi \, \Psi_\phi \\
\mathcal{L}_{K_1} \Psi_\phi &= \mathcal{L}^{(1/2)}_{K_1} \Psi_\phi + \sin T \sin \theta_3 \sin \theta_2 \sin \phi \,\Psi_T - \cos T \cos \theta_3 \sin \theta_2 \sin \phi \, \Psi_{\theta_3 }  \\
&~~ - \cos T \csc \theta_3 \cos \theta_2 \sin \phi \, \Psi_{\theta_2} - \cos T \csc \theta_3 \csc \theta_2 \cos \phi \, \Psi_{\phi}
\end{align*}
The quadratic Casimir \eqref{C2ds4Cas} is related to the spin-vector Laplacian in the following way
\begin{equation}
    \mathcal{C}^{(2)} \Psi_\mu-\nabla^2 \Psi_\mu - \gamma_\mu (\gamma \cdot \Psi)=-\frac{11}{2} \Psi_\mu \, \,.
    \label{C14dimpsi}
\end{equation}
Finally, we note that, when evaluated in the spin-vector representation, the quartic Casimir reduces to a second-order differential operator. However, its precise relation to other quadratic operators remains to be established. We leave a detailed analysis of this relation for future work. Nevertheless, the fact that the quartic Casimir becomes a second-order differential operator, together with the existence in four dimensions of a chiral matrix relating solutions with masses $M$ and $-M$, suggests that $\mathcal{C}^{(4)}\Psi_\mu$ may be expressible in terms of $\nabla^2\Psi_\mu$ plus additional quadratic terms that vanish upon imposing the transverse-traceless (TT) conditions, along the lines of Eq.~\ref{C14dimpsi}.

\end{document}